%% file: JHEP.tex
\documentclass{article}
\usepackage{amsmath}
\usepackage{jheppub}
\usepackage{graphicx} 
\usepackage{feynmf} 
\usepackage{color}

\usepackage{amssymb} 
\usepackage{wasysym}
\usepackage{tikz}
\usepackage{subcaption}
\usepackage[numbers]{natbib}

\definecolor{lmugreen}{RGB}{0, 136, 58}

\usepackage{hyperref}
\hypersetup{
    bookmarks=false,         
pdfstartview={FitH},    
    pdftitle={Resummation of Cosmological Correlators and UV-Regularization},    
    pdfauthor={Matthias Nowinski, Ivo Sachs},     
    pdfsubject={Resummation of Cosmological Correlators and UV-Regularization},   
    pdfcreator={Matthias Nowinski, Ivo Sachs},   
    pdfproducer={Matthias Nowinski, Ivo Sachs}, 
    pdfnewwindow=true,      
    colorlinks=true,       
    linkcolor=lmugreen,          
    citecolor=lmugreen,        
    filecolor=lmuorange,      
    urlcolor=lmugreen           
}

\renewcommand{\d}{\mathrm{d}}
\newcommand{\e}{\mathrm{e}}
\renewcommand{\O}{\mathcal{O}}

\newcommand{\lambdaR}{\lambda_\mathrm{R}}

\DeclareUnicodeCharacter{2212}{\textendash}

\preprint{\vbox{\hbox{LMU-ASC 18/25}}}

\title{Resummation of Cosmological Correlators \\ and their  UV-Regularization}

\author[a]{Matthias Nowinski}
\author[a]{and Ivo Sachs} 

\affiliation[a]{Arnold-Sommerfeld-Center for Theoretical Physics, Ludwig-Maximilians-Universit\"at M\"unchen, Theresienstr. 37, D-80333 Munich, Germany}

\emailAdd{m.nowinski@lmu.de}

\abstract{Cosmological correlators are important observables in cosmology. They are often approximated by de Sitter space correlators. In this paper, we give a first precise diagrammatical computation of higher loop diagrams to all orders for a conformally coupled scalar in four dimensions. We show that, in contrast to flat space, diagrams of necklace topology do not resum using the natural application of de Sitter-invariant regularization and are thus hard to evaluate. We propose a modification to the UV-regularization of the loops, compatible with de Sitter invariance, but much easier to work with. The modified diagrams can be resummed to give a glimpse at non-perturbative effects for de Sitter correlators. Furthermore, they fit nicely with recently proposed cosmological dressing rules.}
\begin{document}

\maketitle

\section{Introduction}

Cosmology is becoming a more and more quantitative field in recent years with numerous observations of the cosmic microwave background (CMB) \cite{collaboration_planck_2020,hinshaw_nine-year_2013} and the large scale structure of the universe (LSS) \cite{cabass_boss_2025} being completed. More precise experiments \cite{dore_cosmology_2015}, as well as additional observational channels are in planning, for example using gravitational waves with LISA \cite{amaro-seoane_elisa_2012} or the so-called 21cm cosmology \cite{pritchard_21_2012}. The observables that all these experiments are aiming to measure are \textit{cosmological correlators}, i.e.\ the statistics of distributions of matter, in the case of the LSS, or temperature fluctuations, in the case of the CMB, at cosmological scale, with the goal of extracting the nature of primordial fluctuations.

By now there is ample evidence that these statistics have a quantum field theoretical origin \cite{Mukhanov:1981xt} within the cosmological era called \textit{inflation} \cite{starobinsky_new_1980,guth_inflationary_1981,linde_new_1982}, where the universe was approximately described a de Sitter geometry. We model de Sitter space in conformal coordinates $(\eta, \vec x)$ using the metric
\begin{align}
    \d s^2 = \frac{1}{\eta^2 H^2} \big( - \d \eta^2 + \d \vec x^2 \big).
\end{align}
To generate predictions for these cosmological correlators, we therefore turn to quantum field theory (QFT) in de Sitter space, and compute quantum mechanical correlation functions of the quantum fields $\phi$, dependent on momentum $\vec k$, of the shape
\begin{align}
    \langle \psi | \phi(\vec k_1) \ldots \phi(\vec k_N) | \psi \rangle. \label{eq:correlator}
\end{align}
As opposed to a typical QFT scattering problem, where we have different in- and out-states, the states left and right of this expectation value are the same, $|\psi\rangle$, and usually given by the time evolution of some initial state. We are using the \textit{Bunch-Davies vacuum}. There are different approaches to evaluate \eqref{eq:correlator}, for example via the \textit{wavefunction of the universe} approach (e.g. \cite{hartle_wave_1983,Maldacena:2002vr,Arkani-Hamed:2015bza,arkani-hamed_cosmological_2019,Goodhew:2020hob,Goodhew:2021oqg,Benincasa:2022gtd}), the \textit{Schwinger-Keldysh} formalism (e.g. \cite{sadovskii_diagram_2023,Chen:2017ryl}), and spectral representation (e.g. \cite{Marolf:2010zp,Cacciatori:2024zrv, pietro_cosmological_2025,Werth:2024mjg}). In this paper, we are using the \textit{shadow action} formalism, which can be derived from the Schwinger-Keldysh perturbation theory by Wick rotating to Euclidean Anti-de Sitter (EAdS) \cite{pietro_analyticity_2022,sleight_ads_2021}  with coordinates $(z,\vec x)$ and metric
\begin{align}
    \d s^2 = \frac{1}{z^2 H^2} \big( \d z^2 + \d \vec x^2 \big). \label{eq:EAdS_metric}
\end{align}
The goal of this paper is to go beyond perturbation theory for these cosmological correlators, by resumming an infinite series of higher loop diagrams. The study of non-perturbative effects is an open problem even in the simpler flat space scenario, however, a lot of work has been directed towards it. We aim to import some of the results from flat space to the area of cosmological correlators. To keep this discussion simple, we consider only conformally coupled scalar fields $\phi$ with a $\lambda \phi^4$ interaction in perfect de Sitter space.

A lot of work on cosmological correlators is done in position space, which lends itself more easily to Mellin-Barnes representation and variable masses of the involved fields. However, since we restrict ourselves to conformally coupled scalars, we can use the machinery of the previous works \cite{chowdhury_subtle_2024,chowdhury_cosmological_2025} for calculations in momentum space. This allows us to compare our results to- and import ideas from flat space more easily, since flat space computations are mostly performed in momentum space.

By "importing" flat space considerations of non-perturbative effects, we mean that we will consider necklace diagrams of $N$ loops in a straight chain in momentum space, as depicted in the figure in equation \eqref{eq:n_loop_diagrams}. The study of these particular diagrams is motivated by large-$N$ models \cite{hooft_planar_1974}. These models have a large number $N$ of scalar fields, which favors a certain type of diagrams. Related discussion in position space can be found in \cite{sachs_bulk_2023} for 4-dimensional Anti-de Sitter space, and in \cite{pietro_cosmological_2025} for de Sitter correlators. In the Anti-de Sitter study, it has already been demonstrated that interpreting resummed necklace diagrams as non-perturbative physics leads to unexpected results, which favors a negative $\lambda \phi^4$ coupling  \cite{sachs_bulk_2023}. We will observe  qualitative differences in the non-perturbative completion for positive and negative couplings in momentum space as well.

In addition, there are quite interesting technical caveats in the momentum space calculation, related to regularization (e.g. \cite{gorbenko_4_2019}). In  previous literature \cite{chowdhury_subtle_2024,chowdhury_cosmological_2025}, an analytical regularization was introduced, which mimics dimensional regularization in flat space in the sense that it changes the power of the loop momentum integrand in order to make the integral convergent, but preserves the de Sitter isometry. The main aspects of the regularization can be seen in the Feynman rules (\ref{eq:bulk_to_boundary_prop}$-$\ref{eq:feynman_rules_odd_coupling}). 

It is then possible to write down some expression for the regularized $N$-loop necklace diagram (see \eqref{eq:N_loop_renormalized_final}) at an integrand level with further integrals still to be performed. These integrals are made quite complicated due to the  regularization, as illustrated in the appendices. In order to have a chance of resumming the diagrams, we therefore introduce a novel regularization scheme in section \ref{sec:new_regularization}. Our new scheme can be viewed as de Sitter invariant modification of analytical regularization \cite{chowdhury_subtle_2024}, where the Feynman rules depend on the topology of the diagram. This prescription is rather natural from the prespective of the "dressing rules" for cosmological correlators, introduced in \cite{chowdhury_cosmological_2025}, where the "auxiliary propagators" distinguish between external and internal momentum. For this reason, our new regularization seems to fit perfectly with the cosmological dressing rules introduced there. While the $N$-loop diagrams are resummable as integrands, we are still left with an integral over the energy, as well as the conformal times of the endpoints of the graph. These integrals can be performed analytically at the 1- and 2-loop level, however, this is not so for the resummed correlator, and we have to resort to a numerical integration. We find that, while the real part of the amplitude does not feature any sizeable cosmological signal \cite{Arkani-Hamed:2015bza} in the shape function (the non-perturbative log-tail is suppressed compared to perturbation theory), the overall sign of the corelation function changes at finite $\lambda\phi^4$-coupling! This can be seen to be related to the tachyonic pathology of the large $N$ model in flat space \cite{Coleman:1974jh}.

Throughout this paper, we make many arguments about necklace diagrams that boil down to what is called "strings" in the computer science sense. We represent properties of the necklace diagrams, for example the field content of the constitutory loops, by a series of "characters", for example $+$ and $-$ or \texttt{M} and \texttt{P}. This notation allows us to focus on the important parts of the given arguments. Since flat space necklace diagrams are very simple, this kind of argument seems to arise only in de Sitter necklace diagrams.

In section \ref{sec:preliminaries}, we state the Feynman rules of the shadow action formalism in momentum space in analytic regularization \cite{chowdhury_subtle_2024}, and repeat some of the computational steps to derive the 1-loop 4-point function of a scalar field, including its renormalization. Building on top of that, we show in section \ref{sec:n_loop} that the same steps that allowed us to compute the 1-loop result generalize in some sense to $N$-loop necklace diagrams. However, we also demonstrate that the remaining integrals are hard to evaluate in general, and highlight some of the main problems. The more lengthy computations of this section are outsourced to appendices \ref{app:N_loop_derivation} and \ref{app:3_loop_details}. Because of these difficulties, we propose a different regularization scheme that is still de Sitter invariant, but produces more manageable expressions, in section \ref{sec:new_regularization}. Using this regularization scheme, we show that it is indeed possible to resum all $N$-loop necklace diagrams. We also discuss interpretations with regard to the new dressing rules formalism, and give directions to proceed further. The conclusions are drawn in section \ref{sec:conclusions}.

\section{Preliminaries} \label{sec:preliminaries}

In this section, we recapitulate some of the previous results for momentum space loops of a conformally coupled scalar field in de Sitter space from \cite{chowdhury_subtle_2024}. We state the Feynman rules of the shadow action formalism and the mechanism for analytical regularization. Then, we compute the 1-loop 4-point Witten diagram and demonstrate how the analytical regularization works, by explicitly renormalizing the 1-loop result.

\subsection{Feynman Rules}

We want to compute correlation functions of a scalar field $\phi$ with an effective mass $m$ and a conformal dimension $\Delta_\pm$ in $d$ space dimensions. Since cosmological correlators \eqref{eq:correlator} are in-in expectation values, it can be shown that they can be computed using Feynman rules with two types of vertices, one we call \textit{time-ordered} and one we call \textit{anti-time-ordered}. Depending on whether a propagator connects two of the same vertices, two different vertices or is an external line, the propagator is different \cite{weinberg_quantum_2005}. These Feynman rules quickly generate many different diagrams even at low orders in perturbation theory, and often some of the diagrams are related by complex conjugation, which can cause unnecessary effort in computations. To evade these complications, a more efficient way of computing correlators was found in \cite{pietro_analyticity_2022}. There, a Wick rotation from de Sitter space to Euclidean Anti-de Sitter space (EAdS), together with a careful reparametrization of the field $\phi$ is performed. As a result, there are no longer two types of vertices, but two types of fields, $\phi_-$ and $\phi_+$. The action of this new theory is sometimes called shadow action and given by \cite{pietro_analyticity_2022}
\begin{align}
    i S &= \int\limits_0^\infty \frac{\d z \d^d x}{z^{d+1}} \Bigg( \sin\Big( \pi \big( \Delta_+ - \frac{d}{2}\big) \Big) \Big( (\partial \phi_+)^2 - m^2 \phi_+^2 \Big) + \sin\Big( \pi \big( \Delta_- - \frac{d}{2}\big) \Big) \Big( (\partial \phi_-)^2 - m^2 \phi_-^2 \Big) \notag\\
    &\hspace{2cm}+ \e^{i\pi \frac{d-1}{2}} V\Big( \e^{-i \frac{\pi}{2} \Delta_+} \phi_+ + \e^{-i \frac{\pi}{2} \Delta_-} \phi_- \Big) + \e^{-i\pi \frac{d-1}{2}} V\Big( \e^{i \frac{\pi}{2} \Delta_+} \phi_+ + \e^{i \frac{\pi}{2} \Delta_-} \phi_- \Big) \Bigg). \label{eq:general_action}
\end{align}
As a remnant of the original two types of vertices, we can see two interaction terms with the same potential $V(\phi)$ in the second line. The first of these comes from the time-ordered vertices, and the second from the anti-time-ordered vertices, which will become important later in section \ref{sec:resummation}.

We will mainly use analytical regularization throughout this paper. As argued in \cite{chowdhury_subtle_2024}, analytical regularization can be interpreted as a modification of the conformal dimensions $\Delta_\pm$ by a small regularization parameter $\kappa > 0$. Therefore, we insert the conformal dimensions
\begin{align}
    \Delta_\pm = \frac{d}{2} \pm \sqrt{\Big(\frac{d}{2}\Big)^2 - \Big( \frac{m}{H} \Big)^2} - \kappa
\end{align}
in $d=3$ space dimensions into the action \eqref{eq:general_action}. Using the potential $V(\phi) = 2\lambda \phi^4/4!$, we then expand the action to second order in small $\kappa$. We use the abbreviation $\mathcal{C}_{\kappa} = 1 - (\pi\kappa)^2/2$ and a rescaling $\phi_\pm \rightarrow \phi_\pm / \sqrt 2$ to get closer to the canonical normalization of the kinetic terms, and find
\begin{align}
    i S &= \int\limits_0^\infty \frac{\d z \d^3 x}{z^{4}} \Bigg( \frac{\mathcal{C}_\kappa}{2} \Big( (\partial \phi_+)^2 - m^2 \phi_+^2 \Big) - \frac{\mathcal{C}_\kappa}{2} \Big( (\partial \phi_-)^2 - m^2 \phi_-^2 \Big) \notag\\
    &\hspace{3cm}+ \frac{\mathcal{C}_{2\kappa}\lambda}{4!} \Big( -\phi_+^4 + 6\phi_-^2 \phi_+^2 - \phi_-^4 \Big) + \frac{\lambda\pi\kappa}{3}\Big( \phi_-\phi_+^3 - \phi_-^3\phi_+ \Big) \Bigg). \label{eq:action_final}
\end{align}
From this action, we deduce the Feynman rules for the vertices. The propagators are a bit more complicated, since they are designed to resemble flat space propagators with a stronger fall-off behavior for large momenta by putting $\kappa$ into the exponent of the denominator of the integrand in \eqref{eq:feynman_rules_propagator-} and \eqref{eq:feynman_rules_propagator+}. To restore invariance under rescaling $z \rightarrow \eta z$, $p, k \rightarrow 1/\lambda p, 1/\lambda k$, we also have to introduce $\kappa$ into the exponent of the conformal times $z$ in front of the integral.

The correlation functions that we compute are evaluated at future infinity. Due to the boundary conditions of the $\phi_\pm$ fields, this implies that the leading contribution is coming from $\phi_-$ correlators \cite{pietro_analyticity_2022}. Therefore, we will only consider external propagators of $\phi_-$ fields. These external propagators have to be normalized correctly, which results in the destinction between bulk-to-boundary propagators \eqref{eq:bulk_to_boundary_prop} and internal bulk-to-bulk propagators \eqref{eq:feynman_rules_propagator-} and \eqref{eq:feynman_rules_propagator+}. In the end, we have the same Feynman rules as \cite{chowdhury_subtle_2024}:
\begin{equation}
\begin{fmffile}{bulk_to_boundary}
    \parbox{200pt}{%
    \begin{fmfgraph*}(80,80)
       \fmfsurround{i1,i2,i3,i4,i5,i6,i7}
       \fmfshift{10 down}{i1}
       \fmfshift{10 down}{i2}
       \fmfshift{5 down}{i3}
       \fmfshift{5 down}{i4}
       \fmfshift{5 down}{i5}
       \fmfshift{5 down}{i6}
       \fmfshift{5 down}{i7}
       \fmfshift{5 right}{i1}
       \fmfshift{5 right}{i2}
       \fmfshift{5 right}{i3}
       \fmfshift{5 right}{i4}
       \fmfshift{5 right}{i5}
       \fmfshift{5 right}{i6}
       \fmfshift{5 right}{i7}
       \fmfright{r1,r2,r3,r4}
       \fmf{dbl_plain,right=0.25,tension=0}{i2,i3,i4,i5}
       \fmf{dashes, label=$\vec k$,tension=0}{i4,r2}
       \fmfv{l=$z$}{r2}
       \fmfforce{(0.7w,0.3h)}{r2}
       \fmfforce{(0.15w,0.8h)}{i4}
       \fmfforce{(0.55w,0.98h)}{i3}
       \fmfforce{(0.9w,0.8h)}{i2}
    \end{fmfgraph*}}
     \hspace{-4cm} = -\frac{(zH)^{1-\kappa}}{\pi C_\kappa} \int\limits_{-\infty}^\infty \d p \frac{\cos(p z)}{(p^2+\vec k^2)^{1+\kappa}} = \bar G_\kappa(\vec k, z) = -\frac{zH}{2k}\e^{-kz} + \O(\kappa),
\end{fmffile}
\label{eq:bulk_to_boundary_prop}
\end{equation}
\vspace{-1.5cm}
\begin{equation}
\begin{fmffile}{phi_minus_prop}
    \parbox{200pt}{
    \begin{fmfgraph*}(80,80)
       \fmfleft{i}
       \fmfright{o}
       \fmf{dashes, label=$\vec k$}{i,o}
       \fmfv{l=$z_1$}{i}
       \fmfv{l=$z_2$}{o}
    \end{fmfgraph*}}
     \hspace{-3cm} = -\frac{(z_1 z_2 H^2)^{1-\kappa}}{\pi C_\kappa} \int\limits_{-\infty}^\infty \d p \frac{\cos(p z_1) \cos(p z_2)}{(p^2+\vec k^2)^{1+\kappa}},
\end{fmffile} \label{eq:feynman_rules_propagator-}
\end{equation}
\vspace{-1.5cm}
\begin{equation}
\begin{fmffile}{phi_plus_prop}
    \parbox{200pt}{
    \begin{fmfgraph*}(80,80)
       \fmfleft{i}
       \fmfright{o}
       \fmf{plain, label=$\vec k$}{i,o}
       \fmfv{l=$z_1$}{i}
       \fmfv{l=$z_2$}{o}
    \end{fmfgraph*}}
     \hspace{-3cm} = \frac{(z_1 z_2 H^2)^{1-\kappa}}{\pi C_\kappa} \int\limits_{-\infty}^\infty \d p \frac{\sin(p z_1) \sin(p z_2)}{(p^2+\vec k^2)^{1+\kappa}},
\end{fmffile} \label{eq:feynman_rules_propagator+}
\end{equation}
\vspace{-0.5cm}
\begin{equation*}
\begin{fmffile}{plus_plus_plus_plus_vert}
    \parbox{200pt}{
    \includegraphics{plus_plus_plus_plus_vert.1.mps}
    }
     \hspace{-5cm} = \frac{\lambda}{4!} C_{2\kappa}, \quad
     \parbox{200pt}{
    \includegraphics{plus_plus_plus_plus_vert.2.mps}
    }
     \hspace{-5cm} = -\frac{6\lambda}{4!} C_{2\kappa}, \quad
     \parbox{200pt}{
    \includegraphics{plus_plus_plus_plus_vert.3.mps}
    }
     \hspace{-5cm} = \frac{\lambda}{4!} C_{2\kappa},
\end{fmffile}
\end{equation*}

\begin{equation}
\begin{fmffile}{plus_plus_plus_minus_vert}
    \parbox{200pt}{
    \includegraphics{plus_plus_plus_minus_vert.1.mps}
    }
     \hspace{-5cm} = -\frac{8\pi\kappa\lambda}{4!}, \quad
     \parbox{200pt}{
    \includegraphics{plus_plus_plus_minus_vert.2.mps}
    }
     \hspace{-5cm} = \frac{8\pi\kappa\lambda}{4!}.
\end{fmffile} \label{eq:feynman_rules_odd_coupling}
\end{equation}
Note that every vertex implicitly comes with an integral over the corresponding Euclidean conformal time $z$ and the measure of EAdS, $1/(zH)^4$, which can be read off from \eqref{eq:EAdS_metric}. Loop momenta $\vec \ell$ are integrated over with the measure $\d^3 \ell/(2\pi)^3$, like in 3-dimensional flat space. 

With these Feynman rules, we now compute the 4-point $\phi_-$ correlation function as an expansion in Witten diagrams, which are Feynman diagrams in EAdS space. The fact that the correlators are evaluated at future infinity is indicated by the symbolical future boundary in a circle around the diagram
\begin{align}
\begin{fmffile}{sample_diagram}
    \langle \phi_- (\vec k_1) \phi_-(\vec k_2) \phi_-(\vec k_3) \phi_-(\vec k_4) \rangle\Big\rvert_{z = 0} = \parbox{45 pt}{
    \includegraphics{sample_diagram.1.mps}
    } \ + \parbox{45 pt}{
    \includegraphics{sample_diagram.2.mps}}
\ +  \ldots  +  \parbox{45pt}{
    \includegraphics{sample_diagram.3.mps}} \ + \parbox{45pt}{
    \includegraphics{sample_diagram.4.mps}} \  + \ldots.
\end{fmffile} \notag
\end{align}
We will only consider connected contributions for the rest of this paper, since they are more theoretically interesting.

\subsection{1-Loop Recap}

Before going all the way to the $N$-loop diagrams, we present the 1-loop computation, as this involves most of the important steps for the general case in a simpler environment. This computation has already been presented before in \cite{chowdhury_subtle_2024}, but we repeat it here to keep this discussion self-contained. 

Before we start, we introduce some abbreviations for the kinematic variables that will often appear. Since we are only considering 4-point functions, we always have four external momenta $\vec k_1, \ldots, \vec k_4$. Combinations of these are
\begin{equation}
\begin{split}
    k_j &= | \vec k_j |, \\
    k_{12} &= k_1 + k_2, \\
    k_{34} &= k_3 + k_4, \\
    \vec k &= \vec k_1 + \vec k_2 = \vec k_3 + \vec k_4, \\
    k &= | \vec k_1 + \vec k_2 | = | \vec k_3 + \vec k_4 | \leq k_{12}, k_{34}.
\end{split} \label{eq:k_notation}
\end{equation}
There are 3 diagrams at 1-loop order:
\begin{align}
\begin{fmffile}{1loop_diagrams_all}
    I_1 = \ \parbox{110pt}{\begin{fmfgraph*}(110,110)
       \fmfsurround{i1,i2,i3,i4,i5,i6,i7,i8}
       \fmfshift{3 left}{i1}
       \fmfshift{3 down}{i3}
       \fmfshift{3 right}{i5}
       \fmfshift{3 up}{i7}
       \fmfshift{4 down}{i2}
       \fmfshift{4 left}{i2}
       \fmfshift{4 down}{i4}
       \fmfshift{4 right}{i4}
       \fmfshift{4 up}{i6}
       \fmfshift{4 right}{i6}
       \fmfshift{4 up}{i8}
       \fmfshift{4 left}{i8}
       \fmf{dbl_plain,right=0.18,tension=0}{i1,i2,i3,i4,i5,i6,i7,i8,i1}
       \fmf{dashes,label=$\vec k_1$,label.side=right,tension=2}{i4,z1}
       \fmf{dashes,label=$\vec k_2$,label.side=right,tension=2}{z1,i6}
       \fmf{dashes,label=$\vec k_3$,label.side=left,tension=2}{i2,z2}
       \fmf{dashes,label=$\vec k_4$,label.side=left,tension=2}{z2,i8}
       \fmf{dashes,left=1,label=\hspace{12mm}\parbox{2cm}{\hspace{3mm}$p_{\mathrm{a}}$\\$\vec k + \vec\ell$}}{z1,z2}
       \fmf{dashes,left=1,label=\hspace{18mm}\parbox{2cm}{\hspace{0mm}$p_{\mathrm{b}}$\\\hspace{15mm}$\vec\ell$}}{z2,z1}
       \fmfforce{(0.30w,0.50h)}{z1}
       \fmfforce{(0.7w,0.50h)}{z2}
       \fmfforce{(0.17w,0.83h)}{i4}
       \fmfforce{(0.17w,0.17h)}{i6}
       \fmfforce{(0.83w,0.17h)}{i8}
       \fmfforce{(0.83w,0.83h)}{i2}
       \fmfforce{(0.5w,0.97h)}{i3}
       \fmfforce{(0.5w,0.04h)}{i7}
       \fmfforce{(0.97w,0.5h)}{i1}
       \fmfforce{(0.04w,0.5h)}{i5}
       \fmfv{l=$z_1$}{z1}
       \fmfv{l=$z_2$}{z2}
    \end{fmfgraph*}}\ +\ 
    \parbox{110pt}{\begin{fmfgraph*}(110,110)
       \fmfsurround{i1,i2,i3,i4,i5,i6,i7,i8}
       \fmfshift{3 left}{i1}
       \fmfshift{3 down}{i3}
       \fmfshift{3 right}{i5}
       \fmfshift{3 up}{i7}
       \fmfshift{4 down}{i2}
       \fmfshift{4 left}{i2}
       \fmfshift{4 down}{i4}
       \fmfshift{4 right}{i4}
       \fmfshift{4 up}{i6}
       \fmfshift{4 right}{i6}
       \fmfshift{4 up}{i8}
       \fmfshift{4 left}{i8}
       \fmf{dbl_plain,right=0.18,tension=0}{i1,i2,i3,i4,i5,i6,i7,i8,i1}
       \fmf{dashes,label=$\vec k_1$,label.side=right,tension=0}{i4,z1}
       \fmf{dashes,label=$\vec k_2$,label.side=right,tension=0}{z1,i6}
       \fmf{dashes,label=$\vec k_3$,label.side=left,tension=0}{i2,z2}
       \fmf{dashes,label=$\vec k_4$,label.side=left,tension=0}{z2,i8}
       \fmf{plain,left=1,label=\hspace{12mm}\parbox{2cm}{\hspace{3mm}$p_{\mathrm{a}}$\\$\vec k + \vec\ell$}}{z1,z2}
       \fmf{plain,left=1,label=\hspace{18mm}\parbox{2cm}{\hspace{0mm}$p_{\mathrm{b}}$\\\hspace{15mm}$\vec\ell$}}{z2,z1}
       \fmfforce{(0.30w,0.50h)}{z1}
       \fmfforce{(0.7w,0.50h)}{z2}
       \fmfv{l=$z_1$}{z1}
       \fmfv{l=$z_2$}{z2}
       \fmfforce{(0.17w,0.83h)}{i4}
       \fmfforce{(0.17w,0.17h)}{i6}
       \fmfforce{(0.83w,0.17h)}{i8}
       \fmfforce{(0.83w,0.83h)}{i2}
       \fmfforce{(0.5w,0.97h)}{i3}
       \fmfforce{(0.5w,0.04h)}{i7}
       \fmfforce{(0.97w,0.5h)}{i1}
       \fmfforce{(0.04w,0.5h)}{i5}
    \end{fmfgraph*}}\ +\ 
    \parbox{110pt}{\begin{fmfgraph*}(110,110)
       \fmfsurround{i1,i2,i3,i4,i5,i6,i7,i8}
       \fmfshift{3 left}{i1}
       \fmfshift{3 down}{i3}
       \fmfshift{3 right}{i5}
       \fmfshift{3 up}{i7}
       \fmfshift{4 down}{i2}
       \fmfshift{4 left}{i2}
       \fmfshift{4 down}{i4}
       \fmfshift{4 right}{i4}
       \fmfshift{4 up}{i6}
       \fmfshift{4 right}{i6}
       \fmfshift{4 up}{i8}
       \fmfshift{4 left}{i8}
       \fmf{dbl_plain,right=0.18,tension=0}{i1,i2,i3,i4,i5,i6,i7,i8,i1}
       \fmf{dashes,label=$\vec k_1$,label.side=right,tension=0}{i4,z1}
       \fmf{dashes,label=$\vec k_2$,label.side=right,tension=0}{z1,i6}
       \fmf{dashes,label=$\vec k_3$,label.side=left,tension=0}{i2,z2}
       \fmf{dashes,label=$\vec k_4$,label.side=left,tension=0}{z2,i8}
       \fmf{dashes,left=1,label=\hspace{12mm}\parbox{2cm}{\hspace{3mm}$p_{\mathrm{a}}$\\$\vec k + \vec\ell$}}{z1,z2}
       \fmf{plain,left=1,label=\hspace{18mm}\parbox{2cm}{\hspace{0mm}$p_{\mathrm{b}}$\\\hspace{15mm}$\vec\ell$}}{z2,z1}
       \fmfforce{(0.30w,0.50h)}{z1}
       \fmfforce{(0.7w,0.50h)}{z2}
       \fmfv{l=$z_1$}{z1}
       \fmfv{l=$z_2$}{z2}
       \fmfforce{(0.17w,0.83h)}{i4}
       \fmfforce{(0.17w,0.17h)}{i6}
       \fmfforce{(0.83w,0.17h)}{i8}
       \fmfforce{(0.83w,0.83h)}{i2}
       \fmfforce{(0.5w,0.97h)}{i3}
       \fmfforce{(0.5w,0.04h)}{i7}
       \fmfforce{(0.97w,0.5h)}{i1}
       \fmfforce{(0.04w,0.5h)}{i5}
    \end{fmfgraph*}}
\end{fmffile} \label{eq:1_loop_diagrams}
\end{align}
We have annotated the internal propagators with both their spacial momentum $\vec \ell \ (+ \vec k)$ and the respective integration variable (or "energy") $p_{\mathrm{a/b}}$ that appears already in the Feynman rules \eqref{eq:feynman_rules_propagator-} and \eqref{eq:feynman_rules_propagator+}. The conformal times of the two vertices are $z_1$ and $z_2$.

In the last of the above diagrams \eqref{eq:1_loop_diagrams}, there is a factor of $\kappa^2$ when compared to the first two diagrams, which comes from the vertices \eqref{eq:feynman_rules_odd_coupling}. Vertices with an odd numbers of $\phi_+$ and $\phi_-$ fields are suppressed by a factor of $\kappa$ each, making the third diagram subleading. For this reason, we will omit this diagram from the following analysis. Diagrams with such \textit{mixed loops}, i.e.\ loops with both $\phi_+$ and $\phi_-$ fields, will be revisited in section \ref{sec:mixed_loops}.

According to the Feynman rules, the sum of the first two diagrams $I_1$ is then
\begin{align}
    I_1 &= \int\limits_0^\infty \frac{\d z_1}{(z_1 H)^4} \frac{\d z_2}{(z_2 H)^4} \bar G_{\kappa}(\vec k_1, z_1) \bar G_{\kappa}(\vec k_2, z_1) \bar G_{\kappa}(\vec k_3, z_2) \bar G_{\kappa}(\vec k_4, z_2) \int\frac{\d^3 \ell}{(2\pi)^3} \int\limits_{-\infty}^\infty \d p_{\mathrm{a}} \d p_{\mathrm{b}} \notag\\
    &\quad \times \Bigg( \frac{2 \cdot 4 \cdot 3 \cdot 4 \cdot 3 \cdot 2}{2!} \Big( \frac{\lambda}{4!} \mathcal{C}_{2\kappa} \Big)^2 \frac{(z_1 z_2 H^2)^{2-2\kappa}}{\pi^2 \mathcal C_\kappa^2} \frac{\cos(p_{\mathrm{a}}z_1) \cos(p_{\mathrm{a}}z_2)}{\big(p_{\mathrm{a}}^2 + (\vec\ell + \vec k)^2\big)^{1+\kappa}} \frac{\cos(p_{\mathrm{b}}z_1) \cos(p_{\mathrm{b}}z_2)}{\big(p_{\mathrm{b}}^2 + \vec\ell^2\big)^{1+\kappa}} \\
    &\quad\quad + \frac{2 \cdot 2 \cdot 2 \cdot 2}{2!} \Big( -\frac{6\lambda}{4!} \mathcal{C}_{2\kappa} \Big)^2 (-1)^2 \frac{(z_1 z_2 H^2)^{2-2\kappa}}{\pi^2 \mathcal C_\kappa^2}\frac{\sin(p_{\mathrm{a}}z_1) \sin(p_{\mathrm{a}}z_2)}{\big(p_{\mathrm{a}}^2 + (\vec\ell + \vec k)^2\big)^{1+\kappa}} \frac{\sin(p_{\mathrm{b}}z_1) \sin(p_{\mathrm{b}}z_2)}{\big(p_{\mathrm{b}}^2 + \vec\ell^2\big)^{1+\kappa}} \Bigg). \notag
\end{align}
We have factored out the integrals and bulk-to-boundary propagators that are common in both relevant diagrams. Then, we have explicitly written the symmetry factors of both diagrams, their vertex factors and finally, their propagators involving either sine or cosine functions. Even though the symmetry and vertex factors are different for both diagrams, their product is the same. This allows us to combine both diagrams by adding their trigonometric functions. We introduce the shorthand $p^{\pm} = p_{\mathrm{a}} \pm p_{\mathrm{b}}$ and $\bar G_\kappa(z_1; z_2) = \bar G_\kappa(k_1, z_1) \bar G_\kappa(k_2, z_1) \bar G_\kappa(k_3, z_2) \bar G_\kappa(k_4, z_2)$ and find
\begin{align}
    I_1 &= \frac{\lambda^2 \mathcal C_{2\kappa}^2}{2\pi^2 \mathcal C_\kappa^2} \int\limits_0^\infty \frac{\d z_1}{(z_1 H)^{2+2\kappa}} \frac{\d z_2}{(z_2 H)^{2+2\kappa}} \bar G_\kappa(z_1; z_2) \notag\\
    &\hspace{1cm} \times \int\frac{\d^3 \ell}{(2\pi)^3} \int\limits_{-\infty}^\infty \d p_{\mathrm{a}} \d p_{\mathrm{b}} \frac{1}{2}\frac{\cos(p^{+}z_1)\cos(p^{+}z_2) + \cos(p^{-}z_1)\cos(p^{-}z_2)}{\big((p^{-} + p^{\mathrm{b}})^2 + (\vec\ell + \vec k)^2\big)^{1+\kappa} \big( p_{\mathrm{b}}^2 + \vec\ell^2 \big)^{1+\kappa}}.
\end{align}
Since the integrand is symmetric in $p_{\mathrm{b}} \rightarrow - p_{\mathrm{b}}$, we can turn the $p^{+}$ inside of the first two cosine functions into $p^{-}$, allowing us to simplify further. Moreover, we introduce 4-component objects\footnote{We do not want to call these objects vectors, since it is not immediately clear how they would transform under de Sitter transformations. However, the intended intuition is that they behave like the 4-momentum in loop integrals in flat space.} $L = (p_{\mathrm{b}}, \vec\ell)$ and $Q = (p^{-}, \vec k)$. The scalar product on these objects is the usual Euclidean 4-dimensional one. Finally, we shift the $p_{\mathrm{a}}$ integral such that it is over $p^{-}$, and combine the integrals over $\vec\ell$ and $p_{\mathrm{b}}$ into one over $L$. With this, we can write
\begin{align}
    I_1 &= \frac{\lambda^2 \mathcal C_{2\kappa}^2}{2\pi^2 \mathcal C_\kappa^2} \int\limits_0^\infty \frac{\d z_1}{(z_1 H)^{2+2\kappa}} \frac{\d z_2}{(z_2 H)^{2+2\kappa}} \bar G_\kappa(z_1; z_2) \int\frac{\d^4 L}{(2\pi)^3} \int\limits_{-\infty}^\infty \d p^{-}\frac{\cos(p^{-}z_1)\cos(p^{-}z_2)}{\big((L + Q)^2\big)^{1+\kappa} \big((L)^2 \big)^{1+\kappa}}. \label{eq:1_loop_step1}
\end{align}
This has decoupled the $L$ integral from all the other integrals and therefore, we can perform this first. Using (e.g. \cite{Smirnov:2012gma})
\begin{align}
    \int \frac{d^4 L}{(2\pi)^3} \frac{1}{\big((L+Q)^2\big)^{1+\kappa} \big(L^2\big)^{1+\kappa}} = \frac{(Q^2)^{-2\kappa}}{8\pi} \underbrace{\frac{\Gamma(1-\kappa)^2 \Gamma(2\kappa)}{\Gamma(2-2\kappa) \Gamma(1+\kappa)^2}}_{=: \Gamma_\kappa} \label{eq:L_integral_identity}
\end{align}
and inserting $Q^2 = \big(p^{-}\big)^2 + \vec k^2$, we finally arrive at
\begin{align}
    I_1 = \frac{\lambda^2 \mathcal C_{2\kappa}^2}{16\pi^3 \mathcal C_\kappa^2} \Gamma_\kappa \int\limits_0^\infty \frac{\d z_1}{(z_1 H)^{2+2\kappa}} \frac{\d z_2}{(z_2 H)^{2+2\kappa}} \bar G_\kappa(z_1; z_2) \int\limits_{-\infty}^\infty \d p_{-}\frac{\cos(p^{-}z_1)\cos(p^{-}z_2)}{\big((p^{-})^2 + \vec k^2\big)^{2\kappa}}.
\end{align}

\subsection{Renormalization}

As expected, the term
\begin{align}
    \Gamma_\kappa = \frac{\Gamma(1-\kappa)^2 \Gamma(2\kappa)}{\Gamma(2-2\kappa) \Gamma(1+\kappa)^2} = \frac{1}{2\kappa} + 1 + 2\kappa + \O(\kappa^2)
\end{align}
contains a $1/\kappa$ pole that diverges in the limit $\kappa \rightarrow 0$. This is simply the UV-divergence of the loop integral. We have to renormalize this by introducing a renormalized coupling $\lambda_\mathrm{R}$ and a counter-term $\delta\lambda$. Further, as in flat space, we should introduce a renormalization scale $\mu$. A convenient parametrization of the bare coupling is given by 
\begin{align}\label{eq:pc}
    \lambda = \lambda_\mathrm{R} \frac{H}{\mu} \mu^{4\kappa} (1 + \delta\lambda).
\end{align}
This turns the result for the 1-loop diagrams into
\begin{align}
    I_1 = \frac{\lambda_\mathrm{R}^2 H^2 \mathcal C_{2\kappa}^2}{16\pi^3 \mu^2 \mathcal C_\kappa^2} \mu^{8\kappa} \Gamma_\kappa \int\limits_0^\infty \frac{\d z_1}{(z_1 H)^{2+2\kappa}} \frac{\d z_2}{(z_2 H)^{2+2\kappa}} \bar G_\kappa(z_1; z_2) \int\limits_{-\infty}^\infty \d p^{-}\frac{\cos(p^{-}z_1)\cos(p^{-}z_2)}{\big((p^{-})^2 + \vec k^2\big)^{2\kappa}}.
\end{align}
At the same time, we now have a new diagram contributing at order $\lambda_\mathrm{R}^2$, which we refer to as "1-loop order". This new diagram is the cross counter-term
\begin{equation}
I_\otimes^1 := \parbox{100 pt}{
\begin{tikzpicture}
  \node[anchor=south west, inner sep=0] (img) at (0,0) {\includegraphics{1loop_counterterm_cross.1.mps}};
  \begin{scope}[x={(img.south east)}, y={(img.north west)}]
    \node at (0.5,0.32) {$z_1$};
    \node at (0.2,0.65) {$\vec k_1$};
    \node at (0.2,0.35) {$\vec k_2$};
    \node at (0.8,0.65) {$\vec k_3$};
    \node at (0.8,0.35) {$\vec k_4$};
  \end{scope}
\end{tikzpicture}
} = \lambdaR \frac{H}{\mu} \mu^{4\kappa} \delta\lambda \int\limits_0^\infty \frac{\d z_1}{(z_1 H)^4} \bar G_\kappa(z_1; z_1) =: \delta\lambda I_\times, \label{eq:counter_term_cross}
\end{equation}
where $I_\times = I_0$ is the tree-level "0-loop" 4-point function diagram, which we call "cross". We can combine the counter-term $I_\otimes^1$ with the 1-loop expression $I_1$ more easily if we introduce a $p^{-}$ integral into the counter-term expression. This can be done using the identity
\begin{align}
    \int\limits_{-\infty}^\infty \d p \cos(p z_1) \cos(p z_2) = \pi \delta(z_1 - z_2) + \pi \delta(z_1 + z_2) \,\hat{=}\, \pi \delta(z_1 - z_2). \label{eq:cosine_delta}
\end{align}
Note that the $z$ integrals only go from $0$ to $\infty$, so the $\delta(z_1 + z_2)$ term is never relevant. We write
\begin{align}
    I_\otimes^1 &= \lambdaR \frac{H}{\mu} \mu^{4\kappa} \delta\lambda \int\limits_0^\infty \frac{\d z_1}{(z_1 H)^2} \frac{\d z_2}{(z_2 H)^2} \delta(z_1 - z_2) \bar G_\kappa(z_1; z_2) \\
    &= \lambdaR \frac{H}{\mu} \mu^{4\kappa} \frac{\delta\lambda}{\pi} \int\limits_0^\infty \frac{\d z_1}{(z_1 H)^2} \frac{\d z_2}{(z_2 H)^2} \bar G_\kappa(z_1; z_2) \int\limits_{-\infty}^\infty \d p^{-} \cos(p^{-} z_1) \cos(p^{-} z_2).
\end{align}
Thus, the sum of the one-loop diagram and counter-term is
\begin{align}
    L_1 = I_1 + I_\oplus^1 &= \frac{\lambdaR H \mu^{4\kappa}}{\mu} \int\limits_0^\infty \frac{\d z_1}{(z_1 H)^2}\frac{\d z_2}{(z_2 H)^2}\bar G_\kappa(z_1; z_2) \notag\\
    &\hspace{0.4cm} \times \int\limits_{-\infty}^\infty \d p^{-} \cos(p^{-} z_1) \cos(p^{-} z_2) \Bigg( \frac{\lambdaR H \mathcal{C}_{2\kappa}^2\mu^{4\kappa}}{16\pi^3 \mu \mathcal{C}_\kappa^2} \frac{\Gamma_\kappa}{\Big(z_1 z_2 H^2 \big((p^{-})^2 + \vec k^2\big)\Big)^{2\kappa}}  + \frac{\delta\lambda}{\pi} \Bigg) \label{eq:one_loop_cancellation}
\end{align}
In view of \eqref{eq:pc} we cancel the divergent term by a counter-term given by
\begin{align}
    \delta\lambda = - \frac{\lambdaR H}{32\pi^2} \frac{1}{\kappa} + \O(\kappa^0).
\end{align}
However, we can do more: Expanding the kinetic variables in the second line of \eqref{eq:one_loop_cancellation}, we see that
\begin{align}
    \frac{\lambdaR H \mathcal{C}_{2\kappa}^2}{16\pi^3 \mu \mathcal{C}_\kappa^2} \Bigg( \frac{1}{2\kappa} + 1 - \log\Big( z_1 z_2 H^2 \frac{(p^{-})^2 + \vec k^2}{\mu^2} \Big) + \O(\kappa) \Bigg) + \frac{\delta\lambda}{\pi} \overset{!}{=} \O(\kappa^0).
\end{align}
We can make a non-minimal subtraction and actually subtract all terms that do not depend on the kinematic variables $z_1, z_2, p^{-},$ and $\vec k$ or $\mu$. This keeps our expressions simpler when we go to higher loop computations. Therefore, we set
\begin{align}
    \delta\lambda = - \frac{\lambdaR H \mathcal C_{2\kappa}^2}{16\pi^2 \mu \mathcal C_\kappa^2} \Big( \frac{1}{2\kappa} + 1\Big). \label{eq:1_loop_counterterm}
\end{align}
This is the final result for the counter-term at 1-loop order. The sum of the diagrams then becomes
\begin{align}
    L_1 &= -\frac{\lambdaR^2 H^2 \mu^{4\kappa} \mathcal C_{2\kappa}^2}{16\pi^3\mu^2 \mathcal C_\kappa^2} \int\limits_0^\infty \frac{\d z_1}{(z_1 H)^2}\frac{\d z_2}{(z_2 H)^2} \bar G_\kappa(z_1; z_2) \notag\\
    &\hspace{1.4cm} \times \int\limits_{-\infty}^\infty \d p^{-} \cos(p^{-} z_1) \cos(p^{-} z_2) \log\Big(z_1 z_2 H^2 \frac{(p^{-})^2 + \vec k^2}{\mu^2}\Big). \label{eq:1_loop_log}
\end{align}
We introduce some general shorthand notation to keep this and future expressions simpler:
\begin{equation}
\begin{split}
    \lambda_{(\mathrm{R})} \mathcal C_{2\kappa} &\xrightarrow{\mathrm{rename}} \lambda_{(\mathrm{R})} \\
    p_{(j)} &= p_{(j)}^-, \\
    C &= \frac{1}{16\pi^3 \mathcal C_\kappa^2}, \\
    \Gamma_\kappa &= \frac{\Gamma(1-\kappa)^2 \Gamma(2\kappa)}{\Gamma(2-2\kappa) \Gamma(1+\kappa)^2} \\
    \bar G_\kappa(z_1; z_{N+1}) &= \bar G_\kappa(k_1, z_1) \bar G_\kappa(k_2, z_1) \bar G_\kappa(k_3, z_{N+1}) \bar G_\kappa(k_4, z_{N+1})
\end{split}
\label{eq:shortcuts}
\end{equation}
Using these, and noting that we can now safely take the $\kappa \rightarrow 0$ limit of \eqref{eq:1_loop_log} since there are no poles anymore, we finally arrive at
\begin{align}
    L_1 &= -\frac{\lambdaR^2 H^2 C}{\mu^2} \int\limits_0^\infty \frac{\d z_1}{(z_1 H)^2}\frac{\d z_2}{(z_2 H)^2} \bar G_0(z_1; z_2) \int\limits_{-\infty}^\infty \d p \cos(p z_1) \cos(p z_2) \log\Big(z_1 z_2 H^2 \frac{p^2 + \vec k^2}{\mu^2}\Big). \label{eq:1_loop_log_final}
\end{align}
The remaining integrals can now be evaluated, but this has already been done in \cite{chowdhury_subtle_2024}. For this discussion, mainly the integral expression is relevant.

\section{$N$-Loop Necklace Diagrams} \label{sec:n_loop}

In this section, we will generalize the results from 1-loop to the $N$-loop necklace diagrams. We are only considering necklace topology diagrams in this work, since, ultimately, they are the easiest to compute. One can also motivate them from a large $N$ model perspective \cite{moshe_quantum_2003}, where they are the leading contribution in the number of scalars (see also \cite{pietro_analyticity_2022, pietro_cosmological_2025}). For now, we will again restrict the discussion to \textit{pure} loops, i.e.\ loops that have either two $\phi_-$ or two $\phi_+$ fields. It is easier to understand \textit{mixed} loops, i.e.\ loops with one $\phi_+$ and one $\phi_-$, as a modification to this simpler case. That will be done in section \ref{sec:mixed_loops}.

\subsection{General Expression for Pure Loops}

For now, we return to the bare coupling $\lambda$ and leave out counter-term diagrams. Renormalization is handled in section \ref{sec:n_loop_renormalization}. There are $2^N$ pure loop diagrams at $N$-loop order, as each of the $N$ loops might be a $\phi_+$ or a $\phi_-$ loop:
\begin{align*}
\begin{fmffile}{n_loopmmm}
    \hspace{-0.5cm}
    \parbox{210pt}{
    \begin{fmfgraph*}(210,210)
       \fmfsurround{i1,i2,i3,i4,i5,i6,i7,i8}
       \fmfshift{5 left}{i1}
       \fmfshift{5 down}{i3}
       \fmfshift{5 right}{i5}
       \fmfshift{5 up}{i7}
       \fmfshift{7 down}{i2}
       \fmfshift{7 left}{i2}
       \fmfshift{7 down}{i4}
       \fmfshift{7 right}{i4}
       \fmfshift{7 up}{i6}
       \fmfshift{7 right}{i6}
       \fmfshift{7 up}{i8}
       \fmfshift{7 left}{i8}
       \fmfleft{l1,l2,l3}
       \fmfshift{40 right}{l3}
       \fmfshift{40 down}{l3}
       \fmf{dbl_plain,right=0.18,force=0}{i1,i2,i3,i4,i5,i6,i7,i8,i1}
       \fmf{dashes,label=$\vec k_1$,label.side=right,force=0}{i4,z1}
       \fmf{dashes,label=$\vec k_2$,label.side=right,force=0}{z1,i6}
       \fmf{dashes,label=$\vec k_3$,label.side=left,force=0}{i2,zN}
       \fmf{dashes,label=$\vec k_4$,label.side=left,force=0}{zN,i8}
       \fmfforce{(0.20w,0.50h)}{z1}
       \fmfforce{(0.35w,0.50h)}{z2}
       \fmfforce{(0.5w,0.50h)}{z3}
       \fmf{dashes,left=1,label=\hspace{12mm}\parbox{2cm}{\hspace{2mm}$p_{1\mathrm{a}}$\\\vspace{2mm}$\vec k + \vec\ell_1$}}{z1,z2}
       \fmf{dashes,left=1,label=\hspace{15mm}\parbox{2cm}{\hspace{0mm}$p_{1\mathrm{b}}$\\\hspace{15mm}$\vec\ell_1$}}{z2,z1}
       \fmf{dashes,left=1,label=\hspace{12mm}\parbox{2cm}{\hspace{2mm}$p_{2\mathrm{a}}$\\\vspace{2mm}$\vec k + \vec\ell_2$}}{z2,z3}
       \fmf{dashes,left=1,label=\hspace{15mm}\parbox{2cm}{\hspace{0mm}$p_{2\mathrm{b}}$\\\hspace{15mm}$\vec\ell_2$}}{z3,z2}
       \fmfv{l=$z_1$}{z1}
       \fmfv{l=$z_2$}{z2}
       \fmfv{l.a=180,l=$z_3$}{z3}
       \fmfforce{(0.80w,0.50h)}{zN}
       \fmfforce{(0.65w,0.50h)}{zN1}
       \fmf{dashes,left=1,label=\hspace{8mm}\parbox{2cm}{\hspace{2mm}$p_{N\mathrm{a}}$\\\vspace{2mm}$\vec k + \vec\ell_N$}}{zN1,zN}
       \fmf{dashes,left=1,label=\hspace{15mm}\parbox{2cm}{\hspace{0mm}$p_{N\mathrm{b}}$\\\hspace{15mm}$\vec\ell_N$}}{zN,zN1}
       \fmfv{l=$z_N$}{zN1}
       \fmfv{l=$z_{N+1}$}{zN}
       \fmfpen{1.5pt}
       \fmf{dots}{z3,zN1}
       \fmfforce{(0.17w,0.83h)}{i4}
       \fmfforce{(0.17w,0.17h)}{i6}
       \fmfforce{(0.83w,0.17h)}{i8}
       \fmfforce{(0.83w,0.83h)}{i2}
       \fmfforce{(0.5w,0.97h)}{i3}
       \fmfforce{(0.5w,0.04h)}{i7}
       \fmfforce{(0.97w,0.5h)}{i1}
       \fmfforce{(0.04w,0.5h)}{i5}
    \end{fmfgraph*}} 
    \quad + \quad
    \parbox{210pt}{
    \begin{fmfgraph*}(210,210)
       \fmfsurround{i1,i2,i3,i4,i5,i6,i7,i8}
       \fmfshift{5 left}{i1}
       \fmfshift{5 down}{i3}
       \fmfshift{5 right}{i5}
       \fmfshift{5 up}{i7}
       \fmfshift{7 down}{i2}
       \fmfshift{7 left}{i2}
       \fmfshift{7 down}{i4}
       \fmfshift{7 right}{i4}
       \fmfshift{7 up}{i6}
       \fmfshift{7 right}{i6}
       \fmfshift{7 up}{i8}
       \fmfshift{7 left}{i8}
       \fmfleft{l1,l2,l3}
       \fmfshift{40 right}{l3}
       \fmfshift{40 down}{l3}
       \fmf{dbl_plain,right=0.18}{i1,i2,i3,i4,i5,i6,i7,i8,i1}
       \fmf{dashes,label=$\vec k_1$,label.side=right}{i4,z1}
       \fmf{dashes,label=$\vec k_2$,label.side=right}{z1,i6}
       \fmf{dashes,label=$\vec k_3$,label.side=left}{i2,zN}
       \fmf{dashes,label=$\vec k_4$,label.side=left}{zN,i8}
       \fmfforce{(0.20w,0.50h)}{z1}
       \fmfforce{(0.35w,0.50h)}{z2}
       \fmfforce{(0.5w,0.50h)}{z3}
       \fmf{plain,left=1,label=\hspace{12mm}\parbox{2cm}{\hspace{2mm}$p_{1\mathrm{a}}$\\\vspace{2mm}$\vec k + \vec\ell_1$}}{z1,z2}
       \fmf{plain,left=1,label=\hspace{15mm}\parbox{2cm}{\hspace{0mm}$p_{1\mathrm{b}}$\\\hspace{15mm}$\vec\ell_1$}}{z2,z1}
       \fmf{dashes,left=1,label=\hspace{12mm}\parbox{2cm}{\hspace{2mm}$p_{2\mathrm{a}}$\\\vspace{2mm}$\vec k + \vec\ell_2$}}{z2,z3}
       \fmf{dashes,left=1,label=\hspace{15mm}\parbox{2cm}{\hspace{0mm}$p_{2\mathrm{b}}$\\\hspace{15mm}$\vec\ell_2$}}{z3,z2}
       \fmfv{l=$z_1$}{z1}
       \fmfv{l=$z_2$}{z2}
       \fmfv{l.a=180,l=$z_3$}{z3}
       \fmfforce{(0.80w,0.50h)}{zN}
       \fmfforce{(0.65w,0.50h)}{zN1}
       \fmf{dashes,left=1,label=\hspace{8mm}\parbox{2cm}{\hspace{2mm}$p_{N\mathrm{a}}$\\\vspace{2mm}$\vec k + \vec\ell_N$}}{zN1,zN}
       \fmf{dashes,left=1,label=\hspace{15mm}\parbox{2cm}{\hspace{0mm}$p_{N\mathrm{b}}$\\\hspace{15mm}$\vec\ell_N$}}{zN,zN1}
       \fmfv{l=$z_N$}{zN1}
       \fmfv{l=$z_{N+1}$}{zN}
       \fmfpen{1.5pt}
       \fmf{dots}{z3,zN1}
       \fmfforce{(0.17w,0.83h)}{i4}
       \fmfforce{(0.17w,0.17h)}{i6}
       \fmfforce{(0.83w,0.17h)}{i8}
       \fmfforce{(0.83w,0.83h)}{i2}
       \fmfforce{(0.5w,0.97h)}{i3}
       \fmfforce{(0.5w,0.04h)}{i7}
       \fmfforce{(0.97w,0.5h)}{i1}
       \fmfforce{(0.04w,0.5h)}{i5}
    \end{fmfgraph*}}
\end{fmffile}
\end{align*}
\vspace{-2cm}
\begin{align}
\begin{fmffile}{n_loop2}
    \hspace{0.4cm}
    + \ldots + \quad \parbox{300pt}{
    \begin{fmfgraph*}(210,210)
       \fmfsurround{i1,i2,i3,i4,i5,i6,i7,i8}
       \fmfshift{5 left}{i1}
       \fmfshift{5 down}{i3}
       \fmfshift{5 right}{i5}
       \fmfshift{5 up}{i7}
       \fmfshift{7 down}{i2}
       \fmfshift{7 left}{i2}
       \fmfshift{7 down}{i4}
       \fmfshift{7 right}{i4}
       \fmfshift{7 up}{i6}
       \fmfshift{7 right}{i6}
       \fmfshift{7 up}{i8}
       \fmfshift{7 left}{i8}
       \fmfleft{l1,l2,l3}
       \fmfshift{40 right}{l3}
       \fmfshift{40 down}{l3}
       \fmf{dbl_plain,right=0.18}{i1,i2,i3,i4,i5,i6,i7,i8,i1}
       \fmf{dashes,label=$\vec k_1$,label.side=right}{i4,z1}
       \fmf{dashes,label=$\vec k_2$,label.side=right}{z1,i6}
       \fmf{dashes,label=$\vec k_3$,label.side=left}{i2,zN}
       \fmf{dashes,label=$\vec k_4$,label.side=left}{zN,i8}
       \fmfforce{(0.20w,0.50h)}{z1}
       \fmfforce{(0.35w,0.50h)}{z2}
       \fmfforce{(0.5w,0.50h)}{z3}
       \fmf{plain,left=1,label=\hspace{12mm}\parbox{2cm}{\hspace{2mm}$p_{1\mathrm{a}}$\\\vspace{2mm}$\vec k + \vec\ell_1$}}{z1,z2}
       \fmf{plain,left=1,label=\hspace{15mm}\parbox{2cm}{\hspace{0mm}$p_{1\mathrm{b}}$\\\hspace{15mm}$\vec\ell_1$}}{z2,z1}
       \fmf{plain,left=1,label=\hspace{12mm}\parbox{2cm}{\hspace{2mm}$p_{2\mathrm{a}}$\\\vspace{2mm}$\vec k + \vec\ell_2$}}{z2,z3}
       \fmf{plain,left=1,label=\hspace{15mm}\parbox{2cm}{\hspace{0mm}$p_{2\mathrm{b}}$\\\hspace{15mm}$\vec\ell_2$}}{z3,z2}
       \fmfv{l=$z_1$}{z1}
       \fmfv{l=$z_2$}{z2}
       \fmfv{l.a=180,l=$z_3$}{z3}
       \fmfforce{(0.80w,0.50h)}{zN}
       \fmfforce{(0.65w,0.50h)}{zN1}
       \fmf{plain,left=1,label=\hspace{8mm}\parbox{2cm}{\hspace{2mm}$p_{N\mathrm{a}}$\\\vspace{2mm}$\vec k + \vec\ell_N$}}{zN1,zN}
       \fmf{plain,left=1,label=\hspace{15mm}\parbox{2cm}{\hspace{0mm}$p_{N\mathrm{b}}$\\\hspace{15mm}$\vec\ell_N$}}{zN,zN1}
       \fmfv{l=$z_N$}{zN1}
       \fmfv{l=$z_{N+1}$}{zN}
       \fmfpen{1.5pt}
       \fmf{dots}{z3,zN1}
       \fmfforce{(0.17w,0.83h)}{i4}
       \fmfforce{(0.17w,0.17h)}{i6}
       \fmfforce{(0.83w,0.17h)}{i8}
       \fmfforce{(0.83w,0.83h)}{i2}
       \fmfforce{(0.5w,0.97h)}{i3}
       \fmfforce{(0.5w,0.04h)}{i7}
       \fmfforce{(0.97w,0.5h)}{i1}
       \fmfforce{(0.04w,0.5h)}{i5}
    \end{fmfgraph*}} \hspace{-2cm} =: I_N \label{eq:n_loop_diagrams}
\end{fmffile}
\end{align}

It turns out that all the important steps that we performed in the 1-loop computation generalize to the $N$-loop case. The key observations are
\begin{itemize}
    \item The products of symmetry factors and vertex factors for all of the above $N$-loop diagrams are the same.
    \item The trigonometric functions from the propagators can be combined into cosines.
    \item The symmetry under $p_{j\mathrm{b}} \rightarrow -p_{j\mathrm{b}}$ lets us write the integrand as a function of $L_j = (p_{j\mathrm{b}}, \vec\ell_j)$, $p_{j}^- = p_{j\mathrm{a}} - p_{j\mathrm{b}}$ and $z_j$.
    \item The integral over $L_j$ factorizes and can be computed using standard techniques.
\end{itemize}
It is not a coincidence that the computation simplifies in this manner. This and the similarity to flat space computations are summarized in the cosmological dressing rules formalism \cite{chowdhury_cosmological_2025}.

As a consequence of these observations, the sum over the $N$-loop diagrams in some sense is the product of $N$ times the 1-loop expression, with some caveats: In flat space, this statement holds true for the final result, where no integrals are left to perform. In EAdS, this product will still be the integrand of all $z_j$ and $p_{j}$ integrals. The details of this derivation can be found in the appendix \ref{app:N_loop_derivation}, and the result in \eqref{eq:app_N_loop_pure_final} is
\begin{align}
    I_N &= \lambda^{N+1} C^N \int\limits_0^\infty \frac{\d z_1}{(z_1H)^{2+2\kappa}} \frac{\d z_2}{(z_2H)^{4\kappa}} \frac{\d z_3}{(z_3H)^{4\kappa}} \ldots \frac{\d z_N}{(z_NH)^{4\kappa}} \frac{\d z_{N+1}}{(z_{N+1}H)^{2+2\kappa}} \bar G_\kappa(z_1; z_{N+1}) \notag\\
    &\hspace{1.8cm}\times \int\limits_{-\infty}^\infty \d p_1 \ldots \d p_N \Gamma_\kappa^N \frac{\cos(p_1 z_1) \cos(p_1 z_2) \ldots \cos(p_N z_N) \cos(p_N z_{N+1})}{\big( p_1^2 + \vec k^2\big)^{2\kappa} \ldots \big( p_N^2 + \vec k^2\big)^{2\kappa}} \label{eq:N_loop_pure_final}
\end{align}
Note the previously defined shortened notation in \eqref{eq:shortcuts}.

\subsection{Renormalization} \label{sec:n_loop_renormalization}

The $N$-loop expression \eqref{eq:N_loop_pure_final} includes a factor of $\Gamma_\kappa^N$, which in turn contains a term proportional to $1/\kappa^N$ that comes from the UV divergence of the loop momentum integrals. As in the 1-loop case, this term has to be canceled via counter-terms. However, for higher loops, there are also sub-leading divergences $1/\kappa^j$ with $j < N$, as well as more counter-term diagrams. To manage the combinatorical complexity of the $N$-loop renormalization, we parametrize the $N$-loop order counter-term as
\begin{align}
    \lambda = \lambdaR \frac{H}{\mu} \mu^{4\kappa} \big( 1 + \delta \lambda + \delta\lambda^2 + \ldots + (\delta\lambda)^N\big). \label{eq:n_loop_counterterm_parametrization}
\end{align}
At this point, we are again open-minded as to what $\delta\lambda$ might be, but it will turn out that if we insert the 1-loop result \eqref{eq:1_loop_counterterm}, this ansatz cancels all UV-divergences and leaves us with only logarithms under the various $p$ and $z$ integrals.

We will now illustrate the combinatorical rationale that goes into the general $N$-loop renormalization with the 2-loop diagram as an example. From now on, we will omit labeling the vertices and momenta in the diagrams, as they are understood to be the same as in the general $N$-loop diagrams \eqref{eq:n_loop_diagrams}. Further, we will only write a single diagram for all possible combinations of $\phi_-$ and $\phi_+$ pure loops, since their topology is the same. This superposition will be indicated by doubly dashed internal propagators.

Since we expect $\delta\lambda$ to be at least of order $\lambdaR^1$, a counter-term diagram at 2-loop order has to be of 1-loop or cross topology in order to be of the same order in $\lambdaR$ as the 2-loop diagram itself. In the 1-loop topology, only one of the two interaction vertices may be a counter-term, since having both would put us at at least order $\lambdaR^4$. For the cross topology, we need a counter-term that is at least of order $\lambdaR^2$. We will indicate this by using the $\delta\lambda^2$ piece of the counter-term \eqref{eq:n_loop_counterterm_parametrization} and draw it in diagrams as two counter-terms right next to each other: "$\otimes\otimes$". Note that this is still only a single vertex. The 3 possible diagrams are therefore

\begin{align}
\begin{fmffile}{2loop_counterterm_1}
    I_\otimes^2 = \parbox{250pt}{
    \includegraphics{2loop_counterterm_1.1.mps}
    } \hspace{-5.2cm} + \parbox{250pt}{
    \includegraphics{2loop_counterterm_1.2.mps}
    } \hspace{-5.2cm} + \parbox{100pt}{
    \includegraphics{2loop_counterterm_1.3.mps}
    }
\end{fmffile}
\end{align}

Since a counter-term is just the normal vertex times $\delta\lambda$, the first two of the above diagrams are actually equal. It does not matter where the extra factor is multiplied in. Further, since all we have done is shift the vertex according to \eqref{eq:n_loop_counterterm_parametrization}, the symmetry factors as well as kinematic dependencies stay the same. We conclude
\begin{align}
    I_\otimes^2 = 2 \delta\lambda I_1 + \delta\lambda^2 I_0,
\end{align}
This discussion generalizes to the $N$-loop case. To make this more clear, we utilize a string-argument\footnote{We use the concept "string" in the computer science sense as a ordered list of "character", which are $\otimes$ and $\ocircle$ here.}. We write any necklace with counter-terms as a string of $\ocircle$ and $\otimes$, where $\ocircle$ symbolizes a superposition of a $\phi_-$ and a $\phi_+$ pure loop, and $\otimes$ symbolizes a counter-term. We write multiple $\otimes$ in succession for a higher order counter-term, just like we have done before. The string will always have length $N$, because if there are $l$ counter-terms, there have to be $N-l$ loops in order to end up at the same order in $\lambdaR$. All the diagrams at 2-loop order from before reduce to $\ocircle\ocircle$, $\otimes\ocircle$, $\ocircle\otimes$ and $\otimes\otimes$ in this new notation. 

Just as $\otimes\ocircle$ and $\ocircle\otimes$ are equal, any two diagrams represented by a string with the same number of $\otimes$ and $\ocircle$ are the same. This is an immediate consequence their topology being the same, their symmetry factors being the same and the convention that a $j$-times repeated $\otimes$ indicates a counter-term $(\delta\lambda)^j$, according to \eqref{eq:n_loop_counterterm_parametrization}. Any diagrams with exactly $l$ counter-terms are therefore simply equal to $(\delta\lambda)^l I_{N-l}$. Now, we just have to count that there are $\binom{N}{l}$ diagrams at $N$-loop order that have $l$ counter-terms. We conclude
\begin{align}
    L_N = I_N + I_\otimes^N = \sum\limits_{l=0}^N \binom{N}{l} (\delta\lambda)^l I_{N-l}. \label{eq:counter_term_combinatorics}
\end{align}
The above equation \eqref{eq:counter_term_combinatorics} already looks like a binomial expansion, but it is not quite clear yet how it can be simplified further. To make this manifest, we rewrite the $N$-loop diagram without counter-terms in \eqref{eq:N_loop_pure_final}, represented by $\ocircle \ldots \ocircle$, as
\begin{align}
    I_N &= \frac{\lambdaR H}{\mu^{1-4\kappa}} \int\limits_0^\infty \frac{\d z_1}{(z_1H)^2} \d z_2 \ldots \d z_N \frac{\d z_{N+1}}{(z_{N+1}H)^2} \bar G_\kappa(z_1; z_{N+1}) \int\limits_{-\infty}^\infty \d p_1 \ldots \d p_N  \cos(p_1 z_1) \ldots \cos(p_N z_{N+1}) \notag\\
    &\quad\quad\quad\times \frac{\lambdaR H C}{\mu} \Gamma_\kappa \Big( z_1 z_2 H^2 \frac{p_1^2 + \vec k^2}{\mu^2}\Big)^{-2\kappa} \ldots \frac{\lambdaR H C}{\mu} \Gamma_\kappa \Big( z_N z_{N+1} H^2 \frac{p_N^2 + \vec k^2}{\mu^2}\Big)^{-2\kappa} \label{eq:N_loop_ct_step1}
\end{align}
Any counter-term diagram $I_N^{\otimes,j}$ represented by the string $\ocircle \ldots \ocircle\otimes\ocircle \ldots \ocircle$ with exactly one $\otimes$ in the $j$-th position will now modify this expression by adding a factor $\delta\lambda$ and removing the $z_j$ and $p_j$ integrals, the cosines that contain $p_j$, and the kinematic block 
\begin{align}
    \frac{\lambdaR H C}{\mu} \Gamma_\kappa \Big(z_j z_{j+1} H^2 \frac{p_j^2 + \vec k^2}{\mu^2}\Big)^{-2\kappa}.
\end{align}
We name the other $z$ and $p$ variables such that the highest $z$ index is still $N+1$ and the highest $p$ index is still $N$. As a consequence, the cosines involving $p_{j-1}$ are now $\cos(p_{j-1} z_{j-1}) \cos(p_{j-1} z_{j+1})$.

Just like how we introduced a $p$ integral for the counter-term at the 1-loop level, using \eqref{eq:cosine_delta}, we can now reinsert a $z_j$ integral by also inserting $\int\d p_j \cos(p_j z_j) \cos(p_j z_{j+1})/\pi$. The appearing delta function allows us to write the cosines for $p_{j-1}$ in the usual form again. In total, we get
\begin{align}
    I_N^{\otimes,j} &= \frac{\lambdaR H}{\mu^{1-4\kappa}} \int\limits_0^\infty \frac{\d z_1}{(z_1H)^2} \d z_2 \ldots \d z_N \frac{\d z_{N+1}}{(z_{N+1}H)^2} \bar G_\kappa(z_1; z_{N+1}) \int\limits_{-\infty}^\infty \d p_1 \ldots \d p_N \cos(p_1 z_1) \ldots \cos(p_N z_{N+1}) \notag\\
    &\quad\times \frac{\lambdaR H C}{\mu} \Gamma_\kappa \Big( z_1 z_2 H^2 \frac{p_1^2 + \vec k^2}{\mu^2}\Big)^{-2\kappa} \ldots \hspace{-4mm} \underbrace{\frac{\delta\lambda}{\pi}}_{j\mathrm{-th \ position}} \hspace{-4mm} \ldots \frac{\lambdaR H C}{\mu} \Gamma_\kappa \Big( z_N z_{N+1} H^2 \frac{p_N^2 + \vec k^2}{\mu^2}\Big)^{-2\kappa}, \label{eq:N_loop_ct_step2}
\end{align}
where the $j$-th position does not have a kinematic block but instead the $\delta\lambda/\pi$.

This pattern continues when we add more counter-terms: Replacing another of the $\ocircle$ by $\otimes$ removes another kinematic block (including the coefficient) and effectively replaces it with $\delta\lambda/\pi$. Since the total expression at the $N$-loop level, $L_N$, includes every chain of length $N$ with any possible combination of $\ocircle$ and $\otimes$, it is easy to see that it equals
\begin{align}
    L_N &= \frac{\lambdaR H}{\mu^{1-4\kappa}} \int\limits_0^\infty \frac{\d z_1}{(z_1H)^2} \d z_2 \ldots \d z_N \frac{\d z_{N+1}}{(z_{N+1}H)^2} \bar G_\kappa(z_1; z_{N+1}) \int\limits_{-\infty}^\infty \d p_1 \ldots \d p_N \cos(p_1 z_1) \ldots \cos(p_N z_{N+1}) \notag\\
    &\hspace{-5mm}\times \Bigg(\frac{\lambdaR H C}{\mu} \Gamma_\kappa \Big( z_1 z_2 H^2 \frac{p_1^2 + \vec k^2}{\mu^2}\Big)^{-2\kappa} + \frac{\delta\lambda}{\pi} \Bigg) \ldots \Bigg(\frac{\lambdaR H C}{\mu} \Gamma_\kappa \Big( z_N z_{N+1} H^2 \frac{p_N^2 + \vec k^2}{\mu^2}\Big)^{-2\kappa} + \frac{\delta\lambda}{\pi} \Bigg). \label{eq:N_loop_renormalized}
\end{align}
Now we have made the form of the renormalized $N$-loop diagram explicitly the $N$-times product of the 1-loop result under all the $z$ and $p$ integrals. Choosing the same $\delta\lambda$ as for the 1-loop case,
\begin{align}
    \delta\lambda = \frac{\pi \lambdaR H C}{\mu} \Big( \frac{1}{2\kappa} + 1 \Big),
\end{align}
we subtract all divergent terms at once, as well as the constant term that comes from the expansion of $\Gamma_\kappa$. Then, we can send $\kappa \rightarrow 0$ to get rid of the regulator, as our result is now finite in this limit. Finally, we end with
\begin{align}
    L_N &= (-1)^N \Big(\frac{\lambdaR H}{\mu}\Big)^{N+1} C^N \int\limits_0^\infty \frac{\d z_1}{(z_1H)^2} \d z_2 \ldots \d z_N \frac{\d z_{N+1}}{(z_{N+1}H)^2} \bar G_0(z_1; z_{N+1}) \int\limits_{-\infty}^\infty \d p_1 \ldots \d p_N \notag\\
    &\hspace{1cm}\times \cos(p_1 z_1) \ldots \cos(p_N z_{N+1}) \log\Big( z_1 z_2 H^2 \frac{p_1^2 + \vec k^2}{\mu^2}\Big) \ldots \log\Big( z_N z_{N+1} H^2 \frac{p_N^2 + \vec k^2}{\mu^2}\Big). \label{eq:N_loop_renormalized_final}
\end{align}
Alternatively to expanding in $\kappa$ before integration, we could also first integrate and then expand. However, this attempt seems ill-fated, as demonstrated in appendix \ref{app:c}. Therefore, we will continue as-is.
Despite the rather simple structure of the integrand of \eqref{eq:N_loop_renormalized_final}, which is just a product over the 1-loop integrands of all involved loops, evaluating this integral turns out to be rather complicated for higher loops. Further, since all logarithms have their own kinematic variables, summing them over $N$ is also not very straightforward. In the following sections, we will show that evaluating the remaining integrals is possible up to 2-loop order. Already at 3-loop, however, certain obstructions arise.

\subsection{2-Loop}

The 2-loop necklace has already been evaluated in \cite{chowdhury_subtle_2024}. However, a careful analysis of the counter-terms has not been performed yet, so only the contribution of the 2-loop diagram itself has been computed there. Nevertheless, we will repeat this calculation here, also to demonstrate why it is simpler than for 3-loop order. We start with \eqref{eq:N_loop_renormalized_final} and split the logarithms to find
\begin{align}
    L_2 &= \Big( \frac{\lambdaR H}{\mu} \Big)^3 C^2 \int\limits_0^\infty \frac{\d z_1}{(z_1H)^2} \d z_2 \frac{\d z_3}{(z_3H)^2} \bar G_0(z_1; z_3) \int\limits_{-\infty}^\infty \d p_1 \d p_2 \cos(p_1 z_1) \cos(p_1 z_2) \cos(p_2 z_2) \cos(p_2 z_3) \notag\\
    &\hspace{3cm} \times \Bigg( \log\big( z_1 z_2 H^2\big) \log\big( z_2 z_3 H^2\big) + \log\big( z_1 z_2 H^2\big) \log\Big( \frac{p_2^2 + \vec k^2}{\mu^2}\Big) \label{eq:2_loop_step1}\\
    &\hspace{3cm}\quad + \log\Big( \frac{p_1^2 + \vec k^2}{\mu^2} \Big) \log\big( z_2 z_3 H^2\big) + \log\Big(\frac{p_1^2 + \vec k^2}{\mu^2}\Big) \log\Big(\frac{p_2^2 + \vec k^2}{\mu^2}\Big) \Bigg). \notag
\end{align}
For the first term that only involves logarithms of $z$, we can perform both $p$ integrals to get delta functions, according to \eqref{eq:cosine_delta}, that set all $z$ equal. However, in order to have a more homogeneous result, we then reintroduce one $p$ integral such that the first integrand contains $\log^2\big(z_1 z_3 H^2\big)$. For the middle two terms that both have one logarithm with $z$ and one with $p$, we perform the $p_1$ or $p_2$ integral, depending on which $p$ does not appear in the logarithms. This gives us again a delta function that sets the $z$ in the other logarithm to be the same. Finally, for the last term, we integrate over $z_2$, since
\begin{align}
    \int\limits_0^\infty \d z \cos(p z) \cos(q z) &= \frac{\pi}{2} \Big( \delta(p+q) + \delta(p-q) \Big) \ \hat{=}\, \pi \delta(p-q), \label{eq:cosine_delta_z}
\end{align}
where in the last step, we have used that all integrands we will encounter are symmetric under $p_j \rightarrow -p_j$, so effectively, $\delta(p_j + p_s) \, \hat{=} \, \delta(p_j - p_s)$. Therefore, just like we can integrate the cosines over $p$, we can also integrate them over $z$ to find a delta function of the other variable. Putting all together, we find
\begin{align}
    L_2 &= \Big( \frac{\lambdaR H}{\mu} \Big)^3 C^2 \pi \int\limits_0^\infty \frac{\d z_1}{(z_1 H)^2} \frac{\d z_3}{(z_3 H)^2} \bar G_0(z_1; z_3) \hspace{-1mm} \int\limits_{-\infty}^\infty \d p \cos(p z_1) \cos(p z_3) \notag\\
    &\quad \times \Bigg( \log^2\big(z_1 z_3 H^2\big) + \Big( \log\big( z_1^2 H^2\big) + \log\big( z_3^2 H^2 \big) \Big) \log\Big( \frac{p^2 + \vec k^2}{\mu^2}\Big) + \log^2\Big( \frac{p^2+\vec k^2}{\mu^2}\Big) \Bigg) \\
    &= \Big( \frac{\lambdaR H}{\mu} \Big)^3 C^2 \pi \int\limits_0^\infty \frac{\d z_1}{(z_1 H)^2} \frac{\d z_3}{(z_3 H)^2} \bar G_0(z_1; z_3) \hspace{-1mm} \int\limits_{-\infty}^\infty \d p \cos(p z_1) \cos(p z_3) \log^2 \Big( z_1 z_3 H^2 \frac{p^2+\vec k^2}{\mu^2}\Big). \label{eq:2_loop_final_form}
\end{align}
We can see that the integrand of the form $\log(\ldots) \log(\ldots)$ in \eqref{eq:N_loop_renormalized_final} reduces to $\log^2(\ldots)$ with a single $p$- and two $z$ integrals. This is the de Sitter analog of the factorization in flat space (e.g. \cite{moshe_quantum_2003}). In de Sitter space we then still need to perform the remaining integrals over $p$ and $z$ which is a rather laborious task.

\subsection{3-Loop} \label{sec:3_loop}

A natural conjecture would now be that the structure of the integrand $\log(\ldots) \log(\ldots) \ldots \log(\ldots)$ reduces to $\log^N(\ldots)$ at all orders. This would then allow us to sum over $N$ and find a geometric series. However, unfortunately, this is not the case, as can be shown already at 3-loop order. Ultimately, it comes down to the combinatorics of the various logarithms not adding up, as well as there being a term where the $z$ and $p$ integrals cannot be performed easily to find delta functions using \eqref{eq:cosine_delta} and \eqref{eq:cosine_delta_z}. Like for 2-loop, we split up the integrand to find
\begin{align}
    L_3 &= -\Big( \frac{\lambdaR H}{\mu} \Big)^4 C^3 \int\limits_0^\infty \frac{\d z_1}{(z_1 H)^2} \d z_2 \d z_3 \frac{\d z_4}{(z_4 H)^2} \bar G_0(z_1; z_4) \notag\\
    & \quad \times\int\limits_{-\infty}^\infty \d p_1 \d p_2 \d p_3 \cos(p_1 z_1) \cos(p_1 z_2) \cos(p_2 z_2) \cos(p_2 z_3) \cos(p_3 z_3) \cos(p_3 z_4) \notag\\
    &\quad \times \Bigg( \log\big( z_1 z_2 H^2\big) \log\big( z_2 z_3 H^2\big) \log\big( z_3 z_4 H^2\big) + \log\big( z_1 z_2 H^2\big) \log\big( z_2 z_3 H^2\big) \log\Big( \frac{p_3^2 + \vec k^2}{\mu^2}\Big) \label{eq:3_loop_step1}\\
    &\quad\quad + \log\big( z_1 z_2 H^2\big)\log\Big( \frac{p_2^2 + \vec k^2}{\mu^2} \Big) \log\big( z_3 z_4 H^2\big) + \log\big( z_1 z_2 H^2\big) \log\Big( \frac{p_2^2 + \vec k^2}{\mu^2} \Big) \log\Big( \frac{p_3^2 + \vec k^2}{\mu^2} \Big) \notag\\
    &\quad\quad + \log\Big( \frac{p_1^2 + \vec k^2}{\mu^2} \Big) \log\big( z_2 z_3 H^2\big) \log\big( z_3 z_4 H^2\big) + \boxed{\log\Big( \frac{p_1^2 + \vec k^2}{\mu^2} \Big) \log\big( z_2 z_3 H^2\big) \log\Big( \frac{p_3^2 + \vec k^2}{\mu^2}\Big)} \notag\\
    &\quad\quad + \log\Big( \frac{p_1^2 + \vec k^2}{\mu^2} \Big) \log\Big( \frac{p_2^2 + \vec k^2}{\mu^2} \Big) \log\big( z_3 z_4 H^2\big) + \log\Big( \frac{p_1^2 + \vec k^2}{\mu^2} \Big) \log\Big( \frac{p_2^2 + \vec k^2}{\mu^2} \Big)\log\Big( \frac{p_3^2 + \vec k^2}{\mu^2} \Big)\Bigg). \notag
\end{align}
In order to get down to only one $p$ and two $z$ integrals, we have to perform two integrals over either one $p$ or $z$ that give us a delta function in the other variable, which we can then use to eliminate one more integral. At first glace, this is possible for every term in \eqref{eq:3_loop_step1}, except the one that is highlighted in a box. For that term, we can only perform the $p_2$ integral that gives us one $\delta(z_2-z_3)$, but then we are out of easy integrals. This consideration is elaborated on further in appendix \ref{app:3_loop_details}. The important takeaway is that $L_3$ is truly not the integral over a logarithm to the third power:
\begin{align}
    L_3 \neq -\Big( \frac{\lambdaR H}{\mu}\Big)^4 C^3 \hspace{-1mm} \int\limits_0^\infty \frac{\d z_1}{(z_1 H)^2} \frac{\d z_4}{(z_4 H)^2} \bar G_0(z_1; z_4) \hspace{-1mm} \int\limits_{-\infty}^\infty \d p \cos(p z_1) \cos(p z_4) \log^3\Big( z_1 z_4 H^2 \frac{p^2 + \vec k^2}{\mu^2} \Big). \label{eq:3_loop_not_generalized}
\end{align}

\subsection{Mixed Loops} \label{sec:mixed_loops}

As promised, we will now also consider diagrams that contain what we call \textit{mixed} loops, i.e.\ loops where both a $\phi_-$ and $\phi_+$ field are propagating. Since our external legs are always $\phi_-$, such diagrams necessarily contain \textit{odd vertices} of the type $\phi_-^3 \phi_+$ or $\phi_- \phi_+^3$ that both come with a factor of $\kappa$ each, according to the Feynman rules \eqref{eq:feynman_rules_odd_coupling}. Due to this factor of $\kappa^2$, mixed loops do not contribute at 1-loop order. The question is how they behave for higher loops.

We can already guess that the result somehow depends on how the mixed loops are distributed. If there are multiple mixed loops next to each other, the vertices in between will be of the $\phi_-^2 \phi_+^2$ type and therefore do not come with an extra $\kappa$. We call this a mixed loop \textit{sub-necklace}. If mixed loops are separated by pure loops, there will be more factors of $\kappa$ coming from the \textit{boundary} of the sub-necklaces, where the odd vertices are suppressed.

This argument is elaborated on in appendix \ref{app:N_loop_derivation}. The full result can be found in \eqref{eq:mixed_loop_renormalized_final}. In this section, we will take a look at the 2-loop and 3-loop cases.

\subsubsection*{2-Loop}

According to \eqref{eq:mixed_loop_renormalized_final}, the mixed two-loop diagrams, including all counter-terms, are equal to
\begin{align}
    I_2^\mathtt{MM} &= -\pi^2 \Big(\frac{\lambdaR H}{\mu}\Big)^3 C^2 \hspace{-1mm} \int\limits_0^\infty \frac{\d z_1 \d z_2 \d z_3}{(z_1 H)^2 (z_3 H)^2}  \bar G_0 (z_1; z_{N+1}) \hspace{-2mm} \int\limits_{-\infty}^\infty \d p_1 \d p_2 \sin(p_1 z_1) \sin(p_1 z_2) \sin(p_2 z_2) \sin(p_2 z_3).
\end{align}
There are no logarithms, since both loops are boundary loops. We now use
\begin{align}
    \int\limits_{-\infty}^\infty \d p \sin(p z_1) \sin(p z_2) &= -\frac{1}{4} \int\limits_{-\infty}^\infty \d p \Big( \e^{i p z_1} - \e^{-i p z_1} \Big) \Big( \e^{i p z_2} - \e^{-i p z_2} \Big) \notag\\
    &= \pi \delta(z_1 - z_2) - \pi \delta(z_1 + z_2) \,\hat{=}\, \pi \delta(z_1 - z_2). \label{eq:sine_delta}
\end{align}
In the last step, we have used again that the $z_j$ are integrated from $0$ to $\infty$ and therefore, $\delta(z_1 + z_2)$ is never relevant. Therefore, we find that the 2-loop mixed expression is proportional to the cross
\begin{align}
    I_2^\mathtt{MM} = -\pi^4 \Big(\frac{\lambdaR H}{\mu}\Big)^3 C^2 \int\limits_0^\infty \frac{\d z}{(z H)^4} \bar G_0(z;z) = -\pi^4 \Big(\frac{\lambdaR H C}{\mu}\Big)^2 I_\times \big\rvert_{\kappa \rightarrow 0}.
\end{align}
We have used the $I_\times$ from \eqref{eq:counter_term_cross} in the limit $\kappa \rightarrow 0$.

\subsubsection*{3-Loop}

For the 3-Loop case, there are multiple diagrams that contribute. Note that in \eqref{eq:mixed_loop_renormalized_final}, we did not sum over all possible combinations yet, because the expressions are qualitatively different for different distributions of the mixed loops. We have the diagrams \texttt{MMM}, \texttt{MMP} and \texttt{PMM} and their counter-terms. We denote their sum by $L_3^{\Sigma \mathtt{M}} = I_3^\mathtt{MMM} + I_3^\mathtt{MMP} + I_3^\mathtt{PMM}$ and find
\begin{align}
    L_3^{\Sigma \mathtt{M}} &= \pi^2 \Big(\frac{\lambdaR H}{\mu}\Big)^{4} C^3 \int\limits_0^\infty \frac{\d z_1}{(z_1 H)^2} \d z_2 \d z_3 \frac{\d z_4}{(z_4 H)^2} \bar G_0(z_1; z_4) \int\limits_{-\infty}^\infty \d p_1 \d p_2 \d p_3 \notag\\
    &\quad \times \Bigg( \sin(p_1 z_1) \sin(p_1 z_2) \sin(p_2 z_2) \sin(p_2 z_3) \sin(p_3 z_3) \sin(p_3 z_4) \log\Big(z_2 z_3 H^2 \frac{p_2^2 + \vec k^2}{\mu^2} \Big) \\
    &\quad\quad + \sin(p_1 z_1) \sin(p_1 z_2) \sin(p_2 z_2) \sin(p_2 z_3) \cos(p_3 z_3) \cos(p_3 z_4) \log\Big(z_3 z_4 H^2 \frac{p_3^2 + \vec k^2}{\mu^2} \Big) \notag\\
    &\quad\quad + \cos(p_1 z_1) \cos(p_1 z_2) \sin(p_2 z_2) \sin(p_2 z_3) \cos(p_3 z_3) \cos(p_3 z_4) \log\Big(z_1 z_2 H^2 \frac{p_1^2 + \vec k^2}{\mu^2} \Big) \Bigg). \notag
\end{align}
Now, we perform the easy $p$ integrals for the boundary mixed loops using \eqref{eq:sine_delta} and get
\begin{align}
    L_3^{\Sigma \mathtt M} &= \pi^4 \Big(\frac{\lambdaR H}{\mu}\Big)^{4} C^3 \int\limits_0^\infty \frac{\d z_1}{(z_1 H)^2}\frac{\d z_4}{(z_4 H)^2} \bar G_0(z_1; z_4) \int\limits_{-\infty}^\infty \d p \notag\\
    &\quad \times \Big( \sin(p z_1) \sin(p z_4) + \cos(p z_1) \cos(p z_4) + \cos(p z_1) \cos(p z_4) \Big) \log\Big(z_1 z_4 H^2 \frac{p^2 + \vec k^2}{\mu^2} \Big). \label{eq:mixed_3_loop_final}
\end{align}
This contains the pure 1-loop result \eqref{eq:1_loop_log}, and a similar term where the cosines are replaced by sines. It is not possible for the mixed loops to cancel the unpleasant terms in \eqref{eq:3_loop_step1} because of their transcendentality: The 3-loop terms involve polylogarithms, while the terms in \eqref{eq:mixed_3_loop_final} have effectively the pure 1-loop kinematic dependence, which only involves logarithms.

\section{Alternative Regularization Scheme} \label{sec:new_regularization}

We have seen that the shadow action formalism combined with a manifestly covariant analytic regularization does not produce necklace diagrams that can be easily resummed. There is no nice pattern in higher loops, leading to  more and more integrals. Furthermore,  taking the shift of the conformal dimensions $\Delta_\pm \rightarrow \Delta_\pm - \kappa$ literally, we obtain mixed loops that introduce combinatorically complicated expressions, where we have to be careful about sub-necklaces and their boundary loops. The $\kappa$-suppressed vertices generating mixed loops \eqref{eq:feynman_rules_odd_coupling} could be interpreted as analogs of evanescent operators in flat space. The latter are operators that appear when dimensionally regularizing the action and vanish after setting the regulator to zero, but can produce finite terms in loop diagrams. These operators are known to be regularization dependent \cite{herrlich_evanescent_1995}, and sometimes can be completely eliminated by choosing a suitable scheme (e.g.\ \cite{fuentes-martin_evanescent_2023}).

In this section, we introduce a new formalism that alleviates all these symptoms, by asking a simple question: Do we renormalize the loop diagram, or the action? Given that the physical observables are correlation functions rather than the Lagrangian, we propose a variant of analytic regularization where the $\kappa$-deformation of the $z$-measure is localized at points where the external energy and momentum flows into the diagram. Crucially, it also completely eliminates the mixed loops. We then show that this scheme reproduces the 1- and 2-loop correlators in analytic regularization. Furthermore, the higher loop diagrams satisfy the de Sitter Ward identities. 

\subsection{Implementation}

The idea is simple. Looking at our propagators \eqref{eq:feynman_rules_propagator-} and \eqref{eq:feynman_rules_propagator+}, the $\kappa$ in
\begin{align}
    \big(z_1 z_2 H^2\big)^{1-\kappa}
\end{align}
has been put there only to make the propagator invariant under rescaling
\begin{equation}
    \vec \ell \rightarrow \frac{1}{\eta} \vec \ell, \quad p \rightarrow \frac{1}{\eta} p, \quad z \rightarrow \eta z.\label{eq:rescaling}
\end{equation}
However, this step causes more problems than it solves. If it was not there, the logarithms in \eqref{eq:N_loop_renormalized_final} would not depend on $z_j$ for $2 \leq j \leq N$, only on the first and last $z_1$ and $z_{N+1}$. We would then be able to perform the integrals over the intermediate $z_j$ using \eqref{eq:cosine_delta_z}. Of course, getting rid of these factors breaks our rescaling invariance. Therefore, we simply propose to evenly distribute all of the $z_j^\kappa$ factors among the external propagators. The resulting theory then does not really possess Feynman rules in the usual sense any more, since the propagators depend on the diagram topology. However, as argued before, it should be sufficient to provide any recipe to compute cosmological correlators. It does not need to be Feynman rules in the classical sense.

Further, we should drop the interpretation that modifying the propagator changes the conformal dimension of the field and just view it as an ad-hoc trick to make our computations well-behaved. As was seen in \cite{chowdhury_subtle_2024}, the pure loop diagrams satisfy the Ward identities on their own, and introducing odd vertices and mixed loops is not needed to restore symmetry. Therefore, we simply drop them. This also eliminates any occurence of $\kappa$ inside of the action \eqref{eq:action_final}, effectively sending $\mathcal C_{\kappa} \rightarrow 1$.

\medskip

Computing the $N$-loop necklace expression in our new approach is now very simple. First, we drop the notion of mixed loops, so we only consider pure loops from \eqref{eq:N_loop_renormalized}. Then, we simply replace every occurrence of $z_j z_{j+1}$ in the derivation with $z_1 z_{N+1}$. That gives us
\begin{align}
    L_N &= (-1)^N \Big(\frac{\lambdaR H}{\mu}\Big)^{N+1} C^N \int\limits_0^\infty \frac{\d z_1}{(z_1 H)^2} \d z_2 \ldots \d z_N \frac{\d z_{N+1}}{(z_{N+1} H)^2} \bar G_0(z_1; z_{N+1}) \int\limits_{-\infty}^\infty \d p_1 \ldots \d p_N \notag\\
    &\hspace{1cm}\times \cos(p_1 z_1) \ldots \cos(p_N z_{N+1}) \log\Big( z_1 z_{N+1} H^2 \frac{p_1^2 + \vec k^2}{\mu^2}\Big) \ldots \log\Big( z_1 z_{N+1} H^2 \frac{p_N^2 + \vec k^2}{\mu^2}\Big) \\
    &= (-1)^N \Big(\frac{\lambdaR H}{\mu}\Big)^{N+1} C^N \pi^{N-1} \int\limits_0^\infty \frac{dz_1}{(z_1 H)^2} \frac{z_{N+1}}{(z_{N+1} H)^2} \bar G_0(z_1; z_{N+1}) \int\limits_{-\infty}^\infty \d p \notag\\
    &\hspace{1cm} \times \cos(p z_1) \cos(p z_{N+1}) \log^N \Big( z_1 z_{N+1} H^2 \frac{p^2 + \vec k^2}{\mu^2} \Big). \label{eq:N_loop_renormalized_new}
\end{align}
We have used the identity \eqref{eq:cosine_delta_z} to integrate out the intermediate $z_j$. It can be immediately read off that for 1-loop, the above result agrees with the previous result \eqref{eq:1_loop_log_final}. For 2-loop, we have computed that the previous result \eqref{eq:2_loop_final_form} coincides with the above \eqref{eq:N_loop_renormalized_new}. 

The differences start at 3-loops, where we have seen in \eqref{eq:3_loop_not_generalized} that the previous formalism fails to produce a simple form, which was the motivation for this new approach. The difference are precisely the terms that we called "obstructions" earlier.

\subsection{Ward Identity}

We have promised that our new approach does not sacrifice any symmetries of the system. To prove this, we now show that the $N$-loop necklace diagrams indeed fulfill the conformal Ward identities, which are the late-time limit of the de Sitter Ward identities \cite{Bzowski:2013sza}. Rotations of the momenta, as well as the aforementioned rescaling \eqref{eq:rescaling} are already manifest symmetries, so we only have to consider the Ward identities of the special conformal transformations. Following the discussion in \cite{chowdhury_subtle_2024}, we define conformal cross ratios
\begin{align}
    u = \frac{k}{k_{12}}, \ v = \frac{k}{k_{34}} \label{eq:uv_def}
\end{align}
and re-scale the correlator by multiplying with $k\, k_1 k_2 k_3 k_4$. We then have to check whether
\begin{align}
    \big( \Delta_u - \Delta_v \big) \big( k\, k_1 k_2 k_3 k_4 L_N \big) \overset{?}{=} 0, \label{eq:ward_identity}
\end{align}
where
\begin{align}
    \Delta_x = x^2 \big(1-x^2\big) \partial_x^2 - 2 x^3 \partial_x.
\end{align}
Since we cannot evaluate \eqref{eq:N_loop_renormalized_new} for general $N$ completely, we have to check the Ward identity under the integral. In order to make this easier, we can eliminate as many integrals as possible. It turns out that the $z$ integrals can actually be performed using the identity
\begin{align}
    \int\limits_0^\infty \d z \ \e^{-A z} \log^n(B z) &= \frac{\d^n}{\d \alpha^n} \underbrace{\int\limits_0^\infty \d z \ \e^{-A z} (B z)^\alpha}_{=B^\alpha \Gamma(\alpha + 1) A^{-\alpha-1}} \big\rvert_{\alpha = 0} \\
    &= \frac{1}{A} \sum\limits_{l=0}^n \binom{n}{l} \log^{n-l}\Big( \frac{B}{A} \Big) \frac{\d^l}{\d \alpha^l} \Gamma(\alpha) \big\rvert_{\alpha=1}. \label{eq:log_exp_integral}
\end{align}
There is no closed algebraic form for the $l$-th derivative of the $\Gamma$ function, so we have to leave it at that. We now insert the bulk to boundary propagator \eqref{eq:bulk_to_boundary_prop} into \eqref{eq:N_loop_renormalized_new} and write the cosines as exponential functions to get
\begin{align}
    L_N &= (-1)^N \Big( \frac{\lambdaR H}{\mu} \Big)^{N+1} \frac{C^N \pi^{N-1}}{64 k_1 k_2 k_3 k_4} \int\limits_0^\infty \d z_1 \d z_{N+1} \int\limits_{-\infty}^\infty \d p \log^N\Big( z_1 z_{N+1} H^2 \frac{p^2 + k^2}{\mu^2} \Big) \notag\\
    &\quad\quad\quad \times \Big( \e^{-(k_{12} + i p) z_1} + \e^{-(k_{12} - i p)z_1} \Big) \Big( \e^{-(k_{34} + i p) z_{N+1}} + \e^{-(k_{34} - i p)z_{N+1}} \Big) \\
    &= (-1)^N \Big( \frac{\lambdaR H}{\mu} \Big)^{N+1} \frac{C^N \pi^{N-1}}{64 k_1 k_2 k_3 k_4} \int\limits_0^\infty \d z_{N+1} \int\limits_{-\infty}^\infty \d p \sum\limits_{l=0}^N \ 1 \notag\\
    &\quad\quad\quad \times \Bigg( \frac{1}{k_{12}+ip} \binom{N}{l} \log^{N-l}\Big( \frac{H^2}{\mu^2} z_{N+1} \frac{p^2+k^2}{k_{12} + ip} \Big) \frac{\d^l}{\d \alpha^l} \Gamma(\alpha) \big\rvert_{\alpha = 1} + (i p \rightarrow -ip) \Bigg) \\
    &= (-1)^N \Big( \frac{\lambdaR H}{\mu} \Big)^{N+1} \frac{C^N \pi^{N-1}}{64 k_1 k_2 k_3 k_4} \int\limits_{-\infty}^\infty \d p \sum\limits_{l=0}^N \sum\limits_{s=0}^{N-l}  \Bigg( \binom{N}{l} \binom{N-l}{s} \frac{1}{k_{12} + ip} \frac{1}{k_{34} + ip} \\
    &\quad\quad\quad \times \log^{N-l-s}\Big( \frac{H^2}{\mu^2} \frac{p^2+k^2}{(k_{12} + ip)(k_{34} + ip)} \Big) \frac{\d^l}{\d \alpha^l} \Gamma(\alpha)\big\rvert_{\alpha = 1} \frac{\d^s}{\d \beta^s} \Gamma(\beta) \big\rvert_{\beta = 1} + (\mathrm{sign \ permutations}) \Bigg). \notag
\end{align}
By "sign permutations", we mean the other terms that come from the different signs in the exponents of the expanded cosines. Now, we express this in terms of $u$ and $v$, as defined in \eqref{eq:uv_def}, by substituting $q = p/k$, and find
\begin{align}
    L_N &= (-1)^N \Big( \frac{\lambdaR H}{\mu} \Big)^{N+1} \frac{C^N \pi^{N-1}}{64 k_1 k_2 k_3 k_4} \int\limits_{-\infty}^\infty \d q \, k \sum\limits_{l=0}^N \sum\limits_{s=0}^{N-l}  \Bigg( \binom{N}{l} \binom{N-l}{s} \frac{1}{k} \frac{u}{1 + i q u} \frac{1}{k} \frac{v}{1 + i q v} \\
    &\quad \times \underbrace{\log^{N-l-s}\Big( \frac{H^2}{\mu^2} u v \frac{q^2 + 1}{(1 + i q u)(1 + i q v)} \Big)}_{=:L^{N-l-s}(q, u, v)} \frac{\d^l}{\d \alpha^l} \Gamma(\alpha)\big\rvert_{\alpha = 1} \frac{\d^s}{\d \beta^s} \Gamma(\beta) \big\rvert_{\beta = 1} + (\mathrm{sign \ permutations}) \Bigg). \notag
\end{align}
We know from \cite{chowdhury_subtle_2024} that the first and second loop order $L_1$ and $L_2$ fulfill the Ward identities. These $L_1$ and $L_2$ involve $L^n(q, u, v)$ with different exponents $n$ and complicated coefficients generated by derivatives of the Gamma function. These coefficients involve the Riemann-Zeta function $\zeta(z)$ at integer values and the Bell-polynomials $B_n(x_1,\ldots, x_n)$, according to
\begin{align}
    \frac{\d^n}{\d \alpha^n} \Gamma(\alpha) \big\rvert_{\alpha=1} = (-1)^n B_n\Big(\gamma_\mathrm{E}, 1! \zeta(2), \ldots, (n-1)!\zeta(n)\Big).
\end{align}
These numbers cannot be generated by acting with derivatives on $L^n(q,u,v)$. It is therefore a natural hypothesis that the individual terms proportional to $L(q, u, v)^{N-l-s}$ fulfill the Ward identities on their own. We will now show that this is indeed the case, which proves that the Ward identity holds for their sum.

For this, we apply the Ward identity \eqref{eq:ward_identity} to a single part involving $L^n(q, u, v)$. Since $\Delta_{u/v}$ involve second derivatives and polynomials in $u$ and $v$, we expect the functional shape
\begin{align}
    (\Delta_u - \Delta_v) &\Big( k\, k_1 k_2 k_3 k_4 f(q, u, v) \Big) = (\Delta_u - \Delta_v) \Bigg(\frac{u}{1+i q u} \frac{v}{1+iqv} L^n(q,u,v) \Bigg) \\
    &= f_1 (q, u, v) L^n(q, u, v) + f_2 (q, u, v) L^{n-1}(q, u, v) + f_3(q, u, v) L^{n-2}(q, u, v)\label{eq:ward_identity_step1}
\end{align}
with some algebraic functions $f_{1/2/3}(q,u,v)$. We now show that this is a total derivative. Therefore, the $q$-integral over this is zero, proving the Ward-identity to hold. 

Integrating $f_1(q,u,v)$ with respect to $q$ to $F_1(q, u, v)$ and multiplying with $L^n(q, u, v)$ allows us to write the first term in \eqref{eq:ward_identity_step1} as a total derivative plus some other term that fits into the shape of \eqref{eq:ward_identity_step1}:
\begin{align}
    \frac{\d}{\d q} \Big( F_1(q, u, v) L^n(q, u, v) \Big) = f_1(q, u, v) L^n(q, u, v) + n F_1(q, u, v) L^{n-1}(q, u, v) \frac{\partial}{\partial q} L(q, u, v).
\end{align}
Now, we repeat this step for the next term and integrate
\begin{align}
    f_2(q,u,v) - n F_1(q, u, v).
\end{align}
Interestingly, this yields exactly $n F_1(q, u, v)$ again. This is where the procedure terminates. After performing the algebra, we find
\begin{align}
    (\Delta_u \hspace{-1mm} - \hspace{-1mm} \Delta_v) \Bigg(\frac{u v}{1+i q u} \frac{L^n(q,u,v)}{1+iqv} \Bigg) \hspace{-1mm} = \frac{\d}{\d q} \Bigg( \hspace{-2mm} -i \frac{ uv(u-v)(1+q^2)}{(1+iqu)^2(1+iqv)^2} \Big( L^n(q, u, v) + n L^{n-1}(q, u, v) \Big) \hspace{-1mm}\Bigg).
\end{align}
The function under the differential on the right hand side falls off quickly for $|q|\rightarrow\infty$, so the boundary terms of the $q$ integral are zero and the Ward identity holds.

\subsection{Resummation} \label{sec:resummation}

Now, we want to demonstrate the true benefit of this new regularization scheme: We can properly resum over all $N$-loop necklace diagrams. We rename $z_{N+1} \rightarrow z_2$ so that all integrals are now over the same variables and combine the constant coefficients in front into
\begin{align}
    K := \frac{\lambdaR H}{\mu} C \pi.
\end{align}
We then sum
\begin{align}
    L_\Sigma &= \sum\limits_{N=0}^\infty L_N \\
    &= \frac{\lambdaR H}{\pi \mu} \int\limits_0^\infty \frac{\d z_1 \d z_2 \bar G_0(z_1; z_2) }{(H z_1)^2 (H z_2)^2}  \hspace{-1mm} \int\limits_{-\infty}^\infty \hspace{-1mm} \d p \cos(p z_1) \cos(p z_2) \sum\limits_{N=0}^\infty (-1)^N K^N \log^N \Big( z_1 z_2 H^2 \frac{p^2 + k^2}{\mu^2} \Big) \label{eq:todo_resum} \\
    &= \frac{\lambdaR H}{16 \pi \mu k_1 k_2 k_3 k_4} \int\limits_0^\infty \d z_1 \d z_2 \, \e^{-k_{12}z_1} \e^{-k_{34}z_2} \hspace{-1mm} \int\limits_{-\infty}^\infty \hspace{-1mm} \d p \cos(p z_1) \cos(p z_2) \frac{1}{1 + K \log\big( z_1 z_2 H^2 \frac{p^2 + k^2}{\mu^2} \big)}. \label{eq:resummed_correlator}
\end{align}
The integrand has a pole at
\begin{align}
    1 + K \log\Big( z_1 z_2 H^2 \frac{p^2 + k^2}{\mu^2} \Big) = 0,
\end{align}
which is always integrated over for small $z_1$ or $z_2$. This pole comes from using the geometry series outside of its area of validity. This pole can be remedied, however, by looking back at the Wick rotation from de Sitter space to EAdS, which is used in \cite{pietro_analyticity_2022} to derive the action that we stated in \eqref{eq:general_action}. 

\begin{figure}
    \centering
    \def\svgwidth{.6\columnwidth}
    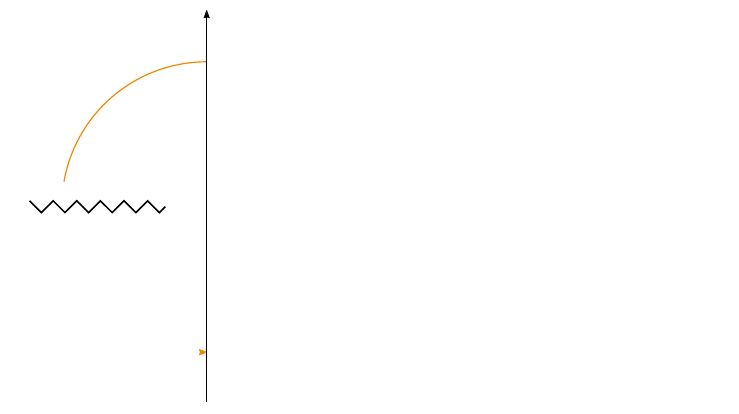 
    \caption{Modified Wick rotation procedure to turn de Sitter correlators into EAdS correlators. We under-rotate the time-ordered $\eta_\mathrm{T}$ and over-rotate the anti-time-ordered $\eta_\mathrm{A}$ with respect to \cite{pietro_analyticity_2022} by multiplying them with $e^{i\epsilon}$ for some small $\epsilon$.}
    \label{fig:wick_rotation}
\end{figure}

To understand this more thoroughly, we will for a moment go back to the Schwinger-Keldysh formalism where there are two types of vertices, time-ordered and anti-time-ordered. As an example, we can consider the time-ordered vertices. According to \cite{weinberg_quantum_2005}, these vertices always come with a factor of $(-i)$ and are integrated over conformal time $\eta_\mathrm{T}$ from $-\infty$ to $0$ along a contour that is slightly above the negative real axis to avoid a branch-cut. The Wick rotation turns the integration contour upwards to align it with the imaginary axis, as illustrated in the left half of figure \ref{fig:wick_rotation}. For clarity, we set $H=1$ in the following. The integral over conformal time then is transformed as
\begin{align}
    -i \int\limits_{-\infty}^0 \frac{\d \eta_\mathrm{T} f(\eta_\mathrm{T})}{(-\eta_\mathrm{T})^{d+1}} &= -i \int\limits_{i \infty}^0 \frac{\d \eta_\mathrm{T} f(\eta_\mathrm{T})}{(-\eta_\mathrm{T})^{d+1}} = (-i)^2 \int\limits_0^\infty \frac{\d z f(i z)}{(-i z)^{d+1}}  = e^{(d-1) \frac{i \pi}{2}} \int\limits_0^\infty \frac{\d z f(iz)}{(z)^{d+1}} . \label{eq:wick_rotation_details}
\end{align}
We have used the reparametrization 
\begin{align}
    \eta_\mathrm{T} = i z.
\end{align}
Since we find a pole in $z_1 z_2$ in the resummed correlator \eqref{eq:resummed_correlator}, we conclude that we should not rotate exactly onto the imaginary axis to avoid hitting the pole directly. Instead of $\eta_\mathrm{T}$ running from $i\infty$ to $0$, we would like it to go from $i\infty - \epsilon$ to $0$. This is equivalent to multiplying $\eta_\mathrm{T}$ by $e^{i\epsilon}$, or equivalently replacing $z \rightarrow z e^{-i\epsilon}$ everywhere in \eqref{eq:wick_rotation_details}. We can also do the same for the anti-time-ordered $\eta_\mathrm{A}$. There, we find that in order to get the same sign of $\epsilon$, we should over-rotate instead of under-rotate, as demonstrated in figure \ref{fig:wick_rotation}. This prescription vanishes everywhere in the limit $\epsilon \rightarrow 0$, except when there is a pole, which is the case in the resummed correlator \eqref{eq:resummed_correlator}. In the end, we effectively replace
\begin{align}
    \frac{1}{1 + K \log\big( z_1 z_2 H^2 \frac{p^2 + k^2}{\mu^2} \big)} \xrightarrow{\mathrm{replace}} \frac{1}{1 + K \log\big( z_1 e^{-i\epsilon} z_2 e^{-i \epsilon} H^2 \frac{p^2 + k^2}{\mu^2} \big)} = \frac{1}{1 + K \log\big( z_1 z_2 H^2 \frac{p^2 + k^2}{\mu^2} \big) - i \tilde\epsilon}.
\end{align}
In the last step, we have rescaled $\epsilon$ to $\tilde\epsilon$. This is equivalent to introducing an ad-hoc $i\epsilon$ prescription to regularize the pole. As a byproduct, our correlator is now imaginary, which should not be the case for a real field. Therefore, we could now add the same term but with a negative $\epsilon$ to take the real part, which according to the Sokhotski-Plemelj theorem is equivalent to taking the principal value of the integral \eqref{eq:resummed_correlator}. Either way, the sign of $\tilde\epsilon$ is arbitrary.

With this prescription, all integrals over $z_1$, $z_2$ and $p$ are now convergent. Unfortunately, they are not solvable in the space of common functions like hypergeometrics, so we have to resort to numerics. For that, we once again rescale the integration variables such that the function has an overall factor of $1/(16 k_1 \, k_2 \, k_3 \, k_4 \, k)$, which we omit from now on because it is always the same. As a result, the quantities only depend on $u$ and $v$. Then, we set the renormalization scale $\mu$ to the canonical choice $H$. To make the numerical integration well-behaved, we further regularize the $p$ integral by multiplying with $e^{-\epsilon p^2}$. Lastly, we cut off the integrals at some value $\Lambda$ for numerical evaluation. We apply this treatment to both the individual $L_N$ and the resummed $L_\Sigma$ to define new quantities
\begin{align}
    \tilde L_N(u,v) &= 16 k_1 \, k_2 \, k_3 \, k_4 \, k L_N \notag\\
    &= \frac{\lambdaR}{\pi} \Big( \hspace{-1mm} -\frac{\lambdaR}{16\pi^2} \Big)^N \hspace{-1mm} \int\limits_0^\Lambda \hspace{-1mm} \d z_1 \d z_2 \hspace{-1mm} \int\limits_{-\Lambda}^\Lambda \hspace{-1mm} \d p \, e^{-\frac{z_1}{u} - \frac{z_2}{v} - \epsilon p^2}\cos(p z_1) \cos(p z_2) \log^N\big( z_1 z_2 (p^2 + 1) \big), \label{eq:tildeLN} \\
    \tilde L_\Sigma(u,v) &= 16 k_1 \, k_2 \, k_3 \, k_4 \, k L_\Sigma = \frac{\lambdaR}{\pi} \int\limits_0^\Lambda \d z_1 \d z_2 \int\limits_{-\Lambda}^\Lambda \d p \frac{e^{-\frac{z_1}{u} - \frac{z_2}{v} - \epsilon p^2}\cos(p z_1) \cos(p z_2)}{1 + \frac{\lambdaR}{16\pi^2} \log\big( z_1 z_2 (p^2 + 1) \big) + i \epsilon}. \label{eq:tildeLSummed}
\end{align}
Unless otherwise stated, we will set $\epsilon = 0.001$ and $\Lambda = 100$ in the following analysis. We have checked that the qualitative features of the following data do not depend on $\epsilon$ or $\Lambda$. In the following, all plots of $\tilde L_N$ and $\tilde L_\Sigma$ over $u$ and $v$ are computed in python on a $101\times101$-value $(u,v)$-grid\footnote{We avoid dividing by $0$ by clamping the values of $u$ and $v$ to be at least $10^{-5}$.} using numerical Monte-Carlo methods with the VEGAS algorithm \cite{lepage_adaptive_2021}. We are using $20$ warmup iterations with $100\,000$ integrand evaluations each, followed by $40$ integration iterations with $1\,000\,000$ integrand evaluations each. With these settings, the error estimation of all shown plots is of order $1\%$ of the values, such that the qualitative features are not impacted by numerical uncertainty. Analytic results are computed directly in python on the same grid.

\begin{figure}[htbp]
  \centering

  \hspace{-5mm}\begin{subfigure}[b]{0.45\textwidth}
    \def\svgwidth{1.2\columnwidth}
    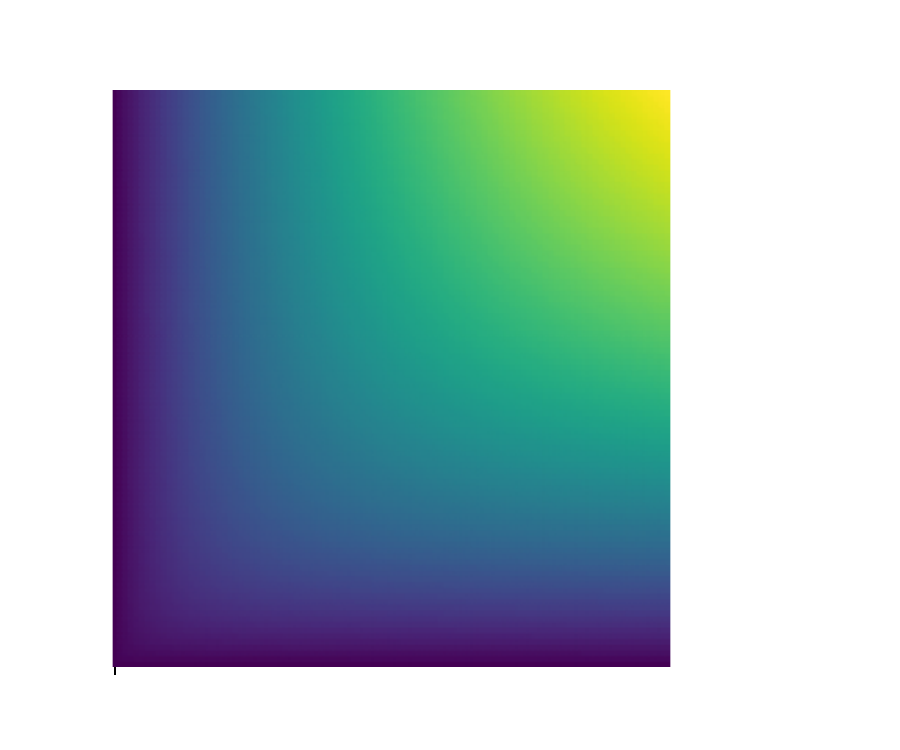
    \caption{Analytic result of $\tilde L_0(u,v)$.}
    \label{fig:l0}
  \end{subfigure}
  \hspace{0.75cm}
  \begin{subfigure}[b]{0.45\textwidth}
    \def\svgwidth{1.2\columnwidth}
    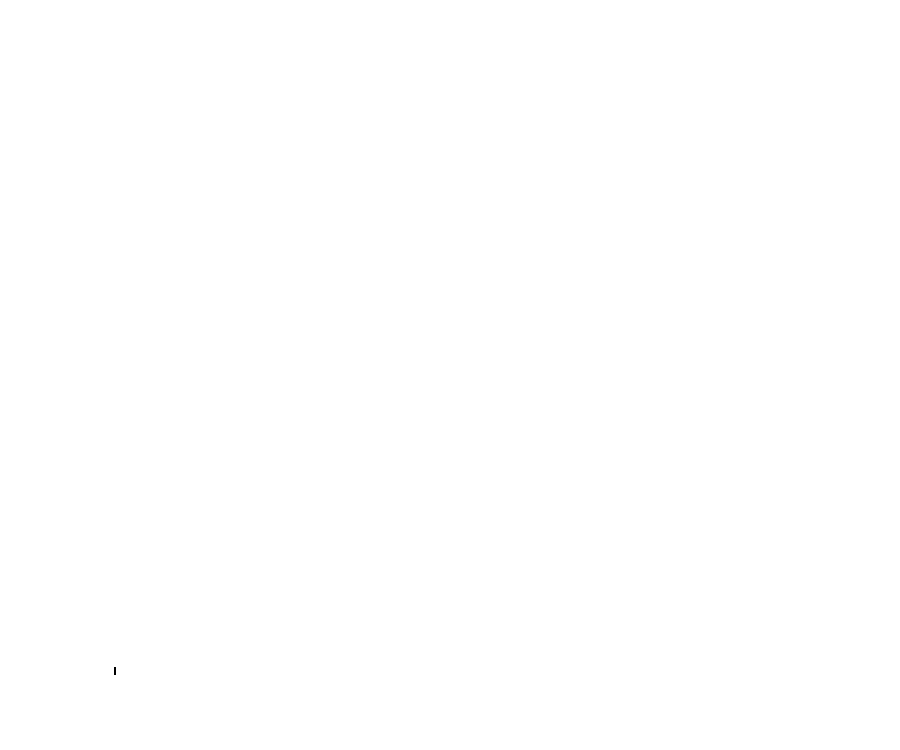
    \caption{Analytic result of $\tilde L_1(u,v)$.}
    \label{fig:l1}
  \end{subfigure}


  \hspace{-5mm}\begin{subfigure}[b]{0.45\textwidth}
    \def\svgwidth{1.2\columnwidth}
    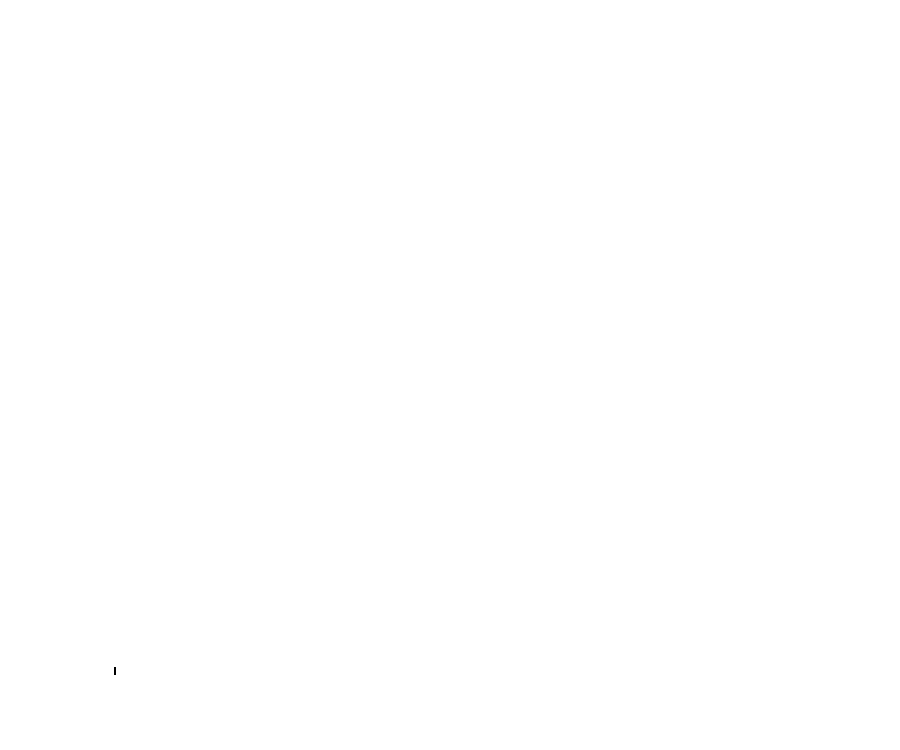
    \caption{Numerical Monte-Carlo result of $\tilde L_2(u,v)$.}
    \label{fig:l2}
  \end{subfigure}
  \hspace{0.75cm}
  \begin{subfigure}[b]{0.45\textwidth}
    \def\svgwidth{1.2\columnwidth}
    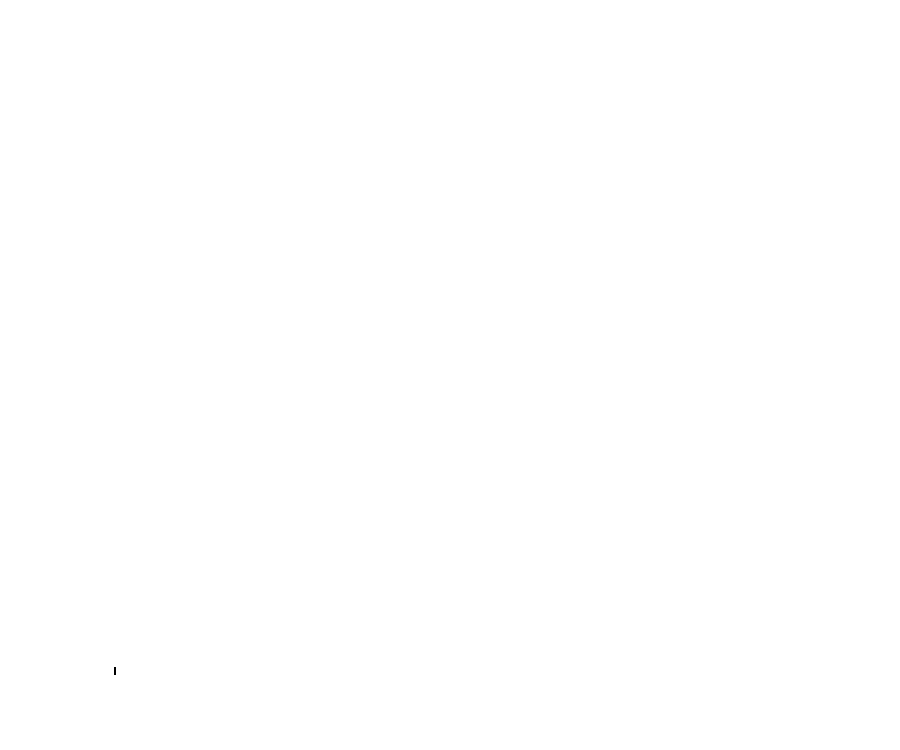
    \caption{Numerical Monte-Carlo result of $\tilde L_3(u,v)$.}
    \label{fig:l3}
  \end{subfigure}

  \caption{Dependence of the perturbative orders $\tilde L_N$ for $N=0,1,2,3$ on the kinetic variables $u$ and $v$. The coupling constant is chosen such that the effective expansion parameter $\lambdaR/(16\pi^2)$ is unity. Orders $N=0,1$ are computed from the analytic expressions that are given in \cite{chowdhury_subtle_2024}. The qualitative features of all shown orders are similar: The value of $\tilde L_N$ approaches $0$ as $u$ or $v$ approach $0$. For all orders, the value is highest for $u = v = 1$, where $\tilde L_N$ is positive. The highest value, as well as the size of the area with large values, increases towards higher orders.}
  \label{fig:lN_numerical}
\end{figure}

The results for the individual orders $\tilde L_N$ for the first few $N=0,1,2,3$ are shown in figure \ref{fig:lN_numerical}. We have chosen to show their values for $\lambdaR = 16\pi^2$, where the effective expansion coefficient is $1$, according to \eqref{eq:tildeLN}. Other $\lambdaR$ can be reached by simply multiplying with $(\lambda / 16\pi^2)^N$. The key observations are
\begin{itemize}
    \item The value of $\tilde L_N(u,v)$ goes to $0$ if $u$ or $v$ approach $0$, for all $N$ that we computed.
    \item For increasing $u$ and $v$, the value of $\tilde L_N(u,v)$ grows positive, and approaches its maximum at $u=v=1$.
    \item For higher the loop order $N$, the  regions around $u = v = 1$ flatten out in the upper right corners.
    \item The maximum value grows with the loop order.
\end{itemize}

\begin{figure}[htbp]
  \centering

  \hspace{-5mm}\begin{subfigure}[b]{0.45\textwidth}
    \def\svgwidth{1.2\columnwidth}
    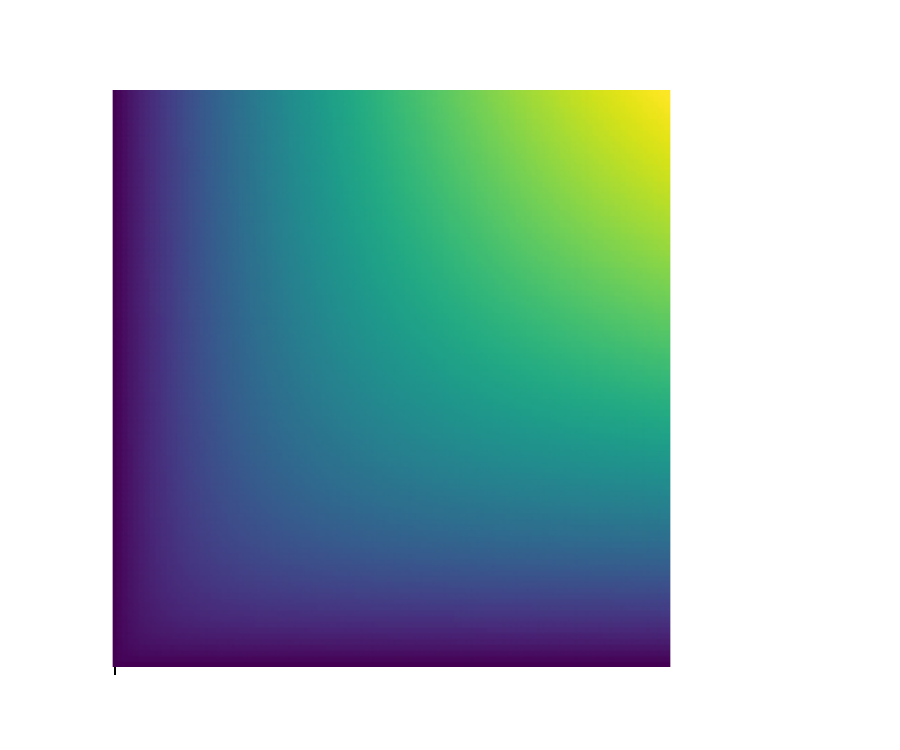
    \caption{Analytic result of the tree level $\tilde L_0(u,v)$.}
    \label{fig:l0lambda1}
  \end{subfigure}
  \hspace{0.75cm}
  \begin{subfigure}[b]{0.45\textwidth}
    \def\svgwidth{1.2\columnwidth}
    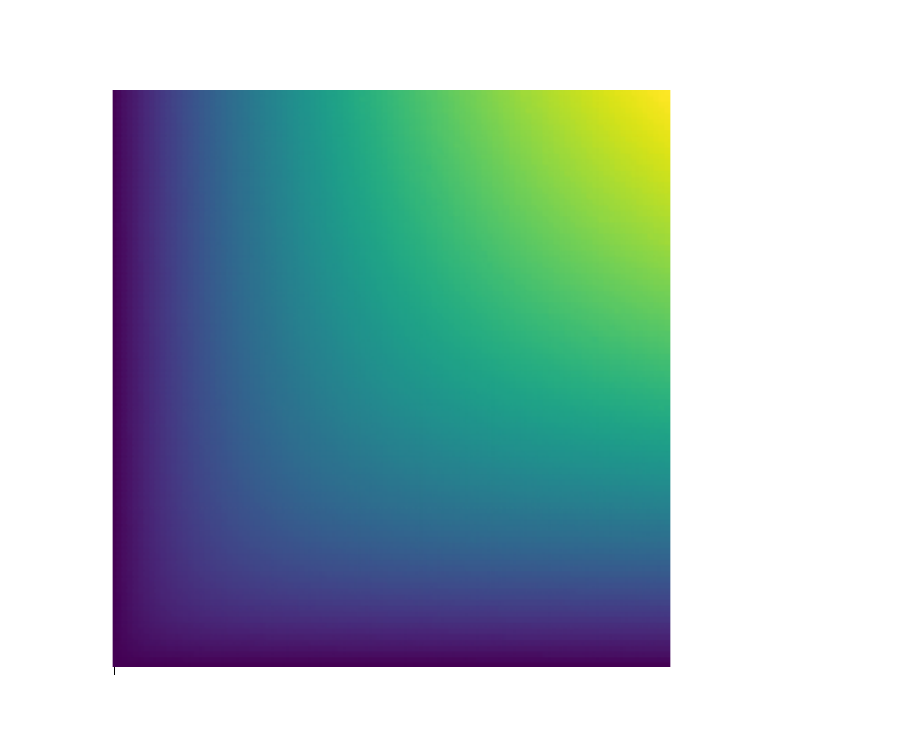
    \caption{Numerical Monte-Carlo result of $\mathrm{Re}\tilde L_\Sigma(u,v)$.}
    \label{fig:reslambda1}
  \end{subfigure}


  \hspace{-5mm}\begin{subfigure}[b]{0.45\textwidth}
    \def\svgwidth{1.2\columnwidth}
    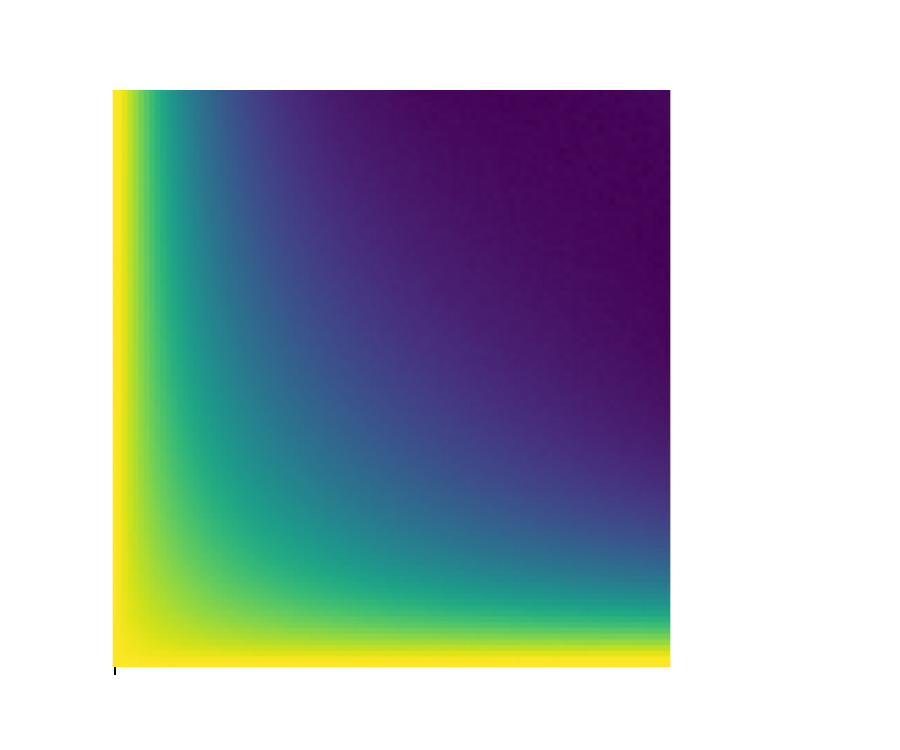
    \caption{Numerical Monte-Carlo result of $\mathrm{Re}\tilde L_\Sigma(u,v)$.}
    \label{fig:reslambda16}
  \end{subfigure}
  \hspace{0.75cm}
  \begin{subfigure}[b]{0.45\textwidth}
    \def\svgwidth{1.2\columnwidth}
    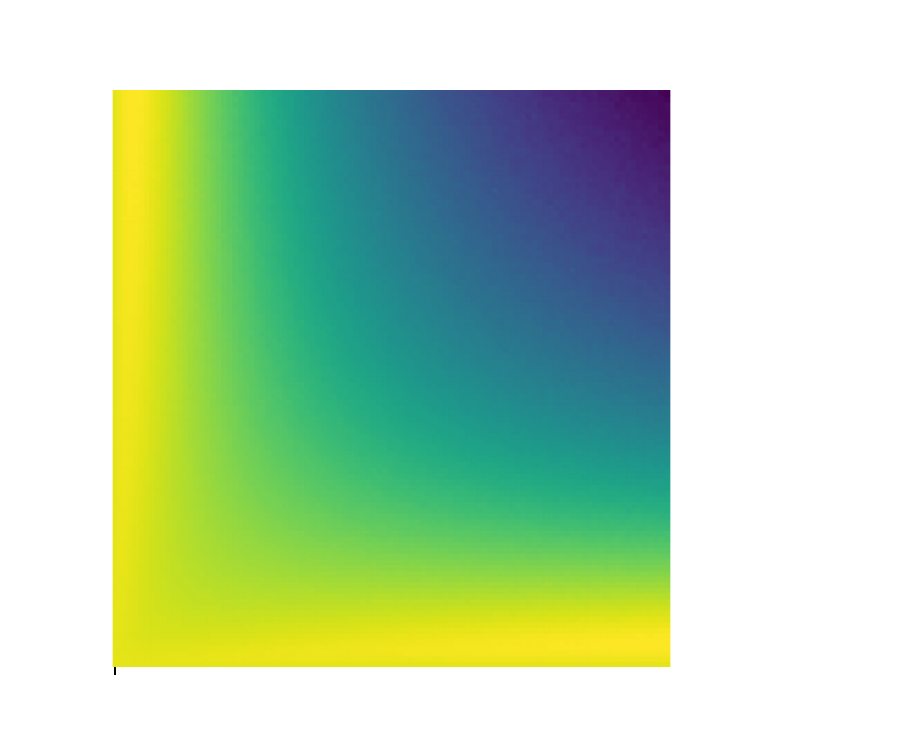
    \caption{Numerical Monte-Carlo result of $\mathrm{Re}\tilde L_\Sigma(u,v)$.}
    \label{fig:reslambda16IMAG}
  \end{subfigure}


  \caption{Dependence of the resummed $\tilde L_\Sigma$ on $u$ and $v$. In (a), the tree level $\tilde L_0$ is given to compare to the resummed function in (b) for small $\lambdaR$. In (b) and (c), the real and imaginary part of the resummed function at large, non-perturbative $\lambdaR / (16\pi^2) = 1$ is shown. Note that, in contrast to small, perturbative $\lambdaR$ and the individual orders $\tilde L_N$, the real part of the non-perturbative resummed result becomes negative for all values where $u \neq 0 \neq v$. Also, the imaginary part is mostly negative. Lastly, the resummed function for large $\lambdaR$ is very flat in the $u = v = 1$ corner, continuing the trend that this happens at higher orders, which has been observed in figure \ref{fig:lN_numerical}. In the sub-figures (a-c), a Gaußian filter has been applied to the data while computing the white contour lines for clarity.}
  \label{fig:resummed_plot}
\end{figure}

The non-perturbative results for $\tilde L_\Sigma$ are shown in figure \ref{fig:resummed_plot}. For the plot in figure \ref{fig:reslambda16}, we have increased the iterations of the integration to $50$ with $2\,000\,000$ integrand evaluations each to reduce noise. The key observations are
\begin{itemize}
    \item Again, the value of $\tilde L_\Sigma$ approaches $0$ if $u$ or $v$ approach $0$, independent of $\lambdaR$.
    \item For small effective expansion coefficients $\lambdaR / (16\pi^2)$, the real part of the resummed $\tilde L_\Sigma$ is very close to the tree level $\tilde L_0$ for all values of $u$ and $v$.
    \item For larger effective expansion coefficients, the real part of $\tilde L_\Sigma$ becomes \textit{negative} for nonzero $u$ and $v$.
    \item The upper right corner of the large $\lambdaR$ plot is very flat, continuing the trend of higher order perturbation theory, as observed in figure  \ref{fig:lN_numerical}.
\end{itemize}

\begin{figure}[htbp]
  \centering

  \hspace{-5mm}\begin{subfigure}[b]{0.45\textwidth}
    \def\svgwidth{1.1\columnwidth}
    \hspace{3mm}
    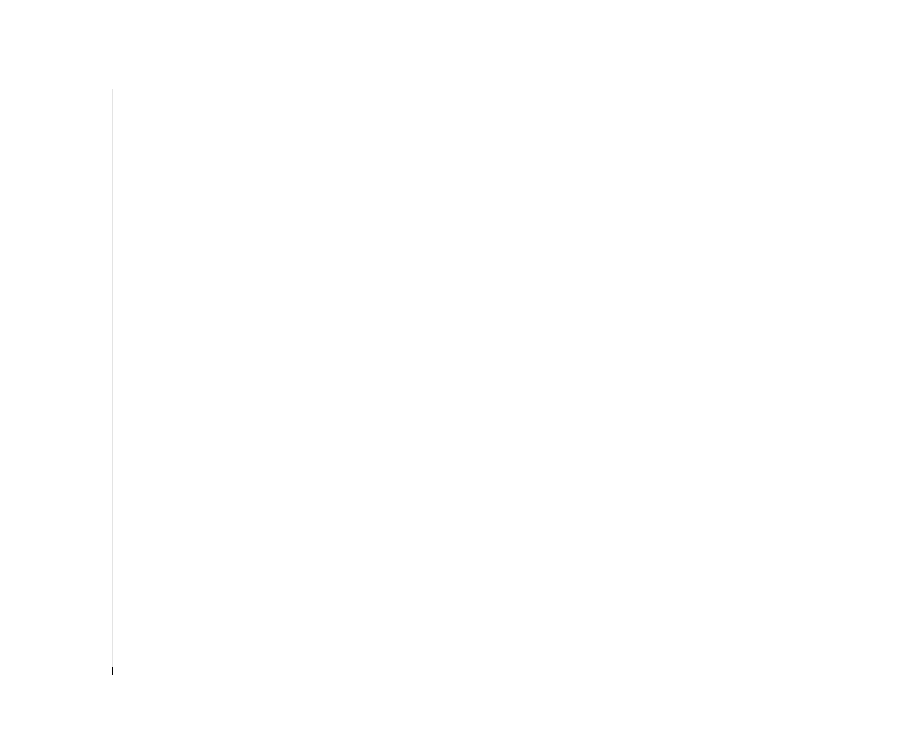
    \caption{Numerical Monte-Carlo result of $\mathrm{Re}\tilde L_\Sigma$ at fixed $u = v = 1$ as a function of $\lambdaR$.}
    \label{fig:lambdascan_wide}
  \end{subfigure}
  \hspace{0.8cm}
  \begin{subfigure}[b]{0.45\textwidth}
    \def\svgwidth{1.1\columnwidth}
    \hspace{2mm}
    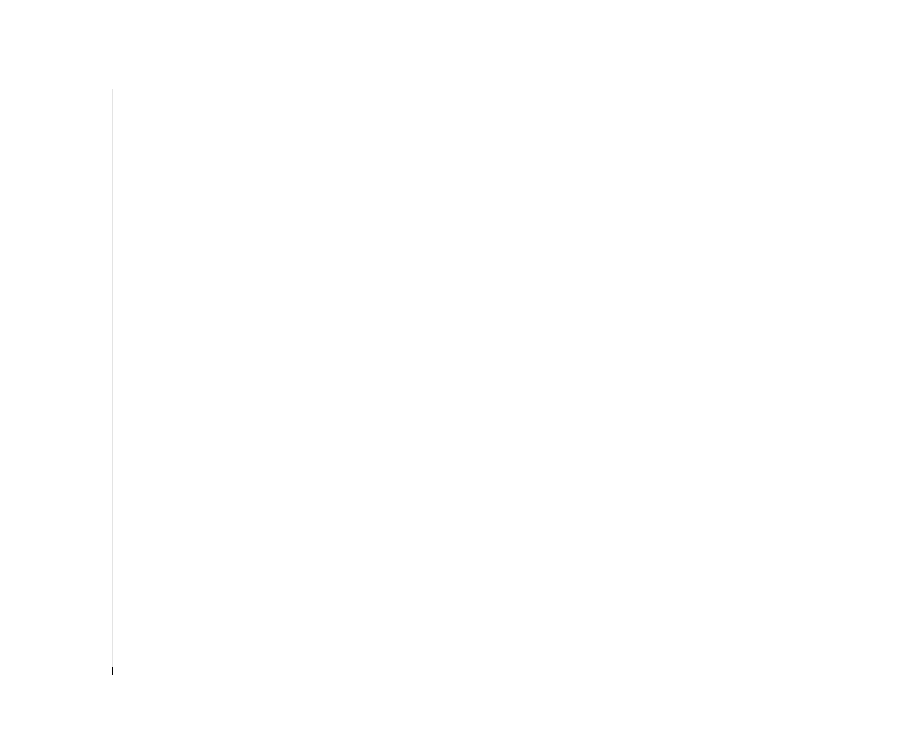
    \caption{Same as (a), but zoomed in to highlight the oscillations and perturbative range.}
    \label{fig:lambdascan_narrow}
  \end{subfigure}


  \caption{Dependence of $\tilde L_\Sigma$ at fixed $u = v = 1$ on the coupling $\lambdaR$. The plot in (b) is a zoomed in version of (a). It can be clearly seen that for large $|\lambdaR|$, the value of $\lambda L_\Sigma$ appoaches the same, negative value. On both sides of $\lambdaR = 0$, there are broad local maxima, at $\lambdaR \approx -100$ and $\lambdaR \approx 40$. For very small $|\lambdaR| \lesssim 10$, $\tilde L_\Sigma$ is approximately linear in $\lambdaR$, as predicted by perturbation theory. However, between this perturbative window and the broad peak on the negative side, there are many rapid oscillations that shrink in amplitude, but increase in frequency when approaching the perturbative window from below. The qualitative features of this plot are invariant under changing $\epsilon$ or $u$ and $v$. However, since the maxima are caused by the pole in \eqref{eq:tildeLSummed}, their hight - but not their positions - depend on $\epsilon$. For smaller $\epsilon$, the amplitude of the oscillations gets larger.}
  \label{fig:lambdascan_plot}
\end{figure}

\noindent Since $\tilde L_\Sigma(u = v = 1)\big\rvert_{\lambdaR = 1}$ is positive and $\tilde L_\Sigma(u = v = 1)\big\rvert_{\lambdaR = 16\pi^2}$ is negative, there is a sign change in between. Analyzing this further, we have shown $\tilde L_\Sigma(u = v = 1)$ as a function of $\lambdaR$ in figure \ref{fig:lambdascan_plot}.

Let us try to understand the qualitative feature of these results: For small $|\lambda_R|\lesssim 10$, the region in the integral \eqref{eq:tildeLSummed}, where the denominator is negative is parametrically small so that the positive contribution dominates. This situation is gradually reversed as $\lambda_R$ grows explaining the negative value of the integral. On the other hand, for negative coupling, with $-100\lesssim\lambda_R\lesssim -13$, the denominator of \eqref{eq:tildeLSummed} has a zero for a non-vanishing values of $p$ which, due to the cosine in the numerator leads to the oscillatory behavior\footnote{Indeed, we have a contribution from the pole around $z_1 z_2 (p^2+1) \approx e^{-\frac{16\pi^2}{\lambdaR}}$, which is a very small value for $\lambdaR > 0$. Thus, the arguments of the cosines in the numerator near the pole is small and the integrand is positive. For $\lambdaR < 0$, the value is large and therefore the sign of the cosines fluctuates rapidly. As $|\lambdaR|$ grows large, the sign does not matter and the exponential suppression of the integrand keeps the integrals bounded.} in figure \ref{fig:lambdascan_plot}. 

As a last point, we want to briefly mention hard cut-off regularization, since this is often used in discussions of the large $N$-model \cite{moshe_quantum_2003}. As derived in \cite{chowdhury_subtle_2024}, we can go from the analytic regularization to the cut-off regularization expression by replacing
\begin{align}
    \log\Big(z_1 z_2 H^2 \frac{p^2 + k^2}{\mu^2}\Big) \xrightarrow{\mathrm{replace}} \log\Big(\frac{p^2 + k^2}{\tilde\Lambda^2}\Big)
\end{align}
in the $N$-loop expression \eqref{eq:N_loop_renormalized_final}, where $\tilde\Lambda$ is the momentum cut-off of the $\d^4L$ loop integration found in \eqref{eq:L_integral_identity}. Since the logarithm on the right hand side does not depend on $z_j$, our arguments for integrating out the interior $z_j$ to get delta functions go through and we arrive at the resummed result
\begin{align}
    \tilde L_\Sigma^{\tilde\Lambda}(u,v) &= \frac{\lambdaR}{\pi} \frac{1}{k^2} \int\limits_0^\Lambda \d z_1 \d z_2 \int\limits_{-\Lambda}^\Lambda \d p \frac{e^{-\frac{z_1}{u} - \frac{z_2}{v} - \epsilon p^2}\cos(p z_1) \cos(p z_2)}{1 + \frac{\lambdaR}{16\pi^2} \log\big( \frac{p^2 + k^2}{\tilde\Lambda} \big) + i \epsilon}. \label{eq:tildeLSummed_cutoff}
\end{align}
We have again stripped off the coefficient $1/(16k_1\,k_2\,k_3\,k_4)$, but without $k$ this time as the rescaling to write the integrand as a function of $u$ and $v$ only does not go through. The reason is that this regularization breaks de Sitter invariance.

\begin{figure}
    \centering
    \def\svgwidth{0.5\columnwidth}
    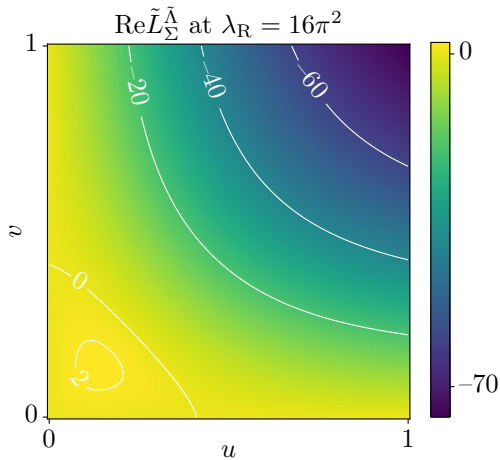
    \caption{Numerical Monte-Carlo result of the resummed cut-off regularized $\mathrm{Re} \tilde L_\Sigma^\Lambda(u, v, k, \tilde\Lambda)$ for as a function of $u$ and $v$ at $k=1$ and $\tilde\Lambda = 10$. While the plot looks somewhat similar to the analytically regularized one in figure \ref{fig:reslambda16}, there are clear differences. Mainly, there is a local maximum for some pair of $(u,v)$ in the bulk of values, i.e.\ for $u \neq 0 \neq v$. Therefore, cut-off regularization does not only break de Sitter invariance, but also produces qualitatively different physical observables.}
    \label{fig:cutoff}
\end{figure}

Using this, the plot in figure \ref{fig:cutoff} looks qualitatively different. In particular it features an extremum in the bulk of the $(u,v)$ region. This shows that choosing a de Sitter isometry violating regulator significantly affects observable predictions. 

\subsection{Analogy to Dressing Rules}

The new regularization scheme is designed to simplify computations as much as possible. Here, "as much as possible" could be interpreted to refer to the simplicity of the flat space necklace computation, where the necklace loops factorize completely \cite{moshe_quantum_2003}. Indeed, it has been demonstrated in \cite{chowdhury_subtle_2024} that up to regularization, de Sitter correlators can be obtained from the equivalent flat space diagrams by adding dressing propagators. These new propagators are added to every vertex and lead to a single common point. An example is the following 2-loop diagram
\begin{align}
\begin{fmffile}{2loop_dressing_rules}
    \parbox{130pt}{\begin{fmfgraph*}(130,130)
       \fmfsurround{i1,i2,i3,i4,i5,i6,i7,i8}
       \fmfshift{2 left}{i1}
       \fmfshift{2 down}{i3}
       \fmfshift{2 right}{i5}
       \fmfshift{2 up}{i7}
       \fmfshift{4 down}{i2}
       \fmfshift{4 left}{i2}
       \fmfshift{4 down}{i4}
       \fmfshift{4 right}{i4}
       \fmfshift{4 up}{i6}
       \fmfshift{4 right}{i6}
       \fmfshift{4 up}{i8}
       \fmfshift{4 left}{i8}
       \fmf{dbl_plain,right=0.18,force=0}{i1,i2,i3,i4,i5,i6,i7,i8,i1}
       \fmf{dashes}{i4,z1}
       \fmf{dashes}{z1,i6}
       \fmf{dashes}{i2,z3}
       \fmf{dashes}{z3,i8}
       \fmf{dbl_dashes,left=1,tension=0.2}{z1,z2,z1}
       \fmf{dbl_dashes,left=1,tension=0.2}{z2,z3,z2}
       \fmfforce{(0.17w,0.83h)}{ic4}
       \fmfforce{(0.17w,0.17h)}{ic6}
       \fmfforce{(0.83w,0.17h)}{ic8}
       \fmfforce{(0.83w,0.83h)}{ic2}
    \end{fmfgraph*}} \quad \hat{=} \quad
    \parbox{130pt}{\begin{fmfgraph*}(130,130)
       \fmfleft{i1,i2}
       \fmfright{o1,o2}
       \fmfshift{10 down}{i1}
       \fmfshift{10 up}{i2}
       \fmfshift{10 down}{o1}
       \fmfshift{10 up}{o2}
       \fmf{plain}{i1,z1}
       \fmf{plain}{z1,i2}
       \fmf{plain}{o1,z3}
       \fmf{plain}{z3,o2}
       \fmf{plain,left=1,tension=0.25}{z1,z2,z1}
       \fmf{plain,left=1,tension=0.25}{z2,z3,z2}
       \fmftop{c}
       \fmf{dashes,tension=0,left=0.4}{z1,c}
       \fmf{dashes,tension=0}{z2,c}
       \fmf{dashes,tension=0,right=0.4}{z3,c}
       \fmfforce{(0.0w,0.75h)}{i1}
       \fmfforce{(0.0w,0.25h)}{i2}
       \fmfforce{(1.0w,0.75h)}{o1}
       \fmfforce{(1.0w,0.25h)}{o2}
       \fmfdot{c}
       \fmfforce{(0.5w,0.95h)}{c}
    \end{fmfgraph*}}
\end{fmffile}
\end{align}
Note that the Feynman rules on the right side are not those given in this paper but those given in \cite{chowdhury_cosmological_2025}. In particular, the dashed propagator on the right is \vspace{-10mm}
\begin{equation}
\begin{fmffile}{dressing_rules_prop}
\vspace{-10mm}
    \parbox{200pt}{
    \begin{fmfgraph*}(80,80)
       \fmfleft{i}
       \fmfright{o}
       \fmfdot{i}
       \fmf{dashes}{i,o}
    \end{fmfgraph*}}
     \hspace{-4cm} \propto \quad \frac{k_\mathrm{ext}}{p_\mathrm{tot}^2 + k_\mathrm{ext}^2} \xrightarrow{k_\mathrm{ext} \rightarrow 0} \# \delta(p_\mathrm{tot}).
\end{fmffile} \label{eq:feynman_rules_dressing_prop}
\end{equation}
Here, $p_\mathrm{tot}$ is the sum over the internal energies connected to this propagator, while $k_\mathrm{ext}$ is the sum of external energies. For the internal vertex in the middle, there are no external energies, so $k_\mathrm{ext} \rightarrow 0$. This could be interpreted as energy conservation holding at the interior vertices. In our necklace computations, energy conservation, i.e., $\delta(p_j - p_{j-1})$ is precisely implemented by removing the factor $(z_j z_{j+1})^\kappa$ in the propagators \eqref{eq:feynman_rules_propagator-} and \eqref{eq:feynman_rules_propagator+}, like we proposed at the start of this section.

The interpretation of this energy-conservation in \cite{chowdhury_cosmological_2025} is that these auxiliary propagators need to be modified in analytical regularization, such that energy conservation does no longer hold. We propose the opposite conclusion: In order to preserve computational simplicity, we should stick to energy conservation at internal indices. As we have seen, we need to put factors of $(z_1 z_{N+1})^{\kappa}$ in place of the $(z_j z_{j+1})^\kappa$ in order to preserve scaling invariance. In the language of dressing rules, this means that those dressing propagators that carry external momentum have to be modified. Still, the diagrams that are relevant for cosmology usually have only few external propagators, so there is still a significant overall simplification for higher order loop diagrams.

\section{Conclusions} \label{sec:conclusions}

In this paper we discussed the resummation of momentum space, scalar bubble diagrams in de Sitter space. We find that, while the overall form of the shape function is comparable with the perturbative result, there is a change of the global sign at finite coupling. This can be traced back to the "tachyonic pathology" of the large N, $O(N)$-model \cite{Coleman:1974jh} which is responsible for the simple zero in the denominator of the energy integral  \eqref{eq:tildeLSummed}. So, while the large $N$ cosmological correlators are not pathological as a 3-dimensional Euclidean conformal field theory, the flat space pathology still leaves an imprint in these correlators. One may wonder if this obsrevation could be used to obtain further details about the primordial (effective) field theory underlying cosmological signals. In the process we did furthermore highlight how choosing a non-de Sitter invariant UV-regulator does qualitatively affect the observable signal at loop-level. On the other hand, as we showed, the natural  application of a de Sitter invariant regularization may obstruct the resummation of necklace diagrams and therefore prevent us form accessing non-perturbative properties in this way. This is the case for the naive application the analytic regularization in \cite{chowdhury_subtle_2024} (and similarly for a $\eta$-dependent cut-off \cite{gorbenko_4_2019}). However, there is a variant of anlaytic regularization which is compatible with resummation and furthermore rather naturally arises form the dressing rules  \cite{chowdhury_cosmological_2025}. 

At small negative $\phi^4$ coupling $\lambda$, perturbation theory agrees well with resummation. However, as $\lambda$ decreases further, the overall sign of the resummed correlator oscillates as a function of $\lambda$. The origin of this can again be explained as a consequence of the pole in the flat space necklace (related to the tachyon) and the specifics of de Sitter propagators combined with the energy integral that is always present in dS (similar to celestial amplitudes). Further decreasing $\lambda$ beyond the Landau pole, we recover the characteristics of large positive $\lambda$. In particular, the space of couplings is compact.  It would be interesting to see if these features can be understood from the position space analysis as well, where the oscillatory region appears to be mapped to the region, in coupling space, where the conformal weights of the dual conformal field theory are all real, while some of them are complex for generic values of $\lambda$ \cite{sachs_bulk_2023}. Perhaps, this can also give further insight about the physics near the Landau pole \cite{Romatschke:2022llf}. 

Finally, we found that the resummed cosmological correlator has an imaginary part, while the correlator is real in perturbation theory. The imaginary part is related to the existence of a pole on the Wick-rotated imaginary (conformal) time axis, which is absent in perturbation theory. One thus has to choose a Wick contour which avoids this pole, giving rise to a residue. It would be interesting to have a physical interpretation of this imaginary part in relation to the large $N$ Tachyon. 

An obvious extension of the present work is to understand how massive fields in the loops affect our conclusions. We hope to report on this in the near future. Another application of our re-summation is address the non-perturbative completion infra red divergent correlators when massless fields are involved (e.g. \cite{Beneke:2012kn,gorbenko_4_2019}).

\section*{Acknowledgments}
We would like to thank Daniel Bockisch, Jonathan Gr\"afe and Weichen Xiao for many helpful discussions. Part of this work was done while I.S. participated in the program on ``Cohomological aspects of Quantum Field Theory" 2025, at Mittag-Leffler Institute in Stockholm, supported by the Swedish Research Council under grant no. 2021-06594. This work was supported in parts by the Excellence Cluster Origins of the DFG under Germany’s
Excellence Strategy EXC-2094 390783311.  

\appendix

\section{Necklace Derivations} \label{app:N_loop_derivation}

In this appendix, we present some of the lengthy derivations of expressions for $N$-loop necklace diagrams. First, we show that all pure $N$-loop necklace diagrams in \eqref{eq:n_loop_diagrams} come with the same overall factor. This allows us to combine the appearing trigonometric functions and finally find a general expression. Afterwards, we generalize our results to also include mixed loops. In the first three parts of this appendix, we are working with the bare coupling $\lambda$ instead of the renormalized $\lambdaR$.

\subsection*{Symmetry Factors}

All of the pure $N$-loop diagrams in \eqref{eq:n_loop_diagrams} can have different symmetry and overall vertex factors. However, we can show that their products are actually all the same. As a starting point, we consider the diagram that only contains $\phi_-$ fields, i.e.\ dashed lines. This diagram has $N+1$ identical vertices, so the order expansion gives us $1/(N+1)!$, but the factor from reordering all vertices cancels this exactly. Then, going from left to right, we have to connect two external legs which gives us $4 \cdot 3$ choices. Every loop has $\binom{4}{2}$ possibilities to pick 2 legs from the next vertex and then 2 possibilities to pair them together with the previous vertex. At the very end, we again have to connect 2 external lines, but only 2 choices remain. All this comes with the vertex factor of $(\lambda / 4!)^{N+1}$ to give a total factor of
\begin{align}
    \frac{(N+1)!}{(N+1)!} \cdot 4 \cdot 3 \left( \binom{4}{2} \cdot 2 \right)^N \cdot 2 \frac{\lambda^{N+1}}{(4!)^{N+1}} = \frac{\lambda^{N+1}}{2^N}. \label{eq:n_loop_symmetry_factor}
\end{align}
Since the diagram only contains a single field with a $\frac{\lambda}{4!} \phi^4$ interaction, the symmetry factor is the same as a normal $\phi^4$ theory in flat space would have. This has been computed in \cite{palmer_general_2002} to be $1/2^N$ as well, which provides a check for our result.

\medskip

We now claim that the other diagrams containing an arbitrary combination of pure loops with $\phi_+$ and $\phi_-$ fields all have the same overall factor. Just as a real necklace can be seen as a string of beads, we can condense the field content information of any necklace diagram to a string of $+$ and $-$, where a loop of $\phi_-$ is symbolized with $-$, and for $\phi_+$ we use $+$. At this point, it is therefore important that we only consider pure loops with only one type of field. The diagrams are then read from left to right, and the $+$ and $-$ are written in a line. For example, the first diagram we considered is $--\ldots-$ with $N$ times $-$. The next shown diagram in \eqref{eq:n_loop_diagrams} would be $+--\ldots-$, and so on. 

If two $-$ or two $+$ come in succession, there is a $\phi_-^4$ or a $\phi_+^4$ vertex in between the loops in the diagram. This gives us a symmetry and vertex factor of $\binom{4}{2} \cdot 2 \cdot \lambda / 4! = \lambda / 2$ for this vertex and the connecting loop the the left of the vertex. If there is a change from $+$ to $-$ or vice versa, the vertex in between in the corresponding diagram has to be of the $\phi_-^2 \phi_+^2$ type. The symmetry factor then is $2 \cdot (- 6) \lambda / 4! = -\lambda / 2$, since there are fewer possibilities to connect the propagators, but also, the vertex factor is different. The only difference to the previous successive case is a sign.

So far, we have only considered the interior loops, because we always looked at two successive loops. Since we only have $\phi_-$ as external legs, we can think of it as though the necklace was padded to both sides with $-$. However, we have to remember to not count this padding as extra loops. Its information is only relevant for deciding what vertices are at the ends. For the left end, there are only two cases: $--$ and $-+$. In the first case, we have a $\phi_-^4$ vertex, which gives us $4 \cdot 3$ possibilities to connect the external legs and and vertex factor of $\lambda/4!$. In case of $-+$, we have $2$ possibilities to connect and a vertex factor of $-6\lambda/4!$. We find either $\lambda / 2$ or $-\lambda / 2$, just as before. On the right end, we find the same situation. We only have to remember to count the external legs as a symmetry of $2$, since the other legs of the last vertex have already been counted. 

In total, we find for every vertex a factor of $\pm \lambda / 2$, together with an overall factor of $2$. As there is exactly one sign for every flip from $-$ to $+$ (now also counting the external padding!), and we always start and end with $-$ because of the external legs, there always is an even number of flips, so the sign will always be positive in the end. Thus, the total factor \eqref{eq:n_loop_symmetry_factor} is correct for any $N$-loop necklace diagram.

\subsection*{Manipulating the Mathematical Expression}

We can now write down the mathematical expression for the sum of diagrams. The computation is very close to the 1-loop case, but we repeat it here for the general case to highlight the dependence of the integrand on the various $z_j$. Note that we are using the abbreviated notation of \eqref{eq:shortcuts}. 

For every loop, there are two possibilities, $+$ and $-$. Therefore, we can factorize the sum over all diagrams as
\begin{align}
    I_N = \frac{\lambda^{N+1}}{2^N} \int\limits_0^\infty \frac{\d z_1}{(z_1 H)^4} \ldots \frac{\d z_{N+1}}{(z_{N+1} H)^4} &\bar G_\kappa(z_1; z_{N+1}) \int \frac{\d^3 \ell_1}{(2\pi)^3} \ldots \frac{\d^3 \ell_N}{(2\pi)^3} \int\limits_{-\infty}^\infty \d p_{1\mathrm{a}} \d p_{1\mathrm{b}} \ldots \d p_{N\mathrm{a}} \d p_{N\mathrm{b}} \notag \\
    \times \prod_{j=1}^N \frac{(z_j z_{j+1} H^2)^{2-2\kappa}}{(\pi C_\kappa)^2} \Bigg( &\frac{\sin(p_{j\mathrm{a}}z_j) \sin(p_{j\mathrm{a}}z_{j+1}) \sin(p_{j\mathrm{b}}z_j) \sin(p_{j\mathrm{b}}z_{j+1})}{\Big(p_{j\mathrm{a}}^2 + (\vec k + \vec\ell_j)^2\Big)^{1+\kappa} \Big(p_{j\mathrm{b}}^2 + \vec\ell_j^2\Big)^{1+\kappa}} \label{eq:IN}\\
    + &\frac{\cos(p_{j\mathrm{a}}z_j) \cos(p_{j\mathrm{a}}z_{j+1}) \cos(p_{j\mathrm{b}}z_j) \cos(p_{j\mathrm{b}}z_{j+1})}{\Big(\underbrace{p_{j\mathrm{a}}^2 + (\vec k + \vec\ell_j)^2}_{=(\vec L_j + \vec Q_j)^2}\Big)^{1+\kappa} \Big(\underbrace{p_{j\mathrm{b}}^2 + \vec\ell_j^2}_{=\vec L_j^2}\Big)^{1+\kappa}}\Bigg) \notag
\end{align}
Just as in the 1-loop computation, we define $p_j^\pm = p_{j\mathrm{a}} \pm p_{j\mathrm{b}}$, $\vec L_j = \big(p_{j\mathrm{b}}, \vec \ell\big)$ and $\vec Q_j = \big( p_j^-, \vec k \big)$. Using these, we can write the second line of the above equation in a way that resembles flat space Feynman diagrams. Using known techniques, we can then evaluate the loop integral over $\d^4 L_j = \d p_{j\mathrm{b}} \d^3 \ell_j$ and again find \eqref{eq:L_integral_identity}. Lastly, we can simplify the trigonometric functions in the second line of \eqref{eq:IN} by using the symmetry of the integrand under $p_{j\mathrm{b}} \rightarrow -p_{j\mathrm{b}}$. Putting everything together and renaming $p^-_j \rightarrow p_j$ as defined in the shortcut notation \eqref{eq:shortcuts}, we find the general expression for $I_N$ to be
\begin{align}
    I_N &= \lambda^{N+1} C^N \int\limits_0^\infty \frac{\d z_1}{(z_1 H)^{2+2\kappa}} \frac{\d z_2}{(z_2 H)^{4\kappa}} \frac{\d z_3}{(z_3 H)^{4\kappa}} \ldots \frac{\d z_N}{(z_N H)^{4\kappa}} \frac{\d z_{N+1}}{(z_{N+1} H)^{2+2\kappa}} \bar G_\kappa(z_1; z_{N+1}) \notag\\
    &\hspace{1.3cm}\times \int\limits_{-\infty}^\infty \d p_1 \ldots \d p_N \Gamma_\kappa^N \frac{\cos(p_1 z_1) \cos(p_1 z_2) \ldots \cos(p_N z_N) \cos(p_N z_{N+1})}{\big( p_1^2 + \vec k^2\big)^{2\kappa} \ldots \big( p_N^2 + \vec k^2\big)^{2\kappa}} \label{eq:app_N_loop_pure_final}
\end{align}

\subsection*{Mixed Loops}

To generalize the previous derivation to include mixed loops, we first analyze what happens if one inserts a single connected sub-necklace of length $M$ into a pure necklace diagram of length $N-M$. By sub-necklace, we mean a chain of subsequent mixed loops. We restrict ourselves to this scenario, because the interesting changes with respect to only pure necklaces happen on the "boundary" of such a mixed sub-necklace. This is because only the vertices at the start and at the end of the mixed sub-necklace have an odd number of $\phi_-$ and $\phi_+$ and are suppressed by $\kappa$, according to \eqref{eq:feynman_rules_odd_coupling}. The insertion of a mixed sub-necklace is illustrated in the following diagrams
\begin{align}
&\begin{fmffile}{mixed_loops_1}
\parbox{110pt}{\begin{fmfgraph*}(110,50)
      \fmfleft{i}
      \fmfright{o}
      \fmf{dots,tension=3.5}{i,v1}
      \fmf{dbl_dashes,left=1}{v1,v2,v1}
      \fmf{dots,tension=3.5}{v2,v3}
      \fmf{dbl_dashes,left=1}{v3,v4,v3}
      \fmf{dots,tension=3.5}{v4,o}
\end{fmfgraph*}}
\end{fmffile}\  \Longrightarrow \ 
\begin{fmffile}{mixed_loops_2}
\parbox{270pt}{\begin{fmfgraph*}(270,50)
      \fmfleft{i}
      \fmfright{o}
      \fmf{dots,tension=3}{i,v1}
      \fmf{dbl_dashes,left=1}{v1,v2,v1}
      \fmf{dots,tension=3}{v2,v3}
      \fmf{dbl_dashes,left=1}{v3,v4,v3}
      \fmf{dashes,left=1}{v4,v5}
      \fmf{plain,left=1}{v5,v4}
      \fmf{dots,tension=3}{v5,v6}
      \fmf{dashes,left=1}{v6,v7}
      \fmf{plain,left=1}{v7,v6}
      \fmf{dbl_dashes,left=1}{v7,v8,v7}
      \fmf{dots,tension=3}{v8,v9}
      \fmf{dbl_dashes,left=1}{v9,v10,v9}
      \fmf{dots,tension=3}{v10,o}
\end{fmfgraph*}}
\end{fmffile} \\
&\vspace{-2.8cm}\hspace{0.5cm}\underbrace{\hspace{2.7cm}}_{(N-M) \times} \hspace{2.3cm}\underbrace{\hspace{2.6cm}}_{(N-M-K) \times} \underbrace{\hspace{2.7cm}}_{M \times} \underbrace{\hspace{2.7cm}}_{K \times} \notag
\end{align}
As defined before, the doubly dashed internal propagators indicate that these lines represent a sum over $\phi_-$ and $\phi_+$ propagators. The symmetry factor of this diagram is computed like that for the pure loop diagram, by representing a diagram by a string of $+$, $-$, and now also \texttt{M} for a mixed loop. Again, we remember to use padding of $-$ on the left and the right of the chain for the external legs. For the next few paragraphs, until we state otherwise, we do not use the abbreviated notation of renaming $\lambda C_{2\kappa} \rightarrow \lambda$.

New are only the combinations \texttt{MM}, $+\mathtt{M}$, $-\mathtt{M}$ and re-orderings of these. The factors associated with each of these combinations can be found in table \ref{tab:mixed_pure_factors}. The overall symmetry factor from the expansion is 1 since we have $1/N! \cdot (N-M)! M! \cdot N! / (N-M)! / M!$ according to the multinomial theorem.
\begin{table}[htbp]
    \centering
    \begin{tabular}{c|c} 
         Combination & Factor \\ \hline
         \texttt{MM} & $-2 \cdot 2 \cdot 6\lambda \mathcal C_{2\kappa} / 4! \vphantom{\Big(}$ \\
         $-\mathtt{M}$ & $3 \cdot 2 \cdot 8 \pi \kappa \lambda / 4!  \vphantom{\big(}$ \\
         $+\mathtt{M}$ & $-3 \cdot 2 \cdot 8 \pi \kappa \lambda / 4!  \vphantom{\Big(}$ \\
         $\mathtt{M}-$ & $3 \cdot 1 \cdot 8 \pi \kappa \lambda / 4!  \vphantom{\big(}$ \\
         $\mathtt{M}+$ & $-3 \cdot 1 \cdot 8 \pi \kappa \lambda / 4!  \vphantom{\Big(}$
    \end{tabular}
    \caption{Symmetry and vertex factors associated with any combination of mixed and pure loops.}
    \label{tab:mixed_pure_factors}
\end{table}
Since every mixed loop sub-chain $\mathtt{MM} \ldots \mathtt{M}$ must end, the overall factor for inserting it is
\begin{align}
    \pm \big( - \lambda \mathcal{C}_{2\kappa} \big)^{M-1} 2 \pi^2 \kappa^2 \lambda^2,
\end{align}
where the sign in front is negative if the mixed-loop sub-necklace is inserted between two different pure loops, for example $- \mathtt{M M} \ldots \mathtt{M} +$, and positive if the pure loops on either side are the same type. With this, it fits perfectly with the earlier observation that $-+$ gets a sign compared to $--$, just now with \texttt{M}s in between. The remaining pure loops and external legs then simply give a total factor of $\mp 2(\lambda \mathcal C_{2\kappa}/2)^{N-M}$.

Computing the complete mathematical expression for any number of mixed loops in a single diagram generalizes from the 1-loop case just as easily as for only pure loops. Therefore, we will only consider the 1-loop mixed diagram $I_1^\mathtt{M}$ and generalize from there. We find
\begin{align}
    I_1^\mathtt{M} \hspace{-1mm} = \hspace{-1mm} -\frac{4 \pi^2 \kappa^2 \lambda^2}{\pi^2 \mathcal C_\kappa^2}  \hspace{-2mm} \int\limits_0^\infty  \hspace{-1mm} \frac{\d z_1 \d z_2 \bar G_\kappa (z_1 ; z_2)}{(z_1 H)^{2+2\kappa}(z_2 H)^{2+2\kappa}} \hspace{-1mm} \int \hspace{-1mm} \frac{\d^3 \ell}{(2\pi)^3} \hspace{-1mm} \int\limits_{-\infty}^\infty \hspace{-1mm} \d p_\mathrm{a} \d p_\mathrm{b} \frac{\cos(p_\mathrm{a} z_1) \cos(p_\mathrm{a} z_2) \sin(p_\mathrm{b} z_1) \sin(p_\mathrm{b} z_2)}{\big( p_\mathrm{a}^2 + (\vec\ell + \vec k^2)^2\big)^{1+\kappa} \big( p_\mathrm{b}^2 + \vec\ell^2\big)^{1+\kappa}}.
\end{align}
The extra sign comes from the relative sign difference of the $\phi_-$ and $\phi_+$ propagators in the Feynman rules \eqref{eq:feynman_rules_propagator-} and \eqref{eq:feynman_rules_propagator+}. Now, we can use that the mixed loop diagram with a $\phi_-$ propagator on top and a $\phi_+$ propagator below is the same as one with the lines reversed. We therefore divide the above expression by $2$ and add the same term with sine and cosine functions interchanged. Then, just as for the pure loops, we use a trigonometric identity
\begin{align}
    &\cos(p_{\mathrm{a}}z_1) \cos(p_{\mathrm{a}}z_2) \sin(p_{\mathrm{b}}z_1) \sin(p_{\mathrm{b}}z_2) + \sin(p_{\mathrm{a}}z_1) \sin(p_{\mathrm{a}}z_2) \cos(p_{\mathrm{b}}z_1) \cos(p_{\mathrm{b}}z_2) \notag\\
    &\hspace{1cm} = \frac{1}{2} \Big( \sin(p^+ z_1) \sin(p^+ z_2) + \sin(p^- z_1) \sin(p^- z_2) \Big) \, \hat{=} \ \sin(p^- z_1) \sin(p^- z_2),
\end{align}
where we once more introduced the objects $p^\pm = p_1 \pm p_2$. In the last step, we used that the integrand is symmetric under $p_\mathrm{b} \rightarrow - p_\mathrm{b}$ to state that both terms are effectively equivalent. Inserting this, we see that the mixed 1-loop expression is equal to the pure 1-loop \eqref{eq:1_loop_step1} times a factor of $-4\pi^2\kappa^2 / \mathcal C_{2\kappa}^2$, and with cosines turned into sines.

This can now be generalized to an arbitrary necklace with an arbitrary number of pure and mixed loops. For this, it is important to realize that so far, we have only inserted one connected mixed sub-necklace. If we insert any number $S$ of unconnected sub-necklaces, their boundaries will generate more $-4\pi^2\kappa^2 / \mathcal C_{2\kappa}^2$ factors for every mixed loop sub-chain. Thus, the general expression, again using $\lambda C_{2\kappa} \rightarrow \lambda$, is
\begin{align}
    I_N^M &= \Big( -\frac{4\pi^2\kappa^2}{\mathcal C_\kappa^2} \Big)^S (-1)^N \lambda^{N+1} C^N \int\limits_0^\infty \frac{\d z_1}{(z_1 H)^2} \d z_2 \d z_3 \ldots \d z_N \frac{\d z_{N+1}}{(z_{N+1} H)^2} \bar G_\kappa(z_1; z_{N+1}) \notag\\
    &\hspace{1.8cm}\times \int\limits_{-\infty}^\infty \d p_1 \ldots \d p_N \Gamma_\kappa^N \frac{\binom{\cos}{\sin}(p_1 z_1) \binom{\cos}{\sin}(p_1 z_2) \ldots \binom{\cos}{\sin}(p_N z_N) \binom{\cos}{\sin}(p_N z_{N+1})}{\big( z_1 z_2 H^2 (p_1^2 + \vec k^2)\big)^{2\kappa} \ldots \big( z_N z_{N+1} H^2 (p_N^2 + \vec k^2)\big)^{2\kappa}}.\label{eq:N_loop_mixed_final}
\end{align}
Here, $\binom{\cos}{\sin}(x)$ is used to denote that a pure loop has cosine functions and a mixed loop sines.

\subsection*{Renormalization of Mixed Loops}

The equation \eqref{eq:N_loop_mixed_final} still contains $\Gamma_\kappa$ that have a pole $1/\kappa$. Renormalization is now done exactly as in the pure loop case, the counter-term does not need to be changed. We employ a string argument again, where \texttt{P} denotes a pure loop, \texttt{M} a mixed loop, and \texttt{X} a counter-term. 

We have to be careful, since we cannot use counter-terms that have already been used to subtract the divergence of a pure loop to renormalize a mixed loop. To illustrate this, we consider the first few loop order explicitly. For example, the 1-loop \texttt{P} is renormalized by \texttt{X}. Therefore, there is no counter-term to pair with the 1-loop \texttt{M}. However, this is also not necessary, since the mixed 1-loop is of order $\kappa$ and so there is no divergence.

At 2-loop level, we have the diagrams
\begin{align}
    \texttt{PP}, \ &\texttt{PX}, \texttt{XP}, \texttt{XX}; \notag\\
    \texttt{MM}, \ &\texttt{MP}, \texttt{PM}, \notag\\
    &\texttt{MX}, \texttt{XM}. \notag
\end{align}
The first line is already accounted for in the pure 2-loop computation. In the second line, the \texttt{MX} counter-term is used to make the \texttt{MP} loop finite and so on, leaving the \texttt{MM} diagram without a counter-term again. Still, this is not a problem, since it is of order $\kappa^0$ anyway.

For 3-loops, we have
\begin{align}
    \texttt{PPP}, \texttt{PPX}, \texttt{PXP}, \texttt{XPP}, \ & \texttt{PXX}, \texttt{XPX}, \texttt{XXP}, \texttt{XXX}; \notag\\
    \texttt{MMM}, \texttt{MMP}, \texttt{MPM}, \texttt{PMM}, \ & \texttt{MPP}, \texttt{PMP}, \texttt{PPM}, \notag\\
    \texttt{MMX}, \texttt{MXM}, \texttt{XMM}, \ & \texttt{MXX}, \texttt{XMX}, \texttt{XXM} \notag\\
    &\texttt{MPX}, \texttt{PMX}, \texttt{PXM}, \notag\\
    &\texttt{MXP}, \texttt{XMP}, \texttt{XPM}. \notag
\end{align}
We have already arranged the expressions in a suggestive way. To explain why, we have to think back to the pure loop diagrams and their counter-terms in the first line. If we treat the expressions \texttt{PPP} and so on as the integrands under all the $z$ and $p$ integrals and after all the trigonometric functions, we realize that their multiplication is associative. Therefore, the result of the $N$-loop renormalization \eqref{eq:N_loop_renormalized} where we found that the expression factorizes into blocks of the form
\begin{align}
    \Bigg(\frac{\lambdaR H C}{\mu} \Gamma_\kappa \Big( z_j z_{j+1} H^2 \frac{p_j^2 + \vec k^2}{\mu^2}\Big)^{-2\kappa} + \frac{\delta\lambda}{\pi} \Bigg) \label{eq:kinematic_loop_factor}
\end{align}
can be understood schematically by writing
\begin{align}
    \texttt{PPP} + \texttt{PPX} + \texttt{PXP} + \texttt{XPP} + \texttt{PXX} + \texttt{XPX} + \texttt{XXP} + \texttt{XXX} = (\texttt{P} + \texttt{X})(\texttt{P} + \texttt{X})(\texttt{P} + \texttt{X}).
\end{align}
In the same sense, we can combine the strings involving \texttt{M} such that they equal
\begin{align}
    \texttt{MM} (\texttt{P} \hspace{-.5mm} + \hspace{-.5mm} \texttt{X}) + (\texttt{P} \hspace{-.5mm} + \hspace{-.5mm} \texttt{X}) \texttt{MM} + \texttt{M} (\texttt{P} \hspace{-.5mm} + \hspace{-.5mm} \texttt{X}) (\texttt{P} \hspace{-.5mm} + \hspace{-.5mm} \texttt{X}) + (\texttt{P} \hspace{-.5mm} + \hspace{-.5mm} \texttt{X}) \texttt{M} (\texttt{P} \hspace{-.5mm} + \hspace{-.5mm} \texttt{X}) + (\texttt{P} \hspace{-.5mm} + \hspace{-.5mm} \texttt{X}) (\texttt{P} \hspace{-.5mm} + \hspace{-.5mm} \texttt{X}) \texttt{M} + \texttt{MMM} + \texttt{MXM} + \texttt{MPM}.
\end{align}
The last three terms need our attention again. Note that \texttt{MPM} has two mixed sub-necklaces of length one and therefore is actually suppressed by order $\kappa$. We can therefore freely use the \texttt{MXM} counter-term to cancel the divergence of \texttt{MMM} instead. From this, we can generalize three important observations
\begin{itemize}
    \item Any diagram containing $\ldots\texttt{PMP}\ldots$ with a mixed sub-necklace of length one is always suppressed after combining it with counter terms for all \texttt{P} loops and \texttt{M} loops in the interior of mixed sub-necklaces (see the last point). Therefore, a mixed sub-necklace of length one does not need a counter-term. The counter-term diagram that replaces the \texttt{M} with \texttt{X} is used to renormalize $\ldots\texttt{PPP}\ldots$ instead.
    \item The boundary \texttt{M} of a longer mixed sub-necklace in a diagram $\ldots\texttt{PMM}\ldots\texttt{MP}\ldots$ do not need to be renormalized, since they bring their own factor of $\kappa$ and are finite. A counter-term diagram that replaces a boundary \texttt{M} with a \texttt{X}, for example $\ldots\texttt{PXM}\ldots\texttt{MP}\ldots$, is used to renormalize $\ldots\texttt{PPM}\ldots\texttt{MP}\ldots$ instead.
    \item A bulk \texttt{M} that sits inside a longer mixed sub-necklace in a diagram $\ldots\texttt{PM} \ldots \texttt{M} \ldots\texttt{MP}\ldots$ is renormalized by the counter-term diagram that replaces this \texttt{M} with \texttt{X}, i.e.\ $\ldots\texttt{PM}\ldots \texttt{X} \ldots\texttt{MP}\ldots$.
\end{itemize}
For the reasons outlined above, the expression for a necklace containing both mixed and pure loops depends on the string, e.g. \texttt{PPMMP}. From now on, we only consider renormalized diagrams, where, given a string of only \texttt{P} and \texttt{M}, we combine this with all counter-term diagrams that are needed to renormalize it according to the above rules. We annotate this renormalized expression by
\begin{align}
    I_N^{\binom{\mathtt P}{\mathtt M} \ldots \binom{\mathtt P}{\mathtt M}},
\end{align}
where the superscript contains the corresponding string.

Splitting the extra factor of $4\pi^2\kappa^2/\mathcal C_\kappa^2$ between the two boundary \texttt{M} loops, these do not get the kinematic factor \eqref{eq:kinematic_loop_factor}, but instead
\begin{align}
    \frac{\lambdaR H C}{\mu \mathcal C_\kappa} 2\pi\kappa\Gamma_\kappa \Big( z_j z_{j+1} H^2 \frac{p_j^2 + \vec k^2}{\mu^2}\Big)^{-2\kappa} \xrightarrow{\kappa \rightarrow 0} \frac{\lambdaR H \pi C}{\mu}.
\end{align}
The corresponding $p_j$ integral can then be performed to give a delta. We will not do this in general, however, in order to write the expression in a more compact form. Also, we can now send $\kappa \rightarrow 0$ again. In total, after renormalization, the general necklace diagram therefore is
\begin{align}
    I_N^{\binom{\mathtt P}{\mathtt M} \ldots \binom{\mathtt P}{\mathtt M}} &= (-1)^N \Big( -\pi^2 \Big)^S \Big(\frac{\lambdaR H}{\mu}\Big)^{N+1} C^N \hspace{-1mm} \int\limits_0^\infty \frac{\d z_1}{(z_1 H)^2} \d z_2 \ldots \d z_N \frac{\d z_{N+1}}{(z_{N+1} H)^2} \bar G_0(z_1; z_{N+1}) \hspace{-1mm} \int\limits_{-\infty}^\infty \hspace{-1mm} \d p_1 \ldots \d p_N \notag\\
    &\hspace{2cm}\times \binom{\cos}{\sin}(p_1 z_1) \binom{\cos}{\sin}(p_1 z_2) \ldots \binom{\cos}{\sin}(p_N z_N) \binom{\cos}{\sin}(p_N z_{N+1}) \label{eq:mixed_loop_renormalized_final}\\
    &\hspace{2cm}\times \binom{\log}{1}\Big(z_1 z_2 H^2 \frac{p_1^2 + \vec k^2}{\mu^2} \Big) \hspace{0.8mm} \ldots \hspace{0.8mm} \binom{\log}{1}\Big(z_N z_{N+1} H^2 \frac{p_N^2 + \vec k^2}{\mu^2} \Big), 
\end{align}
where we choose cosine functions for pure loops and sine functions for mixed loops. The number of connected mixed sub-necklaces is $S$. Further, we omit the logarithm for loops that are at the boundary of a mixed loop sub-necklace.

\section{3-Loop Details} \label{app:3_loop_details}

In this appendix, we present some details for the derivation of the 3-loop result in section \ref{sec:3_loop}. We repeat the expression here for easier reference:
\begin{align}
    L_3 &= - \Big( \frac{\lambdaR H}{\mu} \Big)^4 C^3 \int\limits_0^\infty \frac{\d z_1}{(z_1 H)^2} \d z_2 \d z_3 \frac{\d z_4}{(z_4 H)^2} \bar G_0(z_1; z_4) \notag\\
    &\quad\times \int\limits_{-\infty}^\infty \d p_1 \d p_2 \d p_3 \cos(p_1 z_1) \cos(p_1 z_2) \cos(p_2 z_2) \cos(p_2 z_3) \cos(p_3 z_3) \cos(p_3 z_4) \notag\\
    &\quad \times \Bigg( \log\big( z_1 z_2 H^2\big) \log\big( z_2 z_3 H^2\big) \log\big( z_3 z_4 H^2\big) + \log\big( z_1 z_2 H^2\big) \log\big( z_2 z_3 H^2\big) \log\Big( \frac{p_3^2 + \vec k^2}{\mu^2}\Big) \label{eq:3_loop_step2}\\
    &\quad\quad + \log\big( z_1 z_2 H^2\big)\log\Big( \frac{p_2^2 + \vec k^2}{\mu^2} \Big) \log\big( z_3 z_4 H^2\big) + \log\big( z_1 z_2 H^2\big) \log\Big( \frac{p_2^2 + \vec k^2}{\mu^2} \Big) \log\Big( \frac{p_3^2 + \vec k^2}{\mu^2} \Big) \notag\\
    &\quad\quad + \log\Big( \frac{p_1^2 + \vec k^2}{\mu^2} \Big) \log\big( z_2 z_3 H^2\big) \log\big( z_3 z_4 H^2\big) + \boxed{\log\Big( \frac{p_1^2 + \vec k^2}{\mu^2} \Big) \log\big( z_2 z_3 H^2\big) \log\Big( \frac{p_3^2 + \vec k^2}{\mu^2}\Big)} \notag\\
    &\quad\quad + \log\Big( \frac{p_1^2 + \vec k^2}{\mu^2} \Big) \log\Big( \frac{p_2^2 + \vec k^2}{\mu^2} \Big) \log\big( z_3 z_4 H^2\big) + \log\Big( \frac{p_1^2 + \vec k^2}{\mu^2} \Big) \log\Big( \frac{p_2^2 + \vec k^2}{\mu^2} \Big)\log\Big( \frac{p_3^2 + \vec k^2}{\mu^2} \Big)\Bigg). \notag
\end{align}
As already indicated before, $L_3$ contains two different kinds of terms: Those that can be integrated easily using delta functions coming from \eqref{eq:cosine_delta} or \eqref{eq:cosine_delta_z}, and the boxed term that cannot. We will first consider the terms than can be integrated, i.e., all terms that are not boxed.

Again we use a string-based argument. The individual terms of \eqref{eq:3_loop_step2} may be written as a string of \texttt{Z} and \texttt{P}, where \texttt{Z} stands for a logarithm that has $z_j z_{j+1} H^2$ as an argument and \texttt{P} stands for a logarithm that has $(p_j^2 + \vec k^2)/\mu^2$ as an argument. We write \texttt{Z} and \texttt{P} in the order they appear in \eqref{eq:3_loop_step2}, so the indices of the involved $z_j$ and $p_j$ can be deduced from the position in the string. The integrals over $z_1$ to $z_4$ and $p_1$ to $p_3$, as well as the remaining parts of the metric determinant $(H z_{1/4})^2$ and the cosines are understood. The above expression \eqref{eq:3_loop_step2} may then be more efficiently written as
\begin{align}
    L_3 \, \hat{=}\, \mathtt{ZZZ} + \mathtt{ZZP} + \mathtt{ZPZ} + \mathtt{ZPP} + \mathtt{PZZ} + \mathtt{PZP} + \mathtt{PPZ} + \mathtt{PPP}. \label{eq:3_loop_step3_simplified}
\end{align}
If our string contains a \texttt{Z} at a given position $j$, the integral over $p_j$ can be performed to give \linebreak $\pi\delta(z_j - z_{j+1})$, according to \eqref{eq:cosine_delta}. Further, if two $\mathtt{Z}$ come in succession, these delta functions set the arguments of the corresponding logarithms to be equal. We indicate this by writing $\mathtt{Z}^2$ (or $\mathtt{Z}^3$ in the case of the first term with tree $Z$).

If our string contains a $\mathtt{P}$ at a given position $j$, this does not mean that the $z_j$ integral can be performed to give a delta function, since $z_j$ could also appear in a $\mathtt{Z}$-logarithm at position $j-1$, just how the second term in \eqref{eq:3_loop_step2} $\mathtt{ZZP}$ involves a logarithm containing $z_3$, even though a $\mathtt{P}$ is in the third position. Only if there are two successive $\mathtt{PP}$ in the $j-1$ and $j$-th position, we can produce a $\pi \delta(p_{j-1} - p_j)$, according to \eqref{eq:cosine_delta_z}. In these cases, we write $\mathtt{P}^2$ (or $\mathtt{P}^3$ in the case of the last term with three $\mathtt{P}$). Using these rules, we can turn \eqref{eq:3_loop_step3_simplified} into
\begin{align}
    L_3 \, \hat{=}\, \mathtt{Z}^3 + \mathtt{Z}^2 \mathtt{P} + \mathtt{ZPZ} + \mathtt{ZP}^2 + \mathtt{PZ}^2 + \mathtt{PZP} + \mathtt{P}^2\mathtt{Z} + \mathtt{P}^3. \label{eq:3_loop_step4}
\end{align}
Note that we never perform the first or the last $z$ integrals, i.e., over $z_1$ or $z_4$. This is because they also contain the measure of EAdS and, more importantly, the bulk-to-boundary propagator $\bar G_0(z_1; z_4)$. We indicate this by turning $\mathtt{Z}^n$ into $\mathtt{Z}_1^n$ if it is at the first position after we have applied our rules. Equivalently, we turn $\mathtt{Z}^n$ into $\mathtt{Z}_4^n$ if it is a the last position. Further, since the argument of the logarithm represented by $\mathtt{Z}_{14}$ contains $z_{1/4}^2 H^2$, we pull out a factor of $2^n$ when turning $\mathtt{Z}^n$ into $\mathtt{Z}_{1/4}^n$ to indicate $\log(z_{1/4} H)$ without the squares. In case of the first term $\mathtt{ZZZ}$, we can redistribute $z_1$ or $z_4$ arbitrarily between the logarithms just as we have done for the 2-loop case going from \eqref{eq:2_loop_step1} to \eqref{eq:2_loop_final_form}. We indicate this by writing $\mathtt{Z}_{14}^3$. Since the logarithm there contains $z_1 z_4 H^2$ and no squares, we do not pull out a factor of $2^3$. With this, we turn \eqref{eq:3_loop_step4} into
\begin{align}
    L_3 \, \hat{=}\, \mathtt{Z}_{14}^3 + 4 \mathtt{Z}_1^2 \mathtt{P} + 4 \mathtt{Z}_1\mathtt{PZ}_4 + 2 \mathtt{Z}_1\mathtt{P}^2 + 4 \mathtt{PZ}_4^2 + \mathtt{PZP} + 2 \mathtt{P}^2\mathtt{Z}_4 + \mathtt{P}^3. \label{eq:3_loop_step5}
\end{align}
Inspecting the above \eqref{eq:3_loop_step5}, we notice that, except for the special $\mathtt{PZP}$ term, all terms involve only a single $\mathtt{P}^n$ and no $\mathtt{Z}$ without an index. This indicates that only a single $p$ integral, and the $z_1$ and $z_4$ integrals are left.

Since $\mathtt{Z}_{1/4}$ has a label for which $z_j$ are inside the logarithms, we do not need to depend on the position in the string to encode this information. Therefore, we can move all $\mathtt{Z}_{1/4}$ to the front. Using associativity and logarithm manipulation, this then allows us to add $\mathtt{Z}_1 + \mathtt{Z}_4$ to $\log(z_1 z_4 H^2)$, which we can also encode as $\mathtt{Z}_{14}$. This is consistent with our earlier notation. Similarly, we can simplify
\begin{align}
    &4 \mathtt{Z}_1^2 \mathtt{P} + 4 \mathtt{Z}_1 \mathtt{Z}_4 \mathtt{P} + 4 \mathtt{Z}_4^2 \mathtt{P} \notag\\
    &\quad \ \hat{=} \, 4\log^2\big( z_1 H\big) \log\Big(\frac{p^2 + \vec k^2}{\mu^2} \Big) + 4\log\big( z_1 H\big) \log\big( z_4 H\big) \log\Big(\frac{p^2 + \vec k^2}{\mu^2} \Big) + 4\log^2\big( z_4 H\big) \log\Big(\frac{p^2 + \vec k^2}{\mu^2} \Big) \notag\\
    &\quad= 4 \Big( \log^2(z_1 H) + \log(z_1 H) \log(z_4 H) + \log^2(z_4 H) \Big) \log\Big(\frac{p^2 + \vec k^2}{\mu^2} \Big) \notag\\
    &\quad= 4 \log^2\big( z_1 z_4 H^2\big) \log\Big(\frac{p^2 + \vec k^2}{\mu^2} \Big) - 4 \log(z_1 H)\log(z_4 H) \log\Big(\frac{p^2 + \vec k^2}{\mu^2} \Big) \notag\\
    &\quad \ \hat{=} \, 4 \mathtt{Z}_{14}^2 \mathtt{P} - 4 \mathtt{Z}_1 \mathtt{Z}_4 \mathtt{P},
\end{align}
giving us
\begin{align}
    L_3 \, \hat{=}\, \mathtt{Z}_{14}^3 + 4 \mathtt{Z}_{14}^2 \mathtt{P} - 4 \mathtt{Z}_1\mathtt{Z}_4\mathtt{P} + 2 \mathtt{Z}_{14} \mathtt{P}^2 + \mathtt{P}^3 + \mathtt{PZP}. \label{eq:3_loop_step6}
\end{align}
Finally, we can identify something close to a third order binomial expansion, and combine
\begin{align}
    L_3 \, \hat{=}\, \big(\mathtt{Z}_{14} + \mathtt{P}\big)^3 + \mathtt{Z}_{14}^2 \mathtt{P} - 4 \mathtt{Z}_1\mathtt{Z}_4\mathtt{P} - \mathtt{Z}_{14} \mathtt{P}^2 + \mathtt{PZP}. \label{eq:3_loop_step7}
\end{align}
The $\mathtt{PZP}$ truly cannot be simplified using our rules because the $\mathtt{Z}$ is shielded from turning into $\mathtt{Z}_{1/4}$ by the $\mathtt{P}$ on both sides. The $\mathtt{P}$ also cannot be combined, since they are separated by $\mathtt{Z}$. Therefore, this term exhibits some kind of stalemate and has to be evaluated explicitly.

Remembering the factor of $\pi^2$ we have picked up from the $\delta$ integrals, we can now reconstruct
\begin{align}
    L_3 &= \pi^2\Big( \frac{\lambdaR H}{\mu} \Big)^4 C^3 \hspace{-1mm} \int\limits_0^\infty \hspace{-1mm} \frac{\d z_1}{(z_1 H)^2} \frac{\d z_4}{(z_4 H)^2} \bar G_0(z_1; z_4) \hspace{-1mm} \int\limits_{-\infty}^\infty \hspace{-1mm} \d p  \cos(p z_1) \cos(p z_4) \Bigg( \log^3\Big( z_1 z_4 H^2 \frac{p^2 + \vec k^2}{\mu^2} \Big) \label{eq:3_loop_obstruction}\\
    &\hspace{-2mm} + \Big(\log^2\big( z_1 z_4 H^2 \big) - 4\log(z_1 H) \log(z_4 H) \Big) \log\Big( \frac{p^2 + \vec k^2}{\mu^2} \Big) - \log\big( z_1 z_4 H^2\big) \log^2\Big(\frac{p^2 + \vec k^2}{\mu^2}\Big) \Bigg) + [\mathtt{PZP}]. \notag
\end{align}
The second line of the above result is the obstruction to the apparent pattern that the $N$-loop integrand is simply $\log^N\big(z_1 z_{N+1} (p^2+\vec k^2)/\mu^2\big)$, which holds for the 1- and 2-loop expressions, according to \eqref{eq:1_loop_log_final} and \eqref{eq:2_loop_final_form}. Evaluating all integrals in this obstruction is quite complicated, but it suffices to do the analysis numerically to see that the obstruction indeed is not 0.

\section{Alternative Integration Approach} \label{app:c}

Throughout this paper, we have repeatedly used the identity
\begin{align}
    \int\limits_0^\infty \d z \cos(z p) \cos(z q) = \pi \delta(p + q) + \pi \delta(p-q). \label{eq:app_cos_delta}
\end{align}
If we insist on the analytical regularization before our modifications in section \ref{sec:new_regularization}, then the $z^\kappa$ factors from the propagators \eqref{eq:feynman_rules_propagator-} and \eqref{eq:feynman_rules_propagator+} modify such integrals over $z$. We can now try to perform these integrals before expanding in $\kappa$, which should solve some of the problems we encounter otherwise. For example, the middle $z_2$ integral in the \texttt{PZP} term from appendix \ref{app:3_loop_details} does not actually converge when one tries to perform it first. This is an artifact of expanding in $\kappa$. Putting the $z$ dependence according to \eqref{eq:N_loop_pure_final} into \eqref{eq:app_cos_delta}, we end up with the new integral
\begin{align}
     \int\limits_0^\infty \hspace{-1mm} \d z z^{4\kappa} \cos(z p) \cos(z q) &= \frac{\Gamma(1 \hspace{-.5mm} - \hspace{-.5mm} 4\kappa)}{4} \Big( \big(i(p \hspace{-.5mm} + \hspace{-.5mm} q)\big)^{4\kappa-1} \hspace{-1mm} + \hspace{-1mm} \big(i(p \hspace{-.5mm} - \hspace{-.5mm} q)\big)^{4\kappa-1} \hspace{-1mm} + \hspace{-1mm} \big(\hspace{-1mm}-i(p \hspace{-.5mm} + \hspace{-.5mm} q)\big)^{4\kappa-1} \hspace{-1mm} + \hspace{-1mm} \big(\hspace{-1mm}-i(p \hspace{-.5mm} - \hspace{-.5mm} q)\big)^{4\kappa-1} \Big) \notag\\
    &= \frac{\Gamma(1\hspace{-.5mm} - \hspace{-.5mm}4\kappa) \sin(2\pi\kappa)}{2} \Big( |p+q|^{4\kappa-1} + |p-q|^{4\kappa-1} \Big) \\
    &= \pi \delta_{4\kappa}(p,q),
\end{align}
where we have introduced $\delta_{4\kappa}(p,q)$ such that
\begin{align}
    \delta_{4\kappa}(p,q) = \delta(p+q) + \delta(p-q) + \O(\kappa). \label{eq:delta_kappa_expectation}
\end{align}
The question is whether this new delta-like function can be integrated over $p$ to give something simple, in particular in combination with the functions that actually appear in the loop computations, see e.g.\ \eqref{eq:one_loop_cancellation}. There, the imprtant integral is
\begin{align}
    J(q) = \int\limits_{-\infty}^\infty \d p \delta_{4\kappa}(p,q) \frac{\cos(a p)}{(p^2+k^2)^{2\kappa}}.
\end{align}

\subsection*{Simplified Case}
We can start with $q=0$ and then generalize from there. In this case,
\begin{align}
    J_{q=0} = \int\limits_{-\infty}^\infty \d p \, \delta_{4\kappa}(p, 0) \frac{\cos(ap)}{(p^2+k^2)^{2\kappa}} &= \frac{\Gamma(1-4\kappa) \sin(2\pi\kappa)}{2\pi} 2 \int\limits_0^\infty \d p \, p^{4\kappa-1} \frac{\cos(ap)}{(p^2+k^2)^{2\kappa}} \\
    &= \frac{\Gamma(1-4\kappa)\sin(2\pi\kappa)}{\sqrt\pi \Gamma(2\kappa)} G^{21}_{13}\Big(\frac{a^2k^2}{4} \Big\rvert \begin{matrix}
        1-2\kappa \\ 0,0,\frac{1}{2}
    \end{matrix}\Big) \\
    &= \frac{\Gamma(1-4\kappa)\sin(2\pi\kappa)}{\sqrt\pi \Gamma(2\kappa)} \frac{1}{2\pi i} \int\limits_\gamma \d s \frac{\Gamma^2(-s) \Gamma(2\kappa + s)}{\Gamma\big( \frac{1}{2} + s \big)} \Big( \frac{a^2k^2}{4} \Big)^s \\
    &= - \frac{\Gamma(1-4\kappa)\sin(2\pi\kappa)}{\sqrt\pi \Gamma(2\kappa)} \sum\limits_{n=0}^\infty \frac{\Gamma(2\kappa + n)}{(n!)^2 \Gamma\big(\frac{1}{2} + n\big)} \Big( \frac{a^2k^2}{4} \Big)^n \notag\\
    &\hspace{.5cm} \times \Bigg( \psi(2\kappa + n) - \psi\Big( \frac{1}{2} + n \Big) - 2\psi(1 + n) + \log\Big( \frac{a^2k^2}{4} \Big) \Bigg) \label{eq:q_0_pole_sum}
\end{align}
We have used 3.773.4 of \cite{gradstejn_table_2009} to get the Meijer-G function in the second line, which can also be derived via the Mellin-Barnes representation of the term $1/(p^2+k^2)^{2\kappa}$, as explained later. Then, we have used the contour integral representation of the Meijer-G function with the contour $\gamma$ that runs from $-i \infty - \kappa$ to $i\infty - \kappa$, separating the poles of $\Gamma^2(-s)$ and $\Gamma(2\kappa + s)$. Since the $\Gamma(z)$ function falls off quickly for negative real $z$, we close the contour $\gamma$ towards the right and convert the integral to a sum over poles of second order, which can be found in the last line.

We can now expand this to first order in $\kappa$ to confirm the form \eqref{eq:delta_kappa_expectation}. Doing that, we can immediately observe that for $n \geq 1$, the terms under the sum are regular in $\kappa \rightarrow 0$, so they are of order $\kappa^2$ due to the coefficient in front of the sum. Only the $n=0$ term of the sum contains poles in $\kappa$, so we can write
\begin{align}
    J_{q=0} &= - \frac{\Gamma(1-4\kappa)\sin(2\pi\kappa)}{\sqrt\pi \Gamma(2\kappa)} \frac{\Gamma(2\kappa)}{\Gamma\big(\frac{1}{2}\big)} \Bigg( \psi(2\kappa) - \psi\Big( \frac{1}{2} \Big) - 2\psi(1) + \log\Big( \frac{a^2k^2}{4} \Big) \Bigg) + \O(\kappa^2) \\
    &= 1 - 2\kappa \log(a^2k^2) + \O(\kappa^2) \\
    &= (ak)^{-4\kappa} + \O(\kappa^2).
\end{align}
If \eqref{eq:delta_kappa_expectation} is true, we expect $J_{q=0} = k^{-4\kappa} + \O(\kappa)$. The Factor $a^{-4\kappa}$ is also somewhat expected, as $a$ has the same dimension as $z$ and therefore, it is the simplest coefficient that can correct the unit of $\delta_{4\kappa}$.

\subsection*{Full Case}
We can also compute $J$ the $q\neq0$ case. This works similar to the $q = 0$ case but with one added layer of complexity: In order to work with the shifted $p \pm q$, we use the Mellin-Barnes formula (3.26) from \cite{dubovyk_mellin-barnes_2022}
\begin{align}
    \frac{1}{(A+B)^\lambda} = \frac{1}{2\pi i} \frac{1}{\Gamma(z)} \int\limits_\gamma \d s \Gamma(\lambda + s) \Gamma(-s) A^s B^{-\lambda-s}.
\end{align}
The contour $\gamma$ separates the poles of $\Gamma(\lambda + s)$ and $\Gamma(-s)$. For example, when $\lambda$ is real and positive, a valid contour would be a straight line from $-i\infty - \lambda/2$ to $i\infty + \lambda/2$.

Applying this to $(i(p\pm q))^{4\kappa-1}$, we can see that this allows us to trade the shift by $q$ for an additional integral. The un-shifted integrand can then be integrated like in the first, special case $q=0$. We write
\begin{align}
    J &= \int\limits_{-\infty}^\infty \d p \, \delta_{4\kappa}(p, q) \frac{\cos(ap)}{(p^2+k^2)^{2\kappa}} \\
    &= \frac{\Gamma(1 \hspace{-.5mm} - \hspace{-.5mm} 4\kappa)}{4} \hspace{-1mm} \int\limits_{-\infty}^\infty \hspace{-1mm} \Big( \big(i(p \hspace{-.5mm} + \hspace{-.5mm} q)\big)^{4\kappa-1} \hspace{-1mm} + \hspace{-1mm} \big(i(p \hspace{-.5mm} - \hspace{-.5mm} q)\big)^{4\kappa-1} \hspace{-1mm} + \hspace{-1mm} \big(\hspace{-1mm}-i(p \hspace{-.5mm} + \hspace{-.5mm} q)\big)^{4\kappa-1} \hspace{-1mm} + \hspace{-1mm} \big(\hspace{-1mm}-i(p \hspace{-.5mm} - \hspace{-.5mm} q)\big)^{4\kappa-1} \Big) \frac{\cos(ap)}{(p^2+k^2)^{2\kappa}} \notag \\
    &= \frac{\Gamma(1-4\kappa)}{2\pi} \int\limits_{-\infty}^\infty \d p \, \frac{\cos(ap)}{\big(i(p+q)\big)^{1-4\kappa} \big(p^2+k^2\big)^{2\kappa}} + (q \rightarrow -q) \\
    &= \frac{\Gamma(1-4\kappa)}{2\pi} \int\limits_{-\infty}^\infty \d p \, \frac{\cos(ap)}{(p^2+k^2)^{2\kappa}} \frac{1}{\Gamma(1-4\kappa)} \int\limits_\gamma \frac{\d s}{2\pi i} \, \Gamma(1-4\kappa + s) \Gamma(-s) \frac{(iq)^s}{(ip)^{1 - 4\kappa + s}} \\
    &= \frac{1}{2\pi} \int\limits_\gamma \frac{\d s}{2\pi i} \, \Gamma(1-4\kappa + s) \Gamma(-s) (iq)^s \int\limits_{-\infty}^\infty \d p \, \frac{\cos(ap)}{(ip)^{1-4\kappa+s}(p^2+k^2)^{2\kappa}} \label{eq:full_J_after_mb}
\end{align}
We now turn to compute the $p$ integral
\begin{align}
    \int\limits_{-\infty}^\infty \d p \, \frac{\cos(ap)}{(ip)^{1-4\kappa+s}(p^2+k^2)^{2\kappa}} &= \int\limits_{-\infty}^0 \d p \, \frac{\cos(ap)}{(p^2+k^2)^{2\kappa}} \exp\Big(-(1-4\kappa+s)\underbrace{\log(ip)}_{=\log(p) + \frac{i\pi}{2}}\Big) \notag\\
    &+ \int\limits_0^\infty \d p \, \frac{\cos(ap)}{(p^2+k^2)^{2\kappa}} \exp\Big(-(1-4\kappa+s)\underbrace{\log(ip)}_{=\log(-p) - \frac{i\pi}{2}}\Big) \\
    &= \Big( \e^{i\frac{\pi}{2} (1-4\kappa+s)} + \e^{-i\frac{\pi}{2} (1-4\kappa+s)} \Big) \int\limits_0^\infty \d p \, \frac{\cos(ap)}{p^{1-4\kappa+s}(p^2+k^2)^{2\kappa}} \\
    &= 2\underbrace{\cos\Big(\frac{\pi}{2}(1-4\kappa+s)\Big)}_{=\sin\big(\frac{\pi}{2}(4\kappa-s)\big)} \frac{\sqrt\pi}{2\Gamma(2\kappa)} k^{-s} G^{21}_{13}\Bigg( \frac{a^2k^2}{4} \Bigg\rvert \begin{matrix}
        1-2\kappa+\frac{s}{2} \\ \frac{s}{2}, 0, \frac{1}{2}
    \end{matrix}\Bigg)
\end{align}
We can also write the Meijer G function in terms of more simple hypergeometric functions
\begin{align}
    &G^{21}_{13}\Bigg( z \Bigg\rvert \begin{matrix}
        1-2\kappa+\frac{s}{2} \\ \frac{s}{2}, 0, \frac{1}{2}
    \end{matrix}\Bigg) = \frac{\Gamma\big(-\frac{s}{2}\big)\Gamma(2\kappa)}{\Gamma\big(\frac{1}{2}+\frac{s}{s}\big)} z^\frac{s}{2} {}_1F_2\Big( \begin{matrix}
        2\kappa \\ 1+\frac{s}{2}, \frac{1}{2} + \frac{s}{2}
    \end{matrix} \Big\rvert z \Big) + \frac{\Gamma\big(\frac{s}{2}\big) \Gamma\big(2\kappa - \frac{s}{2}\big)}{\Gamma\big(\frac{1}{2}\big)} {}_1F_2\Big( \begin{matrix}
        2\kappa-\frac{s}{2} \\ 1-\frac{s}{2}, \frac{1}{2}
    \end{matrix} \Big\rvert z \Big) \notag \\
    &\hspace{5mm}= \Gamma\Big(-\frac{s}{2}\Big)\Gamma\Big(2\kappa\Big) \Gamma\Big(1+\frac{s}{2}\Big) z^\frac{s}{2} {}_1\tilde F_2\Big( \begin{matrix}
        2\kappa \\ 1+\frac{s}{2}, \frac{1}{2} + \frac{s}{2}
    \end{matrix} \Big\rvert z \Big) + \Gamma\Big(\frac{s}{2}\Big) \Gamma\Big(2\kappa - \frac{s}{2}\Big) \Gamma\Big(1-\frac{s}{2}\Big) {}_1\tilde F_2\Big( \begin{matrix}
        2\kappa-\frac{s}{2} \\ 1-\frac{s}{2}, \frac{1}{2}
    \end{matrix} \Big\rvert z \Big) \notag
\end{align}
with the regularized hypergeometric function
\begin{align}
    {}_1F_2\Big( \begin{matrix}
        a_1 \\ b_1, b_2
    \end{matrix} \Big\rvert z \Big) = \Gamma(b)\Gamma(c) {}_1 \tilde F_2 \Big( \begin{matrix}
        a_1 \\ b_1, b_2
    \end{matrix} \Big\rvert z \Big). \label{eq:hypergeometric_regularized}
\end{align}
Inserting this into our previous calculation \eqref{eq:full_J_after_mb}, we find
\begin{align}
    J &= \frac{1}{2\sqrt{\pi}} \int\limits_\gamma \frac{\d s}{2\pi i} \Gamma(1-4\kappa+s) \Gamma(-s) \sin\Big( \frac{\pi}{2} (4\kappa-s)\Big) \notag\\
    &\hspace{3cm} \times \Bigg( \Gamma\Big(-\frac{s}{2}\Big) \Gamma\Big(1 + \frac{s}{2}\Big) \Big(\frac{a^2k^2}{4}\Big)^\frac{s}{2} \Big(i\frac{q}{k}\Big)^s {}_1\tilde F_2\Big( \begin{matrix}
        2\kappa \\ 1+\frac{s}{2}, \frac{1}{2} + \frac{s}{2}
    \end{matrix} \Big\rvert \frac{a^2k^2}{4} \Big) \label{eq:full_j_integral}\\
    &\hspace{3.2cm} + \frac{1}{\Gamma(2\kappa)} \Gamma\Big(\frac{s}{2}\Big) \Gamma\Big(2\kappa - \frac{s}{2}\Big) \Gamma\Big(1-\frac{s}{2}\Big) (iq)^s {}_1\tilde F_2\Big( \begin{matrix}
        2\kappa-\frac{s}{2} \\ 1-\frac{s}{2}, \frac{1}{2}
    \end{matrix} \Big\rvert \frac{a^2k^2}{4} \Big) \Bigg) \notag
\end{align}
We close the contour to the right, so we have to sum over the poles of the gamma functions in which $s$ appears with a minus sign (and the one Gamma function in the second line that also has a pole at $s=0$). Note that the regularized hypergeometric function does not have any poles and that the sine function cancels some of the poles in the second line. Further, due to the mixing of $s$ and $s/2$ in the arguments of the various Gamma functions, poles at $s = n$ for even and odd $n$ are of second and first order, respectively, and have to be computed separately.

After a long manipulation we can write the sums over various types of poles of both the first and the second line in \eqref{eq:full_j_integral} as a single sum over $n \in \mathbb{N}_0$
\begin{align}
    J &= (-2\pi i) \Big(-\frac{i}{2\pi^\frac{3}{2}}\Big) \frac{\sin(2\pi\kappa)}{\Gamma(2\kappa)} \sum\limits_{n=0}^\infty \frac{\Gamma(1+2n-4\kappa)}{\Gamma(1+2n)} \Big(-\frac{q^2}{k^2}\Big)^n \notag\\
    &\quad \Bigg( (4z)^n \Gamma(2\kappa) \Big( \log(z) + \psi(2\kappa - n) + \partial_{b_1} + \partial_{b_2} \Big) {}_1\tilde F_2\Big( \begin{matrix}
        2\kappa \\ 1+n, \frac{1}{2} + n
    \end{matrix} \Big\rvert z\Big) \notag\\
    &\quad \quad + \Gamma(2\kappa-n) \Big( \partial_{a_1} + \partial_{b_1} \Big) {}_1 \tilde F_2 \Big( \begin{matrix}
        2\kappa-n \\ 1-n, \frac{1}{2}
    \end{matrix} \Big\rvert z\Big)\Bigg)
\end{align}
We have defined the abbreviation $z = a^2k^2/4$. The derivatives act on the arguments of the hypergeometric functions, according to the notation in \eqref{eq:hypergeometric_regularized}. From here, it is difficult to continue without inserting the definition of the hypergeometric function as
\begin{align}
    {}_1 \tilde F_2 \Big( \begin{matrix}
        a_1 \\ b_1, b_2
    \end{matrix} \Big\rvert z \Big) = \sum\limits_{l=0}^\infty \frac{(a_1)_l}{\Gamma(b_1)\Gamma(b_2)} \frac{z^l}{l!}
\end{align}
using the Pochhammer symbol
\begin{align}
    (x)_y = \frac{\Gamma(x+y)}{\Gamma(x)}.
\end{align}
The derivatives acting on the hypergeometric function than act on the Gamma functions inside, giving us
\begin{align}
    \partial_x \Gamma(x) &= \Gamma(x) \psi(x), \\
    \partial_x (x)_y &= \Big( \psi(x+y) - \psi(x) \Big) (x)_y.
\end{align}
All together, we find
\begin{align}
    J &= - \frac{1}{\sqrt\pi} \frac{\sin(2\pi\kappa)}{\Gamma(2\kappa)} \sum\limits_{n=0}^\infty \frac{\Gamma(1+2n-4\kappa)}{\Gamma(1+2n)} \Big(-\frac{q^2}{k^2}\Big)^n \notag\\
    &\quad \Bigg[ z^n \sum\limits_{l=0}^\infty \frac{\Gamma(2\kappa) (2\kappa)_l}{\Gamma\big(1+n+l\big)\Gamma\big(\frac{1}{2} + n + l\big)} \frac{z^l}{l!} \Bigg(\log z + \psi(2\kappa-n) - \psi(1+n+l) - \psi\Big(\frac{1}{2}+n+l\Big) \Bigg) \notag\\
    &\quad\quad + \sum\limits_{l=0}^\infty \frac{\Gamma(2\kappa-n)(2\kappa-n)_l}{\Gamma\big(1-n+l\big)\Gamma\big(\frac{1}{2} + l\big)} \frac{z^l}{l!} \Bigg( \psi(2\kappa-n+l) - \psi(2\kappa-n) - \psi(1-n+l) \Bigg) \Bigg] \label{eq:full_n_l_sums}
\end{align}
We can now combine the Pochhammer symbol in both sums with the Gamma. Then, we notice that the second sum includes a $1/\Gamma(1-n+l)$, which is $0$ for $l \leq n - 1$, thus deleting the first $n$ terms of the sum. There is one exception to this: The last Polygamma function also has the argument $\psi(1-n+l)$, which is infinite in the first $n$ terms of the sum. This is to be understood such that we take the limit of these particular arguments simultaneously, which gives us
\begin{align}
    \frac{\psi(1-n+l)}{\Gamma(1-n+l)} = (-1)^{n-l} (n-l-1)!
\end{align}
After separating out the first $n$ terms of the second $l$-sum in \eqref{eq:full_n_l_sums}, we can shift this sum using $l - n \rightarrow l$, such that it starts from $0$ again. The fraction with the Gamma functions in front together with the $z^l/l!$ in the second sum then align precisely with those in the first sum, allowing us to write both as one $l$-sum and to cancel the $\psi(2\kappa-n)$ term to get
\begin{align}
    J &= - \frac{1}{\sqrt\pi} \frac{\sin(2\pi\kappa)}{\Gamma(2\kappa)} \sum\limits_{n=0}^\infty \frac{\Gamma(1+2n-4\kappa)}{\Gamma(1+2n)} \Big(-\frac{q^2}{k^2}\Big)^n \notag\\
    &\quad \Bigg[ z^n \sum\limits_{l=0}^\infty \frac{\Gamma(2\kappa + l)}{\Gamma\big(\frac{1}{2}+n+l\big)} \frac{z^l}{l!(n+l)!} \Bigg( \log(z) + \psi(2\kappa+l) - \psi(1+l) - \psi(1+n+l) - \psi\Big(\frac{1}{2}+n+l\Big) \Bigg) \notag\\
    &\quad\quad -(-1)^n \sum\limits_{l=0}^{n-1} \frac{\Gamma(2\kappa + l - n)}{\Gamma\big(\frac{1}{2} + l\big)} \frac{(-z)^l}{l!} (n-l-1)! \Bigg]
\end{align}
Upon close inspection, we notice that the $n=0$ term of the outer $n$-sum is precisely the result from the $q=0$ case in \eqref{eq:q_0_pole_sum}. Note that the last line vanishes for $n=0$, since it is an empty sum. This is expected, since setting $q=0$ in the above equation yields $0$ for all terms with $n\neq0$ because it is a power series in $q^{2n}$.

The sums again seem impossible to evaluate exactly, but we can expand in $\kappa$, just like before. As in the $q=0$ case, only the $l=0$ term of the first $l$-sum has a pole in $\kappa$ and thus is of leading order. For the second $l$-sum, all terms have a pole. We find

\begin{align}
    J &= - \frac{4\pi\kappa^2}{\sqrt\pi} \Big(1 + 2\gamma_\mathrm{E} \kappa + \O(\kappa^2)\Big) \sum\limits_{n=0}^\infty \Big( 1 - 4\kappa \psi(1+2n) + \O(\kappa^2) \Big) \Big(-\frac{q^2}{k^2}\Big)^n \notag\\
    &\quad \Bigg[ z^n \Big( \frac{1}{2\kappa} - \gamma_\mathrm{E} + \O(\kappa) \Big) \frac{1}{\Gamma\big(\frac{1}{2}+n\big) n!} \Bigg( \log(z) - \frac{1}{2\kappa} -\gamma_\mathrm{E} + \O(\kappa) + \gamma_\mathrm{E} - \psi(1+n) - \psi\Big( \frac{1}{2} + n \Big) \Bigg) \notag\\
    &\quad\quad -  (-1)^n \sum\limits_{l=0}^{n-1} \frac{(-1)^{l+n}}{2\kappa \Gamma\big(\frac{1}{2}+n\big)} \frac{(-z)^l}{l!} \frac{(n-l-1)!}{(n-l)!} + \O\big(\kappa^0\big) \Bigg] \\
    &= - \frac{4\pi\kappa^2}{\sqrt\pi} \sum\limits_{n=0}^\infty \notag\\
    &\quad \Bigg[ \frac{1}{\sqrt\pi} (-q^2a^2)^n \frac{1}{(2n)!} \Bigg( -\frac{1}{4\kappa^2} + \frac{1}{2\kappa}\Big( \log\Big(\frac{a^2k^2}{4}\Big) \underbrace{- 2\psi(1+2n) - \psi(1+n) - \psi\Big( \frac{1}{2} + n \Big)}_{=+\log(4)} \Big) \Bigg) \notag\\
    &\quad\quad - \frac{1}{2\kappa \sqrt\pi} \Big(-\frac{q^2}{k^2}\Big)^n \sum\limits_{l=0}^{n-1}  \frac{(a^2k^2)^l}{(2l)!} \frac{1}{n-l} + \O(\kappa)\Bigg] \\
    &= \cos(aq) - 2\kappa \cos(aq) \log(a^2k^2) + 2\kappa \sum\limits_{n=0}^\infty \Big(-\frac{q^2}{k^2}\Big)^n \sum\limits_{l=0}^{n-1} \frac{(a^2k^2)^l}{(2l)!} \frac{1}{n-l} + \O(\kappa^2) \label{eq:j_almost_done}
\end{align}
We are left with only one infinite sum. This sum, however, actually is a Cauchy product (or "convolution") of two series
\begin{align}
    \sum\limits_{n=0}^\infty \Big(-\frac{q^2}{k^2}\Big)^n \sum\limits_{l=0}^{n-1} \frac{(a^2k^2)^l}{(2l)!} \frac{1}{n-l} &= \Big( \sum\limits_{n=0}^\infty \frac{(-q^2a^2)^n}{(2n)!} \Big) \Big( \sum\limits_{n=1}^\infty \frac{\big(-\frac{q^2}{k^2}\big)^n}{n} \Big) \\
    &= \cos(aq) \log\Big( 1 + \frac{q^2}{k^2}\Big).
\end{align}
Inserting this into \eqref{eq:j_almost_done}, we get the desired result
\begin{align}
    J = \cos(aq) \Big( 1 - 2\kappa \log\big(a^2(q^2+k^2)\big) \Big) + \O(\kappa^2) = \frac{\cos(aq)}{\big(a^2(q^2+k^2)\big)^{2\kappa}} + \mathcal{O}\big(\kappa^2\big).
\end{align}
Higher orders seem increasingly hard to compute, since there is no closed general expression of the $n$th derivative of the Gamma function, which would be needed for the $n$th order expansion in $\kappa$.

\subsection*{We Need $J$ to All Orders}
Unfortunately, it seems that we indeed need $J$ to all orders for the $N$-loop computation. This can be seen when we try to compute $N$-loop expression by integrating over the intermediate $z_j$ before expanding in $\kappa$. For this, we parametrize the higher orders of $J$ as
\begin{align}
    J = \frac{\cos(aq)}{\big(a^2(q^2+k^2)\big)^{2\kappa}} \Big( 1 + \kappa^2 f(p, a, q) \Big).
\end{align}
Inserting that into \eqref{eq:N_loop_renormalized} for the 2-loop case, we get
\begin{align}
    I_2 &= \lambda^3 K^2 \int\limits_0^\infty \frac{\d z_1}{z_1^{2+2\kappa}} \frac{\d z_2}{z_2^{4\kappa}} \frac{\d z_3}{z_3^{2+2\kappa}} \bar G_\kappa^4 \int\limits_{-\infty}^\infty \d p_1 \d p_2 \Gamma_\kappa^2 \frac{\cos(p_1 z_1)\cos(p_1 z_2) \cos(p_2 z_2) \cos(p_2 z_3)}{\big(p_1^2 + \vec k^2\big)^{2\kappa} \big(p_2^2 + \vec k^2\big)^{2\kappa}} \\
    &= \lambda^3 K^2 \Gamma_\kappa^2 \int\limits_0^\infty \frac{\d z_1}{z_1^{2+2\kappa}} \frac{\d z_3}{z_3^{2+2\kappa}} \bar G_\kappa^4 \int\limits_{-\infty}^\infty \d p_1 \d p_2 \frac{\cos(p_1 z_1) \cos(p_2 z_3)}{\big(p_1^2 + \vec k^2\big)^{2\kappa} \big(p_2^2 + \vec k^2\big)^{2\kappa}} \int\limits_0^\infty \frac{\d z_2}{z_2^{4\kappa}} \cos(p_1 z_2) \cos(p_2 z_2) \\
    &= \lambda^3 K^2 \Gamma_\kappa^2 \pi \int\limits_0^\infty \frac{\d z_1}{z_1^{2+2\kappa}} \frac{\d z_3}{z_3^{2+2\kappa}} \bar G_\kappa^4 \frac{1}{2} \Bigg( \int\limits_{-\infty}^\infty \d p_1 \frac{\cos(p_1 z_1)}{\big(p_1^2 + \vec k^2\big)^{2\kappa}} \hspace{-.5mm}\int\limits_{-\infty}^\infty \hspace{-.5mm} \d p_2 \, \delta_{4\kappa}(p_1, p_2) \frac{\cos(p_2 z_3)}{\big( p_2^2 + \vec k^2 \big)^{2\kappa}} + (p_1 \leftrightarrow p_2) \Bigg) \notag\\
    &= \lambda^3 K^2 \Gamma_\kappa^2 \pi \int\limits_0^\infty \frac{\d z_1}{z_1^2} \frac{\d z_3}{z_3^2} \bar G_\kappa^4 \frac{1}{2} \Bigg( \int\limits_{-\infty}^\infty \d p \frac{\cos(p z_1) \cos(p z_3)}{z_1^{2\kappa}z_3^{6\kappa} \big( p^2 + \vec k^2 \big)^{4\kappa}} \Big( 1 + \kappa^2 f(p, z_3, \vec k) \Big) + (z_1 \leftrightarrow z_3) \Bigg) \label{eq:2loop_J}
\end{align}
Since to first order in $\kappa$, we can again use \eqref{eq:cosine_delta} to change the indices of the $z_j$ variables in the denominator, after we expand
\begin{align}
    f(p,z,k) = f^{(0)}(p,z,k) + \kappa f^{(1)}(p,z,k) + \ldots,
\end{align}
we know that the first term $f^{(0)}(p,z,k) = 0$ has to be zero or a total derivative, because its contribution to the two loop result has to vanish.
Going to 3 loops - the first order where we do not have a result to compare to - we find
\begin{align}
    I_3 &= \lambda^4 K^3 \int\limits_0^\infty \frac{\d z_1}{z_1^{2+2\kappa}} \frac{\d z_2}{z_2^{4\kappa}} \frac{\d z_3}{z_3^{4\kappa}} \frac{\d z_4}{z_4^{2+2\kappa}} \bar G_\kappa^4 \int\limits_{-\infty}^\infty \d p_1 \d p_2 \d p_3 \Gamma_\kappa^3 \frac{\cos(p_1 z_1)\cos(p_1 z_2)\ldots\cos(p_3 z_3) \cos(p_3 z_4)}{\big(p_1^2 + \vec k^2\big)^{2\kappa} \big(p_2^2 + \vec k^2\big)^{2\kappa} \big(p_3^2 + \vec k^2\big)^{2\kappa}} \notag\\
    &= \lambda^4 K^3 \Gamma_\kappa^3 \int\limits_0^\infty \frac{\d z_1}{z_1^{2+2\kappa}} \frac{\d z_4}{z_4^{2+2\kappa}} \bar G_\kappa^4 \int\limits_{-\infty}^\infty \d p_1 \d p_2 \d p_3 \frac{\cos(p_1 z_1) \cos(p_3 z_4)}{\big(p_1^2 + \vec k^2\big)^{2\kappa} \big(p_2^2 + \vec k^2\big)^{2\kappa} \big(p_3^2 + \vec k^2\big)^{2\kappa}} \notag\\
    &\hspace{4cm} \times \int\limits_0^\infty  \frac{\d z_2}{z_2^{4\kappa}} \cos(p_1 z_2) \cos(p_2 z_2) \int\limits_0^\infty \frac{\d z_3}{z_3^{4\kappa}} \cos(p_2 z_3) \cos(p_3 z_3) \\
    &= \lambda^4 K^3 \pi^2 \Gamma_\kappa^3 \int\limits_0^\infty \frac{\d z_1}{z_1^{2+2\kappa}} \frac{\d z_4}{z_4^{2+2\kappa}} \bar G_\kappa^4 \int\limits_{-\infty}^\infty \d p_2 \frac{1}{\big(p_2^2 + \vec k^2\big)^{2\kappa}} \notag \\
    &\hspace{3cm} \times \underbrace{\int\limits_{-\infty}^\infty \d p_1 \, \delta_{4\kappa}(p_1, p_2) \frac{\cos(p_1 z_1) }{\big(p_1^2 + \vec k^2\big)^{2\kappa}}}_{=\frac{\cos(p_2 z_1)}{(z_1^2(p_2^2+\vec k^2))^{2\kappa}} (1 + \kappa^2 f(p_2, z_1, \vec k))} \underbrace{\int\limits_{-\infty}^\infty \d p_3 \, \delta_{4\kappa}(p_2,p_3) \frac{\cos(p_3 z_4)}{\big(p_3^2 + \vec k^2\big)^{2\kappa}}}_{=\frac{\cos(p_2 z_4)}{(z_4^2(p_2^2+\vec k^2))^{2\kappa}}(1 + \kappa^2 f(p_2, z_4, \vec k))} \\
    &= \lambda^4 K^3 \pi^2 \int\limits_0^\infty \frac{\d z_1}{z_1^2} \frac{\d z_4}{z_4^2} \bar G_\kappa^4 \int\limits_{-\infty}^\infty \d p \frac{\cos(p z_1)\cos(p z_4)}{\Big( z_1 z_4 \big(p^2 + \vec k^2\big)\Big)^{6\kappa}} \Gamma_\kappa^3 \Big( 1 + \kappa^2 \big( f(p, z_1, \vec k) + f(p, z_4, \vec k) \big) \Big)
\end{align}
Since the zeroth order in $\kappa$ of $f(p,z,k)$ is zero, only the first order can contribute, which in total comes with a factor of $\kappa^3$. The contribution to the interesting finite term of the diagram thus has to be multiplied with $1/(2\kappa)^3$ from $\Gamma_\kappa^3$ and contributes with an overall coefficient (ignoring the integrals and $z$-dependent terms that should be the same at all orders) of $\lambda_\mathrm{R}^4 K^3 \pi^2 / 8$. 

This same term also contributes to the counter-term with 2-loop topology. We get the contribution by replacing $\lambda^3 \rightarrow \lambda_\mathrm{R}^4 (-\pi K / (2\kappa))$ in \eqref{eq:2loop_J}. The contribution of $f^{(1)}(p,z,k)$ to the finite order again comes with the highest order pole in $\kappa$, namely $1/(4\kappa)^2$ from $\Gamma_\kappa^2$. However, because of the extra factor of $1/2$ in \eqref{eq:2loop_J} that comes from symmetrizing the terms in $z_1$ and $z_3$, the coefficient is off by a factor of $1/2$ when compared to the 3-loop contribution. This means that the $f^{(1)}(p,z,k)$ term does not conveniently cancel from the sum of $I_3$ and two-loop-topology counter term. The other counter-terms at this order do not contain $J$ and so there is no hope of them contributing to the cancellation.

We can therefore conclude that integrating out the $z_j$ before expanding in $\kappa$ is not feasible and therefore, the alternative regularization scheme is the best solution. In the language of this appendix, the alternative regularization scheme is the replacement
\begin{align}
    \delta_{4\kappa}(p,q) \rightarrow \delta_0(p,q).
\end{align}

\input{JHEP.bbl}

\end{document}

%% file: eta_wick_rotation.pdf_tex
\begingroup%
  \makeatletter%
  \providecommand\color[2][]{%
    \errmessage{(Inkscape) Color is used for the text in Inkscape, but the package 'color.sty' is not loaded}%
    \renewcommand\color[2][]{}%
  }%
  \providecommand\transparent[1]{%
    \errmessage{(Inkscape) Transparency is used (non-zero) for the text in Inkscape, but the package 'transparent.sty' is not loaded}%
    \renewcommand\transparent[1]{}%
  }%
  \providecommand\rotatebox[2]{#2}%
  \newcommand*\fsize{\dimexpr\f@size pt\relax}%
  \newcommand*\lineheight[1]{\fontsize{\fsize}{#1\fsize}\selectfont}%
  \ifx\svgwidth\undefined%
    \setlength{\unitlength}{354.33070866bp}%
    \ifx\svgscale\undefined%
      \relax%
    \else%
      \setlength{\unitlength}{\unitlength * \real{\svgscale}}%
    \fi%
  \else%
    \setlength{\unitlength}{\svgwidth}%
  \fi%
  \global\let\svgwidth\undefined%
  \global\let\svgscale\undefined%
  \makeatother%
  \begin{picture}(1,0.56)%
    \lineheight{1}%
    \setlength\tabcolsep{0pt}%
    \put(0,0){\includegraphics[width=\unitlength,page=1]{eta_wick_rotation.pdf}}%
    \put(0.36322501,0.45173749){\makebox(0,0)[t]{\lineheight{1.25}\smash{\begin{tabular}[t]{c}$\eta$\end{tabular}}}}%
    \put(0,0){\includegraphics[width=\unitlength,page=2]{eta_wick_rotation.pdf}}%
    \put(0.96322504,0.45173749){\makebox(0,0)[t]{\lineheight{1.25}\smash{\begin{tabular}[t]{c}$\eta$\end{tabular}}}}%
    \put(0,0){\includegraphics[width=\unitlength,page=3]{eta_wick_rotation.pdf}}%
    \put(0.05601912,0.34490492){\makebox(0,0)[t]{\lineheight{1.25}\smash{\begin{tabular}[t]{c}$\eta_\mathrm{T}$\end{tabular}}}}%
    \put(0,0){\includegraphics[width=\unitlength,page=4]{eta_wick_rotation.pdf}}%
    \put(0.6560191,0.34490657){\makebox(0,0)[t]{\lineheight{1.25}\smash{\begin{tabular}[t]{c}$\eta_\mathrm{T}$\end{tabular}}}}%
    \put(0.0560179,0.20296895){\makebox(0,0)[t]{\lineheight{1.25}\smash{\begin{tabular}[t]{c}$\eta_\mathrm{A}$\end{tabular}}}}%
    \put(0.65601788,0.20297056){\makebox(0,0)[t]{\lineheight{1.25}\smash{\begin{tabular}[t]{c}$\eta_\mathrm{A}$\end{tabular}}}}%
  \end{picture}%
\endgroup%

%% file: L0l16plot.pdf_tex
\begingroup%
  \makeatletter%
  \providecommand\color[2][]{%
    \errmessage{(Inkscape) Color is used for the text in Inkscape, but the package 'color.sty' is not loaded}%
    \renewcommand\color[2][]{}%
  }%
  \providecommand\transparent[1]{%
    \errmessage{(Inkscape) Transparency is used (non-zero) for the text in Inkscape, but the package 'transparent.sty' is not loaded}%
    \renewcommand\transparent[1]{}%
  }%
  \providecommand\rotatebox[2]{#2}%
  \newcommand*\fsize{\dimexpr\f@size pt\relax}%
  \newcommand*\lineheight[1]{\fontsize{\fsize}{#1\fsize}\selectfont}%
  \ifx\svgwidth\undefined%
    \setlength{\unitlength}{432bp}%
    \ifx\svgscale\undefined%
      \relax%
    \else%
      \setlength{\unitlength}{\unitlength * \real{\svgscale}}%
    \fi%
  \else%
    \setlength{\unitlength}{\svgwidth}%
  \fi%
  \global\let\svgwidth\undefined%
  \global\let\svgscale\undefined%
  \makeatother%
  \begin{picture}(1,0.83333333)%
    \lineheight{1}%
    \setlength\tabcolsep{0pt}%
    \put(0,0){\includegraphics[width=\unitlength,page=1]{L0l16plot.pdf}}%
    \put(0.12806015,0.04398555){\makebox(0,0)[t]{\lineheight{1.25}\smash{\begin{tabular}[t]{c}0\end{tabular}}}}%
    \put(0,0){\includegraphics[width=\unitlength,page=2]{L0l16plot.pdf}}%
    \put(0.74193064,0.04398555){\makebox(0,0)[t]{\lineheight{1.25}\smash{\begin{tabular}[t]{c}1\end{tabular}}}}%
    \put(0.435,0.02621163){\makebox(0,0)[t]{\lineheight{1.25}\smash{\begin{tabular}[t]{c}$u$\end{tabular}}}}%
    \put(0,0){\includegraphics[width=\unitlength,page=3]{L0l16plot.pdf}}%
    \put(0.1087963,0.0860393){\makebox(0,0)[rt]{\lineheight{1.25}\smash{\begin{tabular}[t]{r}0\end{tabular}}}}%
    \put(0,0){\includegraphics[width=\unitlength,page=4]{L0l16plot.pdf}}%
    \put(0.1087963,0.72136223){\makebox(0,0)[rt]{\lineheight{1.25}\smash{\begin{tabular}[t]{r}1\end{tabular}}}}%
    \put(0.07999494,0.4125){\rotatebox{90}{\makebox(0,0)[t]{\lineheight{1.25}\smash{\begin{tabular}[t]{c}$v$\end{tabular}}}}}%
    \put(0,0){\includegraphics[width=\unitlength,page=5]{L0l16plot.pdf}}%
    \put(0.435,0.74722222){\makebox(0,0)[t]{\lineheight{1.25}\smash{\begin{tabular}[t]{c}$\tilde L_0$ at $\lambda_\mathrm{R} = 16\pi^2$\end{tabular}}}}%
    \put(0.1595043,0.68994746){\color[rgb]{1,1,1}\rotatebox{-89.603981}{\makebox(0,0)[lt]{\lineheight{1.25}\smash{\begin{tabular}[t]{l}10\end{tabular}}}}}%
    \put(0.26350858,0.69190127){\color[rgb]{1,1,1}\rotatebox{-85.095766}{\makebox(0,0)[lt]{\lineheight{1.25}\smash{\begin{tabular}[t]{l}30\end{tabular}}}}}%
    \put(0.41109642,0.68249551){\color[rgb]{1,1,1}\rotatebox{-70.08279}{\makebox(0,0)[lt]{\lineheight{1.25}\smash{\begin{tabular}[t]{l}50\end{tabular}}}}}%
    \put(0.63224502,0.68412607){\color[rgb]{1,1,1}\rotatebox{-43.03024}{\makebox(0,0)[lt]{\lineheight{1.25}\smash{\begin{tabular}[t]{l}70\end{tabular}}}}}%
    \put(0,0){\includegraphics[width=\unitlength,page=6]{L0l16plot.pdf}}%
    \put(0.83203704,0.08287218){\makebox(0,0)[lt]{\lineheight{1.25}\smash{\begin{tabular}[t]{l}0\end{tabular}}}}%
    \put(0,0){\includegraphics[width=\unitlength,page=7]{L0l16plot.pdf}}%
    \put(0.83203704,0.69238242){\makebox(0,0)[lt]{\lineheight{1.25}\smash{\begin{tabular}[t]{l}75\end{tabular}}}}%
    \put(0,0){\includegraphics[width=\unitlength,page=8]{L0l16plot.pdf}}%
  \end{picture}%
\endgroup%

%% file: L1l16plot.pdf_tex
\begingroup%
  \makeatletter%
  \providecommand\color[2][]{%
    \errmessage{(Inkscape) Color is used for the text in Inkscape, but the package 'color.sty' is not loaded}%
    \renewcommand\color[2][]{}%
  }%
  \providecommand\transparent[1]{%
    \errmessage{(Inkscape) Transparency is used (non-zero) for the text in Inkscape, but the package 'transparent.sty' is not loaded}%
    \renewcommand\transparent[1]{}%
  }%
  \providecommand\rotatebox[2]{#2}%
  \newcommand*\fsize{\dimexpr\f@size pt\relax}%
  \newcommand*\lineheight[1]{\fontsize{\fsize}{#1\fsize}\selectfont}%
  \ifx\svgwidth\undefined%
    \setlength{\unitlength}{432bp}%
    \ifx\svgscale\undefined%
      \relax%
    \else%
      \setlength{\unitlength}{\unitlength * \real{\svgscale}}%
    \fi%
  \else%
    \setlength{\unitlength}{\svgwidth}%
  \fi%
  \global\let\svgwidth\undefined%
  \global\let\svgscale\undefined%
  \makeatother%
  \begin{picture}(1,0.83333333)%
    \lineheight{1}%
    \setlength\tabcolsep{0pt}%
    \put(0,0){\includegraphics[width=\unitlength,page=1]{L1l16plot.pdf}}%
    \put(0.12806015,0.04398555){\makebox(0,0)[t]{\lineheight{1.25}\smash{\begin{tabular}[t]{c}0\end{tabular}}}}%
    \put(0,0){\includegraphics[width=\unitlength,page=2]{L1l16plot.pdf}}%
    \put(0.74193064,0.04398555){\makebox(0,0)[t]{\lineheight{1.25}\smash{\begin{tabular}[t]{c}1\end{tabular}}}}%
    \put(0.435,0.02621163){\makebox(0,0)[t]{\lineheight{1.25}\smash{\begin{tabular}[t]{c}$u$\end{tabular}}}}%
    \put(0,0){\includegraphics[width=\unitlength,page=3]{L1l16plot.pdf}}%
    \put(0.1087963,0.0860393){\makebox(0,0)[rt]{\lineheight{1.25}\smash{\begin{tabular}[t]{r}0\end{tabular}}}}%
    \put(0,0){\includegraphics[width=\unitlength,page=4]{L1l16plot.pdf}}%
    \put(0.1087963,0.72136223){\makebox(0,0)[rt]{\lineheight{1.25}\smash{\begin{tabular}[t]{r}1\end{tabular}}}}%
    \put(0.07999494,0.4125){\rotatebox{90}{\makebox(0,0)[t]{\lineheight{1.25}\smash{\begin{tabular}[t]{c}$v$\end{tabular}}}}}%
    \put(0,0){\includegraphics[width=\unitlength,page=5]{L1l16plot.pdf}}%
    \put(0.435,0.74722222){\makebox(0,0)[t]{\lineheight{1.25}\smash{\begin{tabular}[t]{c}$\tilde L_1$ at $\lambda_\mathrm{R} = 16\pi^2$\end{tabular}}}}%
    \put(0.13472989,0.690213){\color[rgb]{1,1,1}\rotatebox{-89.573956}{\makebox(0,0)[lt]{\lineheight{1.25}\smash{\begin{tabular}[t]{l}30\end{tabular}}}}}%
    \put(0.2144663,0.68606032){\color[rgb]{1,1,1}\rotatebox{-84.330269}{\makebox(0,0)[lt]{\lineheight{1.25}\smash{\begin{tabular}[t]{l}90\end{tabular}}}}}%
    \put(0.33876781,0.69035808){\color[rgb]{1,1,1}\rotatebox{-69.683282}{\makebox(0,0)[lt]{\lineheight{1.25}\smash{\begin{tabular}[t]{l}130\end{tabular}}}}}%
    \put(0.46777981,0.68966875){\color[rgb]{1,1,1}\rotatebox{-54.38518}{\makebox(0,0)[lt]{\lineheight{1.25}\smash{\begin{tabular}[t]{l}150\end{tabular}}}}}%
    \put(0,0){\includegraphics[width=\unitlength,page=6]{L1l16plot.pdf}}%
    \put(0.83203704,0.08287218){\makebox(0,0)[lt]{\lineheight{1.25}\smash{\begin{tabular}[t]{l}0\end{tabular}}}}%
    \put(0,0){\includegraphics[width=\unitlength,page=7]{L1l16plot.pdf}}%
    \put(0.83203704,0.72413493){\makebox(0,0)[lt]{\lineheight{1.25}\smash{\begin{tabular}[t]{l}170\end{tabular}}}}%
    \put(0,0){\includegraphics[width=\unitlength,page=8]{L1l16plot.pdf}}%
  \end{picture}%
\endgroup%

%% file: L2l16plot.pdf_tex
\begingroup%
  \makeatletter%
  \providecommand\color[2][]{%
    \errmessage{(Inkscape) Color is used for the text in Inkscape, but the package 'color.sty' is not loaded}%
    \renewcommand\color[2][]{}%
  }%
  \providecommand\transparent[1]{%
    \errmessage{(Inkscape) Transparency is used (non-zero) for the text in Inkscape, but the package 'transparent.sty' is not loaded}%
    \renewcommand\transparent[1]{}%
  }%
  \providecommand\rotatebox[2]{#2}%
  \newcommand*\fsize{\dimexpr\f@size pt\relax}%
  \newcommand*\lineheight[1]{\fontsize{\fsize}{#1\fsize}\selectfont}%
  \ifx\svgwidth\undefined%
    \setlength{\unitlength}{432bp}%
    \ifx\svgscale\undefined%
      \relax%
    \else%
      \setlength{\unitlength}{\unitlength * \real{\svgscale}}%
    \fi%
  \else%
    \setlength{\unitlength}{\svgwidth}%
  \fi%
  \global\let\svgwidth\undefined%
  \global\let\svgscale\undefined%
  \makeatother%
  \begin{picture}(1,0.83333333)%
    \lineheight{1}%
    \setlength\tabcolsep{0pt}%
    \put(0,0){\includegraphics[width=\unitlength,page=1]{L2l16plot.pdf}}%
    \put(0.12806015,0.04398555){\makebox(0,0)[t]{\lineheight{1.25}\smash{\begin{tabular}[t]{c}0\end{tabular}}}}%
    \put(0,0){\includegraphics[width=\unitlength,page=2]{L2l16plot.pdf}}%
    \put(0.74193064,0.04398555){\makebox(0,0)[t]{\lineheight{1.25}\smash{\begin{tabular}[t]{c}1\end{tabular}}}}%
    \put(0.435,0.02621163){\makebox(0,0)[t]{\lineheight{1.25}\smash{\begin{tabular}[t]{c}$u$\end{tabular}}}}%
    \put(0,0){\includegraphics[width=\unitlength,page=3]{L2l16plot.pdf}}%
    \put(0.1087963,0.0860393){\makebox(0,0)[rt]{\lineheight{1.25}\smash{\begin{tabular}[t]{r}0\end{tabular}}}}%
    \put(0,0){\includegraphics[width=\unitlength,page=4]{L2l16plot.pdf}}%
    \put(0.1087963,0.72136223){\makebox(0,0)[rt]{\lineheight{1.25}\smash{\begin{tabular}[t]{r}1\end{tabular}}}}%
    \put(0.07999494,0.4125){\rotatebox{90}{\makebox(0,0)[t]{\lineheight{1.25}\smash{\begin{tabular}[t]{c}$v$\end{tabular}}}}}%
    \put(0,0){\includegraphics[width=\unitlength,page=5]{L2l16plot.pdf}}%
    \put(0.435,0.74722222){\makebox(0,0)[t]{\lineheight{1.25}\smash{\begin{tabular}[t]{c}$\tilde L_2$ at $\lambda_\mathrm{R} = 16\pi^2$\end{tabular}}}}%
    \put(0.14664551,0.70229405){\color[rgb]{1,1,1}\rotatebox{-88.740106}{\makebox(0,0)[lt]{\lineheight{1.25}\smash{\begin{tabular}[t]{l}200\end{tabular}}}}}%
    \put(0.23655359,0.70042316){\color[rgb]{1,1,1}\rotatebox{-82.613039}{\makebox(0,0)[lt]{\lineheight{1.25}\smash{\begin{tabular}[t]{l}400\end{tabular}}}}}%
    \put(0.33422172,0.68796233){\color[rgb]{1,1,1}\rotatebox{-72.232381}{\makebox(0,0)[lt]{\lineheight{1.25}\smash{\begin{tabular}[t]{l}500\end{tabular}}}}}%
    \put(0,0){\includegraphics[width=\unitlength,page=6]{L2l16plot.pdf}}%
    \put(0.83203704,0.08287218){\makebox(0,0)[lt]{\lineheight{1.25}\smash{\begin{tabular}[t]{l}0\end{tabular}}}}%
    \put(0,0){\includegraphics[width=\unitlength,page=7]{L2l16plot.pdf}}%
    \put(0.83203704,0.69645157){\makebox(0,0)[lt]{\lineheight{1.25}\smash{\begin{tabular}[t]{l}650\end{tabular}}}}%
    \put(0,0){\includegraphics[width=\unitlength,page=8]{L2l16plot.pdf}}%
  \end{picture}%
\endgroup%

%% file: L3l16plot.pdf_tex
\begingroup%
  \makeatletter%
  \providecommand\color[2][]{%
    \errmessage{(Inkscape) Color is used for the text in Inkscape, but the package 'color.sty' is not loaded}%
    \renewcommand\color[2][]{}%
  }%
  \providecommand\transparent[1]{%
    \errmessage{(Inkscape) Transparency is used (non-zero) for the text in Inkscape, but the package 'transparent.sty' is not loaded}%
    \renewcommand\transparent[1]{}%
  }%
  \providecommand\rotatebox[2]{#2}%
  \newcommand*\fsize{\dimexpr\f@size pt\relax}%
  \newcommand*\lineheight[1]{\fontsize{\fsize}{#1\fsize}\selectfont}%
  \ifx\svgwidth\undefined%
    \setlength{\unitlength}{432bp}%
    \ifx\svgscale\undefined%
      \relax%
    \else%
      \setlength{\unitlength}{\unitlength * \real{\svgscale}}%
    \fi%
  \else%
    \setlength{\unitlength}{\svgwidth}%
  \fi%
  \global\let\svgwidth\undefined%
  \global\let\svgscale\undefined%
  \makeatother%
  \begin{picture}(1,0.83333333)%
    \lineheight{1}%
    \setlength\tabcolsep{0pt}%
    \put(0,0){\includegraphics[width=\unitlength,page=1]{L3l16plot.pdf}}%
    \put(0.12806015,0.04398555){\makebox(0,0)[t]{\lineheight{1.25}\smash{\begin{tabular}[t]{c}0\end{tabular}}}}%
    \put(0,0){\includegraphics[width=\unitlength,page=2]{L3l16plot.pdf}}%
    \put(0.74193064,0.04398555){\makebox(0,0)[t]{\lineheight{1.25}\smash{\begin{tabular}[t]{c}1\end{tabular}}}}%
    \put(0.435,0.02621163){\makebox(0,0)[t]{\lineheight{1.25}\smash{\begin{tabular}[t]{c}$u$\end{tabular}}}}%
    \put(0,0){\includegraphics[width=\unitlength,page=3]{L3l16plot.pdf}}%
    \put(0.1087963,0.0860393){\makebox(0,0)[rt]{\lineheight{1.25}\smash{\begin{tabular}[t]{r}0\end{tabular}}}}%
    \put(0,0){\includegraphics[width=\unitlength,page=4]{L3l16plot.pdf}}%
    \put(0.1087963,0.72136223){\makebox(0,0)[rt]{\lineheight{1.25}\smash{\begin{tabular}[t]{r}1\end{tabular}}}}%
    \put(0.07999494,0.4125){\rotatebox{90}{\makebox(0,0)[t]{\lineheight{1.25}\smash{\begin{tabular}[t]{c}$v$\end{tabular}}}}}%
    \put(0,0){\includegraphics[width=\unitlength,page=5]{L3l16plot.pdf}}%
    \put(0.435,0.74722222){\makebox(0,0)[t]{\lineheight{1.25}\smash{\begin{tabular}[t]{c}$\tilde L_3$ at $\lambda_\mathrm{R} = 16\pi^2$\end{tabular}}}}%
    \put(0.14095278,0.7126247){\color[rgb]{1,1,1}\rotatebox{-88.808901}{\makebox(0,0)[lt]{\lineheight{1.25}\smash{\begin{tabular}[t]{l}1000\end{tabular}}}}}%
    \put(0.27085592,0.70071727){\color[rgb]{1,1,1}\rotatebox{-77.603203}{\makebox(0,0)[lt]{\lineheight{1.25}\smash{\begin{tabular}[t]{l}2000\end{tabular}}}}}%
    \put(0,0){\includegraphics[width=\unitlength,page=6]{L3l16plot.pdf}}%
    \put(0.83203704,0.08287218){\makebox(0,0)[lt]{\lineheight{1.25}\smash{\begin{tabular}[t]{l}0\end{tabular}}}}%
    \put(0,0){\includegraphics[width=\unitlength,page=7]{L3l16plot.pdf}}%
    \put(0.83203704,0.70619022){\makebox(0,0)[lt]{\lineheight{1.25}\smash{\begin{tabular}[t]{l}2600\end{tabular}}}}%
    \put(0,0){\includegraphics[width=\unitlength,page=8]{L3l16plot.pdf}}%
  \end{picture}%
\endgroup%

%% file: L0l1plot.pdf_tex
\begingroup%
  \makeatletter%
  \providecommand\color[2][]{%
    \errmessage{(Inkscape) Color is used for the text in Inkscape, but the package 'color.sty' is not loaded}%
    \renewcommand\color[2][]{}%
  }%
  \providecommand\transparent[1]{%
    \errmessage{(Inkscape) Transparency is used (non-zero) for the text in Inkscape, but the package 'transparent.sty' is not loaded}%
    \renewcommand\transparent[1]{}%
  }%
  \providecommand\rotatebox[2]{#2}%
  \newcommand*\fsize{\dimexpr\f@size pt\relax}%
  \newcommand*\lineheight[1]{\fontsize{\fsize}{#1\fsize}\selectfont}%
  \ifx\svgwidth\undefined%
    \setlength{\unitlength}{432bp}%
    \ifx\svgscale\undefined%
      \relax%
    \else%
      \setlength{\unitlength}{\unitlength * \real{\svgscale}}%
    \fi%
  \else%
    \setlength{\unitlength}{\svgwidth}%
  \fi%
  \global\let\svgwidth\undefined%
  \global\let\svgscale\undefined%
  \makeatother%
  \begin{picture}(1,0.83333333)%
    \lineheight{1}%
    \setlength\tabcolsep{0pt}%
    \put(0,0){\includegraphics[width=\unitlength,page=1]{L0l1plot.pdf}}%
    \put(0.12806015,0.04398555){\makebox(0,0)[t]{\lineheight{1.25}\smash{\begin{tabular}[t]{c}0\end{tabular}}}}%
    \put(0,0){\includegraphics[width=\unitlength,page=2]{L0l1plot.pdf}}%
    \put(0.74193064,0.04398555){\makebox(0,0)[t]{\lineheight{1.25}\smash{\begin{tabular}[t]{c}1\end{tabular}}}}%
    \put(0.435,0.02621163){\makebox(0,0)[t]{\lineheight{1.25}\smash{\begin{tabular}[t]{c}$u$\end{tabular}}}}%
    \put(0,0){\includegraphics[width=\unitlength,page=3]{L0l1plot.pdf}}%
    \put(0.1087963,0.0860393){\makebox(0,0)[rt]{\lineheight{1.25}\smash{\begin{tabular}[t]{r}0\end{tabular}}}}%
    \put(0,0){\includegraphics[width=\unitlength,page=4]{L0l1plot.pdf}}%
    \put(0.1087963,0.72136223){\makebox(0,0)[rt]{\lineheight{1.25}\smash{\begin{tabular}[t]{r}1\end{tabular}}}}%
    \put(0.07999494,0.4125){\rotatebox{90}{\makebox(0,0)[t]{\lineheight{1.25}\smash{\begin{tabular}[t]{c}$v$\end{tabular}}}}}%
    \put(0,0){\includegraphics[width=\unitlength,page=5]{L0l1plot.pdf}}%
    \put(0.435,0.74722222){\makebox(0,0)[t]{\lineheight{1.25}\smash{\begin{tabular}[t]{c}$\tilde L_0$ at $\lambda_\mathrm{R} = 1$\end{tabular}}}}%
    \put(0.18635062,0.69605488){\color[rgb]{1,1,1}\rotatebox{-88.909342}{\makebox(0,0)[lt]{\lineheight{1.25}\smash{\begin{tabular}[t]{l}0.1\end{tabular}}}}}%
    \put(0.27258202,0.69916278){\color[rgb]{1,1,1}\rotatebox{-84.461174}{\makebox(0,0)[lt]{\lineheight{1.25}\smash{\begin{tabular}[t]{l}0.2\end{tabular}}}}}%
    \put(0.38472408,0.69825789){\color[rgb]{1,1,1}\rotatebox{-73.642662}{\makebox(0,0)[lt]{\lineheight{1.25}\smash{\begin{tabular}[t]{l}0.3\end{tabular}}}}}%
    \put(0.53629387,0.69891358){\color[rgb]{1,1,1}\rotatebox{-54.646347}{\makebox(0,0)[lt]{\lineheight{1.25}\smash{\begin{tabular}[t]{l}0.4\end{tabular}}}}}%
    \put(0,0){\includegraphics[width=\unitlength,page=6]{L0l1plot.pdf}}%
    \put(0.83203704,0.08287218){\makebox(0,0)[lt]{\lineheight{1.25}\smash{\begin{tabular}[t]{l}0.0\end{tabular}}}}%
    \put(0,0){\includegraphics[width=\unitlength,page=7]{L0l1plot.pdf}}%
    \put(0.83203704,0.72453885){\makebox(0,0)[lt]{\lineheight{1.25}\smash{\begin{tabular}[t]{l}0.5\end{tabular}}}}%
    \put(0,0){\includegraphics[width=\unitlength,page=8]{L0l1plot.pdf}}%
  \end{picture}%
\endgroup%

%% file: resl1plot.pdf_tex
\begingroup%
  \makeatletter%
  \providecommand\color[2][]{%
    \errmessage{(Inkscape) Color is used for the text in Inkscape, but the package 'color.sty' is not loaded}%
    \renewcommand\color[2][]{}%
  }%
  \providecommand\transparent[1]{%
    \errmessage{(Inkscape) Transparency is used (non-zero) for the text in Inkscape, but the package 'transparent.sty' is not loaded}%
    \renewcommand\transparent[1]{}%
  }%
  \providecommand\rotatebox[2]{#2}%
  \newcommand*\fsize{\dimexpr\f@size pt\relax}%
  \newcommand*\lineheight[1]{\fontsize{\fsize}{#1\fsize}\selectfont}%
  \ifx\svgwidth\undefined%
    \setlength{\unitlength}{432bp}%
    \ifx\svgscale\undefined%
      \relax%
    \else%
      \setlength{\unitlength}{\unitlength * \real{\svgscale}}%
    \fi%
  \else%
    \setlength{\unitlength}{\svgwidth}%
  \fi%
  \global\let\svgwidth\undefined%
  \global\let\svgscale\undefined%
  \makeatother%
  \begin{picture}(1,0.83333333)%
    \lineheight{1}%
    \setlength\tabcolsep{0pt}%
    \put(0,0){\includegraphics[width=\unitlength,page=1]{resl1plot.pdf}}%
    \put(0.12715171,0.04398555){\makebox(0,0)[t]{\lineheight{1.25}\smash{\begin{tabular}[t]{c}0\end{tabular}}}}%
    \put(0,0){\includegraphics[width=\unitlength,page=2]{resl1plot.pdf}}%
    \put(0.74192612,0.04398555){\makebox(0,0)[t]{\lineheight{1.25}\smash{\begin{tabular}[t]{c}1\end{tabular}}}}%
    \put(0.435,0.02621163){\makebox(0,0)[t]{\lineheight{1.25}\smash{\begin{tabular}[t]{c}$u$\end{tabular}}}}%
    \put(0,0){\includegraphics[width=\unitlength,page=3]{resl1plot.pdf}}%
    \put(0.1087963,0.08509911){\makebox(0,0)[rt]{\lineheight{1.25}\smash{\begin{tabular}[t]{r}0\end{tabular}}}}%
    \put(0,0){\includegraphics[width=\unitlength,page=4]{resl1plot.pdf}}%
    \put(0.1087963,0.72135756){\makebox(0,0)[rt]{\lineheight{1.25}\smash{\begin{tabular}[t]{r}1\end{tabular}}}}%
    \put(0.07999494,0.4125){\rotatebox{90}{\makebox(0,0)[t]{\lineheight{1.25}\smash{\begin{tabular}[t]{c}$v$\end{tabular}}}}}%
    \put(0,0){\includegraphics[width=\unitlength,page=5]{resl1plot.pdf}}%
    \put(0.435,0.74722222){\makebox(0,0)[t]{\lineheight{1.25}\smash{\begin{tabular}[t]{c}$\mathrm{Re}\tilde L_\Sigma$ at $\lambda_\mathrm{R} = 1$\end{tabular}}}}%
    \put(0.18439639,0.69592713){\color[rgb]{1,1,1}\rotatebox{-88.794085}{\makebox(0,0)[lt]{\lineheight{1.25}\smash{\begin{tabular}[t]{l}0.1\end{tabular}}}}}%
    \put(0.26873671,0.69868257){\color[rgb]{1,1,1}\rotatebox{-84.547581}{\makebox(0,0)[lt]{\lineheight{1.25}\smash{\begin{tabular}[t]{l}0.2\end{tabular}}}}}%
    \put(0.37874434,0.69607718){\color[rgb]{1,1,1}\rotatebox{-74.046964}{\makebox(0,0)[lt]{\lineheight{1.25}\smash{\begin{tabular}[t]{l}0.3\end{tabular}}}}}%
    \put(0.52919863,0.69286319){\color[rgb]{1,1,1}\rotatebox{-55.197868}{\makebox(0,0)[lt]{\lineheight{1.25}\smash{\begin{tabular}[t]{l}0.4\end{tabular}}}}}%
    \put(0,0){\includegraphics[width=\unitlength,page=6]{resl1plot.pdf}}%
    \put(0.83203704,0.08287218){\makebox(0,0)[lt]{\lineheight{1.25}\smash{\begin{tabular}[t]{l}0.0\end{tabular}}}}%
    \put(0,0){\includegraphics[width=\unitlength,page=7]{resl1plot.pdf}}%
    \put(0.83203704,0.71618678){\makebox(0,0)[lt]{\lineheight{1.25}\smash{\begin{tabular}[t]{l}0.5\end{tabular}}}}%
    \put(0,0){\includegraphics[width=\unitlength,page=8]{resl1plot.pdf}}%
  \end{picture}%
\endgroup%

%% file: resl16plot.pdf_tex
\begingroup%
  \makeatletter%
  \providecommand\color[2][]{%
    \errmessage{(Inkscape) Color is used for the text in Inkscape, but the package 'color.sty' is not loaded}%
    \renewcommand\color[2][]{}%
  }%
  \providecommand\transparent[1]{%
    \errmessage{(Inkscape) Transparency is used (non-zero) for the text in Inkscape, but the package 'transparent.sty' is not loaded}%
    \renewcommand\transparent[1]{}%
  }%
  \providecommand\rotatebox[2]{#2}%
  \newcommand*\fsize{\dimexpr\f@size pt\relax}%
  \newcommand*\lineheight[1]{\fontsize{\fsize}{#1\fsize}\selectfont}%
  \ifx\svgwidth\undefined%
    \setlength{\unitlength}{432bp}%
    \ifx\svgscale\undefined%
      \relax%
    \else%
      \setlength{\unitlength}{\unitlength * \real{\svgscale}}%
    \fi%
  \else%
    \setlength{\unitlength}{\svgwidth}%
  \fi%
  \global\let\svgwidth\undefined%
  \global\let\svgscale\undefined%
  \makeatother%
  \begin{picture}(1,0.83333333)%
    \lineheight{1}%
    \setlength\tabcolsep{0pt}%
    \put(0,0){\includegraphics[width=\unitlength,page=1]{resl16plot.pdf}}%
    \put(0.12806015,0.04398555){\makebox(0,0)[t]{\lineheight{1.25}\smash{\begin{tabular}[t]{c}0\end{tabular}}}}%
    \put(0,0){\includegraphics[width=\unitlength,page=2]{resl16plot.pdf}}%
    \put(0.74193064,0.04398555){\makebox(0,0)[t]{\lineheight{1.25}\smash{\begin{tabular}[t]{c}1\end{tabular}}}}%
    \put(0.435,0.02621163){\makebox(0,0)[t]{\lineheight{1.25}\smash{\begin{tabular}[t]{c}$u$\end{tabular}}}}%
    \put(0,0){\includegraphics[width=\unitlength,page=3]{resl16plot.pdf}}%
    \put(0.1087963,0.0860393){\makebox(0,0)[rt]{\lineheight{1.25}\smash{\begin{tabular}[t]{r}0\end{tabular}}}}%
    \put(0,0){\includegraphics[width=\unitlength,page=4]{resl16plot.pdf}}%
    \put(0.1087963,0.72136223){\makebox(0,0)[rt]{\lineheight{1.25}\smash{\begin{tabular}[t]{r}1\end{tabular}}}}%
    \put(0.07999494,0.4125){\rotatebox{90}{\makebox(0,0)[t]{\lineheight{1.25}\smash{\begin{tabular}[t]{c}$v$\end{tabular}}}}}%
    \put(0,0){\includegraphics[width=\unitlength,page=5]{resl16plot.pdf}}%
    \put(0.435,0.74722222){\makebox(0,0)[t]{\lineheight{1.25}\smash{\begin{tabular}[t]{c}$\mathrm{Re}\tilde L_\Sigma$ at $\lambda_\mathrm{R} = 16\pi^2$\end{tabular}}}}%
    \put(0.14835464,0.70556287){\color[rgb]{1,1,1}\rotatebox{-89.172552}{\makebox(0,0)[lt]{\lineheight{1.25}\smash{\begin{tabular}[t]{l}−10\end{tabular}}}}}%
    \put(0.21564237,0.70006048){\color[rgb]{1,1,1}\rotatebox{-81.225334}{\makebox(0,0)[lt]{\lineheight{1.25}\smash{\begin{tabular}[t]{l}−30\end{tabular}}}}}%
    \put(0.295368,0.70238044){\color[rgb]{1,1,1}\rotatebox{-67.142824}{\makebox(0,0)[lt]{\lineheight{1.25}\smash{\begin{tabular}[t]{l}−40\end{tabular}}}}}%
    \put(0,0){\includegraphics[width=\unitlength,page=6]{resl16plot.pdf}}%
    \put(0.83203704,0.11658916){\makebox(0,0)[lt]{\lineheight{1.25}\smash{\begin{tabular}[t]{l}−45\end{tabular}}}}%
    \put(0,0){\includegraphics[width=\unitlength,page=7]{resl16plot.pdf}}%
    \put(0.83203704,0.72453885){\makebox(0,0)[lt]{\lineheight{1.25}\smash{\begin{tabular}[t]{l}0\end{tabular}}}}%
    \put(0,0){\includegraphics[width=\unitlength,page=8]{resl16plot.pdf}}%
  \end{picture}%
\endgroup%

%% file: resl16IMAGplot.pdf_tex
\begingroup%
  \makeatletter%
  \providecommand\color[2][]{%
    \errmessage{(Inkscape) Color is used for the text in Inkscape, but the package 'color.sty' is not loaded}%
    \renewcommand\color[2][]{}%
  }%
  \providecommand\transparent[1]{%
    \errmessage{(Inkscape) Transparency is used (non-zero) for the text in Inkscape, but the package 'transparent.sty' is not loaded}%
    \renewcommand\transparent[1]{}%
  }%
  \providecommand\rotatebox[2]{#2}%
  \newcommand*\fsize{\dimexpr\f@size pt\relax}%
  \newcommand*\lineheight[1]{\fontsize{\fsize}{#1\fsize}\selectfont}%
  \ifx\svgwidth\undefined%
    \setlength{\unitlength}{432bp}%
    \ifx\svgscale\undefined%
      \relax%
    \else%
      \setlength{\unitlength}{\unitlength * \real{\svgscale}}%
    \fi%
  \else%
    \setlength{\unitlength}{\svgwidth}%
  \fi%
  \global\let\svgwidth\undefined%
  \global\let\svgscale\undefined%
  \makeatother%
  \begin{picture}(1,0.83333333)%
    \lineheight{1}%
    \setlength\tabcolsep{0pt}%
    \put(0,0){\includegraphics[width=\unitlength,page=1]{resl16IMAGplot.pdf}}%
    \put(0.12806015,0.04398555){\makebox(0,0)[t]{\lineheight{1.25}\smash{\begin{tabular}[t]{c}0\end{tabular}}}}%
    \put(0,0){\includegraphics[width=\unitlength,page=2]{resl16IMAGplot.pdf}}%
    \put(0.74193064,0.04398555){\makebox(0,0)[t]{\lineheight{1.25}\smash{\begin{tabular}[t]{c}1\end{tabular}}}}%
    \put(0.435,0.02621163){\makebox(0,0)[t]{\lineheight{1.25}\smash{\begin{tabular}[t]{c}$u$\end{tabular}}}}%
    \put(0,0){\includegraphics[width=\unitlength,page=3]{resl16IMAGplot.pdf}}%
    \put(0.1087963,0.0860393){\makebox(0,0)[rt]{\lineheight{1.25}\smash{\begin{tabular}[t]{r}0\end{tabular}}}}%
    \put(0,0){\includegraphics[width=\unitlength,page=4]{resl16IMAGplot.pdf}}%
    \put(0.1087963,0.72136223){\makebox(0,0)[rt]{\lineheight{1.25}\smash{\begin{tabular}[t]{r}1\end{tabular}}}}%
    \put(0.07999494,0.4125){\rotatebox{90}{\makebox(0,0)[t]{\lineheight{1.25}\smash{\begin{tabular}[t]{c}$v$\end{tabular}}}}}%
    \put(0,0){\includegraphics[width=\unitlength,page=5]{resl16IMAGplot.pdf}}%
    \put(0.435,0.74722222){\makebox(0,0)[t]{\lineheight{1.25}\smash{\begin{tabular}[t]{c}$\mathrm{Im}\tilde L_\Sigma$ at $\lambda_\mathrm{R} = 16\pi^2$\end{tabular}}}}%
    \put(0.17978589,0.65558431){\color[rgb]{1,1,1}\rotatebox{86.993098}{\makebox(0,0)[lt]{\lineheight{1.25}\smash{\begin{tabular}[t]{l}0\end{tabular}}}}}%
    \put(0.25636299,0.70671919){\color[rgb]{1,1,1}\rotatebox{-85.324405}{\makebox(0,0)[lt]{\lineheight{1.25}\smash{\begin{tabular}[t]{l}−20\end{tabular}}}}}%
    \put(0.37226939,0.68286001){\color[rgb]{1,1,1}\rotatebox{-70.43322}{\makebox(0,0)[lt]{\lineheight{1.25}\smash{\begin{tabular}[t]{l}−40\end{tabular}}}}}%
    \put(0.54970494,0.68286724){\color[rgb]{1,1,1}\rotatebox{-48.297054}{\makebox(0,0)[lt]{\lineheight{1.25}\smash{\begin{tabular}[t]{l}−60\end{tabular}}}}}%
    \put(0,0){\includegraphics[width=\unitlength,page=6]{resl16IMAGplot.pdf}}%
    \put(0.83203704,0.09147785){\makebox(0,0)[lt]{\lineheight{1.25}\smash{\begin{tabular}[t]{l}−75\end{tabular}}}}%
    \put(0,0){\includegraphics[width=\unitlength,page=7]{resl16IMAGplot.pdf}}%
    \put(0.83203704,0.69862066){\makebox(0,0)[lt]{\lineheight{1.25}\smash{\begin{tabular}[t]{l}0\end{tabular}}}}%
    \put(0,0){\includegraphics[width=\unitlength,page=8]{resl16IMAGplot.pdf}}%
  \end{picture}%
\endgroup%

%% file: LambdascanWideplot.pdf_tex
\begingroup%
  \makeatletter%
  \providecommand\color[2][]{%
    \errmessage{(Inkscape) Color is used for the text in Inkscape, but the package 'color.sty' is not loaded}%
    \renewcommand\color[2][]{}%
  }%
  \providecommand\transparent[1]{%
    \errmessage{(Inkscape) Transparency is used (non-zero) for the text in Inkscape, but the package 'transparent.sty' is not loaded}%
    \renewcommand\transparent[1]{}%
  }%
  \providecommand\rotatebox[2]{#2}%
  \newcommand*\fsize{\dimexpr\f@size pt\relax}%
  \newcommand*\lineheight[1]{\fontsize{\fsize}{#1\fsize}\selectfont}%
  \ifx\svgwidth\undefined%
    \setlength{\unitlength}{432bp}%
    \ifx\svgscale\undefined%
      \relax%
    \else%
      \setlength{\unitlength}{\unitlength * \real{\svgscale}}%
    \fi%
  \else%
    \setlength{\unitlength}{\svgwidth}%
  \fi%
  \global\let\svgwidth\undefined%
  \global\let\svgscale\undefined%
  \makeatother%
  \begin{picture}(1,0.83333333)%
    \lineheight{1}%
    \setlength\tabcolsep{0pt}%
    \put(0,0){\includegraphics[width=\unitlength,page=1]{LambdascanWideplot.pdf}}%
    \put(0.125,0.04398466){\makebox(0,0)[t]{\lineheight{1.25}\smash{\begin{tabular}[t]{c}−300\end{tabular}}}}%
    \put(0,0){\includegraphics[width=\unitlength,page=2]{LambdascanWideplot.pdf}}%
    \put(0.25416667,0.04398466){\makebox(0,0)[t]{\lineheight{1.25}\smash{\begin{tabular}[t]{c}−200\end{tabular}}}}%
    \put(0,0){\includegraphics[width=\unitlength,page=3]{LambdascanWideplot.pdf}}%
    \put(0.38333335,0.04398466){\makebox(0,0)[t]{\lineheight{1.25}\smash{\begin{tabular}[t]{c}−100\end{tabular}}}}%
    \put(0,0){\includegraphics[width=\unitlength,page=4]{LambdascanWideplot.pdf}}%
    \put(0.51249999,0.04398466){\makebox(0,0)[t]{\lineheight{1.25}\smash{\begin{tabular}[t]{c}0\end{tabular}}}}%
    \put(0,0){\includegraphics[width=\unitlength,page=5]{LambdascanWideplot.pdf}}%
    \put(0.64166669,0.04398466){\makebox(0,0)[t]{\lineheight{1.25}\smash{\begin{tabular}[t]{c}100\end{tabular}}}}%
    \put(0,0){\includegraphics[width=\unitlength,page=6]{LambdascanWideplot.pdf}}%
    \put(0.77083333,0.04398466){\makebox(0,0)[t]{\lineheight{1.25}\smash{\begin{tabular}[t]{c}200\end{tabular}}}}%
    \put(0,0){\includegraphics[width=\unitlength,page=7]{LambdascanWideplot.pdf}}%
    \put(0.89999997,0.04398466){\makebox(0,0)[t]{\lineheight{1.25}\smash{\begin{tabular}[t]{c}300\end{tabular}}}}%
    \put(0.51249999,-0.0015667){\makebox(0,0)[t]{\lineheight{1.25}\smash{\begin{tabular}[t]{c}$\lambdaR$\end{tabular}}}}%
    \put(0,0){\includegraphics[width=\unitlength,page=8]{LambdascanWideplot.pdf}}%
    \put(0.1087963,0.12423699){\makebox(0,0)[rt]{\lineheight{1.25}\smash{\begin{tabular}[t]{r}−200\end{tabular}}}}%
    \put(0,0){\includegraphics[width=\unitlength,page=9]{LambdascanWideplot.pdf}}%
    \put(0.1087963,0.27725033){\makebox(0,0)[rt]{\lineheight{1.25}\smash{\begin{tabular}[t]{r}−100\end{tabular}}}}%
    \put(0,0){\includegraphics[width=\unitlength,page=10]{LambdascanWideplot.pdf}}%
    \put(0.1087963,0.43026366){\makebox(0,0)[rt]{\lineheight{1.25}\smash{\begin{tabular}[t]{r}0\end{tabular}}}}%
    \put(0,0){\includegraphics[width=\unitlength,page=11]{LambdascanWideplot.pdf}}%
    \put(0.1087963,0.58327696){\makebox(0,0)[rt]{\lineheight{1.25}\smash{\begin{tabular}[t]{r}100\end{tabular}}}}%
    \put(-0.02158608,0.41250055){\rotatebox{90}{\makebox(0,0)[t]{\lineheight{1.25}\smash{\begin{tabular}[t]{c}$\tilde L_\Sigma(u = v = 1)$\end{tabular}}}}}%
    \put(0,0){\includegraphics[width=\unitlength,page=12]{LambdascanWideplot.pdf}}%
    \put(0.51249999,0.74722222){\makebox(0,0)[t]{\lineheight{1.25}\smash{\begin{tabular}[t]{c}$\tilde L_\Sigma(\lambdaR)$ at $u = v = 1$\end{tabular}}}}%
  \end{picture}%
\endgroup%

%% file: LambdascanNarrowplot.pdf_tex
\begingroup%
  \makeatletter%
  \providecommand\color[2][]{%
    \errmessage{(Inkscape) Color is used for the text in Inkscape, but the package 'color.sty' is not loaded}%
    \renewcommand\color[2][]{}%
  }%
  \providecommand\transparent[1]{%
    \errmessage{(Inkscape) Transparency is used (non-zero) for the text in Inkscape, but the package 'transparent.sty' is not loaded}%
    \renewcommand\transparent[1]{}%
  }%
  \providecommand\rotatebox[2]{#2}%
  \newcommand*\fsize{\dimexpr\f@size pt\relax}%
  \newcommand*\lineheight[1]{\fontsize{\fsize}{#1\fsize}\selectfont}%
  \ifx\svgwidth\undefined%
    \setlength{\unitlength}{432bp}%
    \ifx\svgscale\undefined%
      \relax%
    \else%
      \setlength{\unitlength}{\unitlength * \real{\svgscale}}%
    \fi%
  \else%
    \setlength{\unitlength}{\svgwidth}%
  \fi%
  \global\let\svgwidth\undefined%
  \global\let\svgscale\undefined%
  \makeatother%
  \begin{picture}(1,0.83333333)%
    \lineheight{1}%
    \setlength\tabcolsep{0pt}%
    \put(0,0){\includegraphics[width=\unitlength,page=1]{LambdascanNarrowplot.pdf}}%
    \put(0.125,0.04398466){\makebox(0,0)[t]{\lineheight{1.25}\smash{\begin{tabular}[t]{c}−60\end{tabular}}}}%
    \put(0,0){\includegraphics[width=\unitlength,page=2]{LambdascanNarrowplot.pdf}}%
    \put(0.34642859,0.04398466){\makebox(0,0)[t]{\lineheight{1.25}\smash{\begin{tabular}[t]{c}−40\end{tabular}}}}%
    \put(0,0){\includegraphics[width=\unitlength,page=3]{LambdascanNarrowplot.pdf}}%
    \put(0.56785714,0.04398466){\makebox(0,0)[t]{\lineheight{1.25}\smash{\begin{tabular}[t]{c}−20\end{tabular}}}}%
    \put(0,0){\includegraphics[width=\unitlength,page=4]{LambdascanNarrowplot.pdf}}%
    \put(0.78928573,0.04398466){\makebox(0,0)[t]{\lineheight{1.25}\smash{\begin{tabular}[t]{c}0\end{tabular}}}}%
    \put(0.51249999,-0.0015667){\makebox(0,0)[t]{\lineheight{1.25}\smash{\begin{tabular}[t]{c}$\lambdaR$\end{tabular}}}}%
    \put(0,0){\includegraphics[width=\unitlength,page=5]{LambdascanNarrowplot.pdf}}%
    \put(0.1087963,0.12423699){\makebox(0,0)[rt]{\lineheight{1.25}\smash{\begin{tabular}[t]{r}−200\end{tabular}}}}%
    \put(0,0){\includegraphics[width=\unitlength,page=6]{LambdascanNarrowplot.pdf}}%
    \put(0.1087963,0.27725033){\makebox(0,0)[rt]{\lineheight{1.25}\smash{\begin{tabular}[t]{r}−100\end{tabular}}}}%
    \put(0,0){\includegraphics[width=\unitlength,page=7]{LambdascanNarrowplot.pdf}}%
    \put(0.1087963,0.43026366){\makebox(0,0)[rt]{\lineheight{1.25}\smash{\begin{tabular}[t]{r}0\end{tabular}}}}%
    \put(0,0){\includegraphics[width=\unitlength,page=8]{LambdascanNarrowplot.pdf}}%
    \put(0.1087963,0.58327696){\makebox(0,0)[rt]{\lineheight{1.25}\smash{\begin{tabular}[t]{r}100\end{tabular}}}}%
    \put(-0.01811386,0.41250055){\rotatebox{90}{\makebox(0,0)[t]{\lineheight{1.25}\smash{\begin{tabular}[t]{c}$\tilde L_\Sigma(u = v = 1)$\end{tabular}}}}}%
    \put(0,0){\includegraphics[width=\unitlength,page=9]{LambdascanNarrowplot.pdf}}%
    \put(0.51249999,0.74722222){\makebox(0,0)[t]{\lineheight{1.25}\smash{\begin{tabular}[t]{c}$\tilde L_\Sigma(\lambdaR)$ at $u = v = 1$\end{tabular}}}}%
  \end{picture}%
\endgroup%

%% file: cutoffresl16plot.pdf_tex
\begingroup%
  \makeatletter%
  \providecommand\color[2][]{%
    \errmessage{(Inkscape) Color is used for the text in Inkscape, but the package 'color.sty' is not loaded}%
    \renewcommand\color[2][]{}%
  }%
  \providecommand\transparent[1]{%
    \errmessage{(Inkscape) Transparency is used (non-zero) for the text in Inkscape, but the package 'transparent.sty' is not loaded}%
    \renewcommand\transparent[1]{}%
  }%
  \providecommand\rotatebox[2]{#2}%
  \newcommand*\fsize{\dimexpr\f@size pt\relax}%
  \newcommand*\lineheight[1]{\fontsize{\fsize}{#1\fsize}\selectfont}%
  \ifx\svgwidth\undefined%
    \setlength{\unitlength}{432bp}%
    \ifx\svgscale\undefined%
      \relax%
    \else%
      \setlength{\unitlength}{\unitlength * \real{\svgscale}}%
    \fi%
  \else%
    \setlength{\unitlength}{\svgwidth}%
  \fi%
  \global\let\svgwidth\undefined%
  \global\let\svgscale\undefined%
  \makeatother%
  \begin{picture}(1,0.83333333)%
    \lineheight{1}%
    \setlength\tabcolsep{0pt}%
    \put(0,0){\includegraphics[width=\unitlength,page=1]{cutoffresl16plot.pdf}}%
    \put(0.13093923,0.04110534){\makebox(0,0)[t]{\lineheight{1.25}\smash{\begin{tabular}[t]{c}0\end{tabular}}}}%
    \put(0,0){\includegraphics[width=\unitlength,page=2]{cutoffresl16plot.pdf}}%
    \put(0.74480972,0.04110531){\makebox(0,0)[t]{\lineheight{1.25}\smash{\begin{tabular}[t]{c}1\end{tabular}}}}%
    \put(0.43787907,0.02333255){\makebox(0,0)[t]{\lineheight{1.25}\smash{\begin{tabular}[t]{c}$u$\end{tabular}}}}%
    \put(0,0){\includegraphics[width=\unitlength,page=3]{cutoffresl16plot.pdf}}%
    \put(0.11167538,0.08316022){\makebox(0,0)[rt]{\lineheight{1.25}\smash{\begin{tabular}[t]{r}0\end{tabular}}}}%
    \put(0,0){\includegraphics[width=\unitlength,page=4]{cutoffresl16plot.pdf}}%
    \put(0.11167538,0.71848315){\makebox(0,0)[rt]{\lineheight{1.25}\smash{\begin{tabular}[t]{r}1\end{tabular}}}}%
    \put(0.08287402,0.40962092){\rotatebox{90}{\makebox(0,0)[t]{\lineheight{1.25}\smash{\begin{tabular}[t]{c}$v$\end{tabular}}}}}%
    \put(0,0){\includegraphics[width=\unitlength,page=5]{cutoffresl16plot.pdf}}%
    \put(0.43787907,0.74955122){\makebox(0,0)[t]{\lineheight{1.25}\smash{\begin{tabular}[t]{c}$\mathrm{Re}\tilde L^{\tilde\Lambda}_\Sigma$ at $\lambdaR = 16\pi^2$\end{tabular}}}}%
    \put(0.16568931,0.16030285){\color[rgb]{1,1,1}\rotatebox{-57.144999}{\makebox(0,0)[lt]{\lineheight{1.25}\smash{\begin{tabular}[t]{l}2\end{tabular}}}}}%
    \put(0.16700856,0.31963951){\color[rgb]{1,1,1}\rotatebox{-32.387421}{\makebox(0,0)[lt]{\lineheight{1.25}\smash{\begin{tabular}[t]{l}0\end{tabular}}}}}%
    \put(0.25979539,0.68890068){\color[rgb]{1,1,1}\rotatebox{-76.799943}{\makebox(0,0)[lt]{\lineheight{1.25}\smash{\begin{tabular}[t]{l}−20\end{tabular}}}}}%
    \put(0.38411292,0.69680181){\color[rgb]{1,1,1}\rotatebox{-69.036569}{\makebox(0,0)[lt]{\lineheight{1.25}\smash{\begin{tabular}[t]{l}−40\end{tabular}}}}}%
    \put(0.5467607,0.68801197){\color[rgb]{1,1,1}\rotatebox{-51.628564}{\makebox(0,0)[lt]{\lineheight{1.25}\smash{\begin{tabular}[t]{l}−60\end{tabular}}}}}%
    \put(0,0){\includegraphics[width=\unitlength,page=6]{cutoffresl16plot.pdf}}%
    \put(0.83203704,0.13536941){\makebox(0,0)[lt]{\lineheight{1.25}\smash{\begin{tabular}[t]{l}−70\end{tabular}}}}%
    \put(0,0){\includegraphics[width=\unitlength,page=7]{cutoffresl16plot.pdf}}%
    \put(0.83203704,0.70483723){\makebox(0,0)[lt]{\lineheight{1.25}\smash{\begin{tabular}[t]{l}0\end{tabular}}}}%
    \put(0,0){\includegraphics[width=\unitlength,page=8]{cutoffresl16plot.pdf}}%
  \end{picture}%
\endgroup%

%% file: JHEP.bbl
\providecommand{\href}[2]{#2}\begingroup\raggedright\endgroup